%% file: master.tex
\documentclass[runningheads, envcountsame,a4paper,orivec]{llncs}

\usepackage{amssymb}
\usepackage{latexsym}
\usepackage{stmaryrd}
\usepackage{amsmath}
\usepackage{amsfonts,bm}

\usepackage{microtype}
\usepackage{url}
\usepackage{mathtools}
\usepackage{setspace}
\usepackage{graphicx,graphics}
\usepackage{subcaption}

\usepackage[linesnumbered,ruled,vlined]{algorithm2e}
\SetKw{Break}{break}
\SetKw{DownTo}{down to}
\SetKw{Where}{where}

\usepackage{cite}
\usepackage{paralist}
\usepackage{etoolbox}

\usepackage{wrapfig}

\usepackage{times}

\usepackage{tikz}
\usetikzlibrary{automata,arrows,intersections,positioning,calc,shapes,patterns}

\input{commands}

\usepackage{thm-restate}

\usepackage{hyperref}

\title{Multi-objective Robust Strategy Synthesis for\\ Interval Markov Decision Processes}

\author{Ernst Moritz Hahn\inst{1,2}\and Vahid Hashemi\inst{1}\and Holger Hermanns\inst{1}\and \\ Morteza Lahijanian\inst{3}\and Andrea Turrini\inst{2} 
	} 
\institute{
	Saarland University, Saarland Informatics Campus, Saarbr{\"u}cken, Germany
	\and
	State Key Laboratory of Computer Science, ISCAS, Beijing, China
	\and
	Department of Computer Science, University of Oxford, Oxford, UK
}
\titlerunning{Multi-objective Robust Strategy Synthesis for Interval MDPs}
\authorrunning{E.\,M.~Hahn et~al.}

\begin{document}
\maketitle

\setcounter{footnote}{0}

\begin{abstract}
Interval Markov decision processes (\IMDP{}s) generalise
classical MDPs by having interval-valued transition probabilities.
They provide a powerful modelling tool for probabilistic systems with
an additional variation or uncertainty that prevents the knowledge of
the exact transition probabilities. In this paper, we consider the
problem of multi-objective robust strategy synthesis for interval MDPs, where
the aim is to find a robust strategy that guarantees the satisfaction
of multiple properties at the same time in face of the transition
probability uncertainty. We first show that this problem is
\pspace-hard. Then, we provide a value iteration-based decision
algorithm to approximate the Pareto set of achievable points.
We finally demonstrate the practical effectiveness of our proposed approaches by applying them on several 
case studies using a prototypical tool.

\end{abstract}

\input{introduction}
\input{preliminaries}
\input{multi-objective-queries}
\input{strategy-synthesis}

\input{case-studies.tex}

\input{conclusion}

%\clearpage

\iftrue
\subsubsection*{Acknowledgements.}
This work is supported 
by the ERC Advanced Investigators Grant 695614 (POWVER),
by the CAS/SAFEA International Partnership Program for Creative Research Teams, 
by the National Natural Science Foundation of China (Grants No.\ 61550110506 and 61650410658),
by the Chinese Academy of Sciences Fellowship for International Young Scientists, 
and 
by the CDZ project CAP (GZ 1023).
\fi

%
%\bigskip
%\noindent Thank you for reading these instructions carefully. We look forward to receiving your electronic files!
% \clearpage
\bibliography{biblio}
\bibliographystyle{abbrv}
\clearpage
\input{appendix}

\end{document}

%% file: commands.tex
\renewcommand{\epsilon}{\varepsilon}
\renewcommand{\phi}{\varphi}

\newenvironment{myproof}{\begin{proof}}{\qed\end{proof}}
\newtheorem{assumption}{Assumption}

\newenvironment{myexample}{\begin{example}}{\hfill$\lozenge$\end{example}}
\iftrue
	\usepackage{todonotes}
	\definecolor{dark_purple}{rgb}{0.1, 0.0, 0.4}
	\definecolor{dark_green}{rgb}{0.0,0.2,0.5}
	\definecolor{dark_red}{rgb}{0.85,0, 0}
	\newcommand{\ml}[1]{\todo[inline,color=teal!70!black!5,bordercolor=white]{\textcolor{teal!70!black!80}{\textbf{Morteza:}\textmd{\;#1}}}} 
	\newcommand{\hh}[1]{\todo[inline,color=red!5,bordercolor=white]{\textcolor{red}{\textbf{Holger:}\textmd{\;#1}}}}
	\newcommand{\vh}[1]{\todo[inline,color=purple!5,bordercolor=white]{\textcolor{purple}{\textbf{Vahid:}\textmd{\;#1}}}}
	\newcommand{\emh}[1]{\todo[inline,color=dark_purple!5,bordercolor=white]{\textcolor{dark_purple}{\textbf{Moritz:}\textmd{\;#1}}}}
	\newcommand{\at}[1]{\todo[inline,color=red!40!black!5,bordercolor=white,caption={AT}]{\textcolor{red!40!black!80}{\textbf{Andrea:}\textmd{\;#1}}}}      
	\newcommand{\td}[1]{\todo[color=dark_red!10,bordercolor=white]{\textcolor{dark_red}{\textbf{ToDo:}\textmd{\;#1}}}} 
\else
	\newcommand{\ml}[1]{}
	\newcommand{\hh}[1]{}
	\newcommand{\vh}[1]{}
	\newcommand{\emh}[1]{}
	\newcommand{\at}[1]{}
	\newcommand{\td}[1]{}
\fi

\DeclarePairedDelimiter{\abs}{\lvert}{\rvert}

\newcommand{\Prob}[1][\imdp]{\textnormal{Pr}_{#1}}

\DeclareMathOperator*{\opt}{opt}

\newcommand{\bigO}[1]{\mathcal{O}(#1)}

\newcommand{\paretoCurve}[1]{\mathcal{P}_{#1}}

\newcommand{\reals}{\mathbb{R}}
\newcommand{\posreals}{\reals_{\geq 0}}
\newcommand{\rationals}{\mathbb{Q}}

\newcommand{\intervalSet}{\mathbb{I}}

\newcommand{\interval}[2]{[#1,#2]}
\newcommand{\setcond}[2]{\{\, #1 \mid #2 \,\}}
\newcommand{\setnocond}[1]{\{#1\}}

\newcommand{\dwc}[1]{\mathnormal{#1\hspace{-.1cm}\downarrow}}

\newcommand{\boundary}[1]{\mathnormal{\partial\hspace{-.03cm}#1}}

\newcommand{\vct}[1]{\mathbf{#1}}

\newcommand{\gd}{\rho} % generic distribution
\newcommand{\dirac}[1]{\delta_{#1}}

\newcommand{\Disc}[1]{\mathrm{Disc}(#1)}

\newcommand{\Supp}[1]{\mathrm{Supp}(#1)}

\newcommand{\mdp}[1][M]{\mathsf{#1}}
\newcommand{\imdp}[1][M]{\mathcal{#1}}

\newcommand{\last}[1]{\mathit{last}(#1)}

\newcommand{\acronym}[1]{\textsl{#1}}

\newcommand{\MC}{\acronym{MC}}
\newcommand{\IMC}{\acronym{IMC}}

\newcommand{\MDP}{\acronym{MDP}}
\newcommand{\IMDP}{\acronym{IMDP}}

\newcommand{\sigmaalgebra}{\mathcal{B}}

\newcommand{\tra}[1]{{}\mathchoice%
	{\stackrel{#1}{\longrightarrow}}
	{\mathop {\smash\longrightarrow}\limits^{\vrule width 0pt height 0pt depth 4pt\smash{#1}}}
	{\stackrel{#1}{\longrightarrow}}
	{\stackrel{#1}{\longrightarrow}}
	{}}

\newcommand{\States}{S}
\newcommand{\GoalSet}{T}
\newcommand{\ActionSet}{\mathcal{A}}
\newcommand{\StateActionSet}[1]{\ActionSet(#1)}
\newcommand{\InitState}{\bar{s}}

\newcommand{\intTransitionProbability}{\mathit{I}}

\newcommand{\rew}{\mathtt{r}}
\newcommand{\naturals}{\mathbb{N}}
\newcommand{\extNaturals}{\mathbb{N}\cup\{\infty\}}
\newcommand{\unitRationals}{\interval{0}{1}\cap \rationals}

\newcommand{\accumPathRew}[3][\mathtt{r}]{{#1}[#3](#2)}
\newcommand{\weightedAccumPathRew}[4]{#1 \cdot {#2}[#3](#4)}

\newcommand{\fixeduncertainty}[2]{\mathfrak{h}^{#1}_{#2}}
\newcommand{\optimalfixeduncertainty}[2]{\bar{\mathfrak{h}}^{#1}_{#2}}
\newcommand{\elementfixeduncertainty}[3]{\mathfrak{h}^{#1}_{#2#3}}

\newcommand{\Pat}{\mathit{\xi}}
\newcommand{\Fpat}{\mathit{FPaths}}
\newcommand{\Infpat}{\mathit{IPaths}}
\newcommand{\Ifpat}[1]{\mathit{Paths}^{#1}}    
\newcommand{\pspace}{\textbf{PSPACE}}
\newcommand{\Str}{\Sigma}
\newcommand{\Env}{\Pi}

\newcommand{\env}{\pi}
\newcommand{\str}{\sigma}
\newcommand{\uncertainty}[2]{\mathcal{H}^{#1}_{#2}}
\newcommand{\size}[1]{\left\vert#1\right\vert}

\newcommand{\reachPred}[3]{[#2]^{\leq #1}_{#3}}
\newcommand{\inducedMC}[1][]{\imdp#1\mathord{\downharpoonright}_{\str#1}}
\newcommand{\rewPred}[4][]{[#2]_{\sim_{#1}#4}^{\leq #3}}
\newcommand{\rewPredMax}[3]{[#1]^{\leq #2}_{\leq #3}}
\newcommand{\rewPredMin}[3]{[#1]^{\leq #2}_{\geq #3}}
\newcommand{\ExpRew}[2]{\textnormal{$\mathit{ExpTot}_{\imdp}^{\str,#2}[#1]$}}
\newcommand{\boundReach}[2]{\Diamond^{\leq #1}\; #2}
\newcommand{\objective}[3]{[#2]^{\leq #1}_{#3}}
\newcommand{\inducedMCBar}{\bar{\imdp}\!\downharpoonright_{\bar{\str}}}
\newcommand{\ExpRewBar}[2]{\textnormal{$\mathit{ExpTot}_{\bar{\imdp}}^{\bar{\str},#2}[#1]$}}
\newcommand{\SEC}{E_\mathcal{M}}
\newcommand{\ExpRewK}[2]{\textnormal{$\mathit{ExpTot}_{\imdp}^{{\str_k},#2}[#1]$}}

\newcommand{\functionDot}{\,\cdot\,}

\newcommand{\paths}{\mathit{Paths}}

\newcommand{\quantity}{\mathit{qnt}}

\newcommand{\placeArrownw}[2]{\draw[->,thick,red] ($(-2.75,-2.75)+0.5*(#1,#2)+(+0.2,-0.2)$) -- ($(-2.75,-2.75)+0.5*(#1,#2)+(-0.2,+0.2)$);}
\newcommand{\placeArrowne}[2]{\draw[->,thick,red] ($(-2.75,-2.75)+0.5*(#1,#2)+(-0.2,-0.2)$) -- ($(-2.75,-2.75)+0.5*(#1,#2)+(+0.2,+0.2)$);}
\newcommand{\placeArrowsw}[2]{\draw[->,thick,red] ($(-2.75,-2.75)+0.5*(#1,#2)+(+0.2,+0.2)$) -- ($(-2.75,-2.75)+0.5*(#1,#2)+(-0.2,-0.2)$);}
\newcommand{\placeArrowse}[2]{\draw[->,thick,red] ($(-2.75,-2.75)+0.5*(#1,#2)+(-0.2,+0.2)$) -- ($(-2.75,-2.75)+0.5*(#1,#2)+(+0.2,-0.2)$);}

%% file: introduction.tex
\section{Introduction}
\label{sec:introduction}

\emph{Interval Markov Decision Processes} (\IMDP{}s) extend the classical \emph{Markov Decision Processes} (\MDP{}s) by including uncertainty over the transition probabilities.  
Instead of a single value for the probability of taking a transition, \IMDP{}s allow ranges of possible probability values given as closed intervals of the reals.  
Thereby, \IMDP{}s  provide a powerful modelling tool for probabilistic systems with an additional variation or uncertainty concerning the knowledge of exact transition probabilities.  
They are especially useful to represent realistic stochastic systems that, for instance, evolve in unknown environments with bounded behaviour or do not preserve the Markov property.

Since their introduction (under the name of bounded-parameter \MDP{}s)~\cite{GivanLD00}, \IMDP{}s have been receiving a lot of attention in the formal verification community.  
They are particularly viewed as the appropriate abstraction model for uncertain systems with large state spaces, including continuous dynamical systems, for the purpose of analysis, verification, and control synthesis. 
Several model checking and control synthesis techniques have been developed~\cite{WolffTM12,PuggelliLSS13,PuggelliThesis2014} causing a boost in the applications of \IMDP{}s, ranging from verification of continuous stochastic systems (e.g.,~\cite{Lahijanian:TAC:2015}) to robust strategy synthesis for robotic systems (e.g.,~\cite{WolffTM12,luna:wafr:2014,luna:icra:2014,luna:aaai:2014}).

In recent years, there has been an increasing interest in multi-objective strategy synthesis for probabilistic systems~\cite{RRS15,CMH06,EKNPY12,KNPQ13,Mouaddib04,OgryczakPW13,PernyWGH13,FKNPQ11,FKP12}.  
Here, the goal is first to provide a complete trade-off analysis of several, possibly conflicting, quantitative properties and then to synthesise a strategy that guarantees the user's desired behaviour. 
Such properties, for instance, ask to ``find a robot strategy that maximises $p_\text{safe}$, the probability of successfully completing  a track by safely maneuvering between obstacles, while minimising $t_\text{travel}$, the total expected travel time''.  
This example has competing objectives: maximising $p_\text{safe}$, which requires the robot to be conservative, and minimising $t_\text{travel}$, which causes the robot to be reckless.  
In such contexts, the interest is in the \emph{Pareto curve} of the possible solution points: 
the set of all pairs of $(p_\text{safe},t_\text{travel})$ for which an increase in the value of $p_\text{safe}$ must induce an increase in the value of $t_\text{travel}$, and vice versa.  
Given a point on the curve, the computation of the corresponding strategy is asked.

Existing multi-objective synthesis frameworks are limited to \MDP{} models of probabilistic systems.  
The algorithms use iterative methods (similar to value iteration) for the computation of the Pareto curve and rely on reductions to linear programming for strategy synthesis. 

As discussed above, \MDP{}s, however, are constrained to single-valued transition probabilities, posing severe limitations for many real-world systems.  

In this paper, we present a novel technique for multi-objective strategy synthesis for \IMDP{}s. 
Our aim is to synthesise a robust strategy that guarantees the satisfaction of the multi-objective property, despite the additional uncertainty over the transition probabilities in these models.  
Our approach views the uncertainty as making adversarial choices among the available transition probability distributions induced by the intervals, as the system evolves along state transitions. 
We refer to this as the \emph{controller synthesis} semantics. 
We first analyse the problem complexity, proving that it is \pspace-hard and then develop a value iteration-based decision algorithm to approximate the Pareto curve. 
In order to show the effectiveness of our approach, we present promising results on a variety of case studies, obtained by prototypical implementations of all algorithms.

\paragraph*{Related work.} 
Related work can be grouped into two main categories: 
uncertain Markov model formalisms and model checking/synthesis algorithms.

Firstly, from the modelling viewpoint, various probabilistic modelling formalisms with uncertain transitions are studied in the literature. 
Interval Markov Chains (\IMC{}s)~\cite{DBLP:conf/lics/JonssonL91,DBLP:journals/rc/KozineU02} or abstract Markov chains~\cite{DBLP:conf/spin/FecherLW06} extend standard discrete-time Markov Chains (\MC{}s) with interval uncertainties. 
They do not feature the non-deterministic choices of transitions. 
Uncertain \MDP{}s~\cite{PuggelliLSS13} allow more general sets of distributions to be associated with each transition, not only those described by intervals. 
They usually are restricted to \emph{rectangular uncertainty sets} requiring that the uncertainty is linear and independent for any two transitions of any two states. %Our general algorithmic approach can be easily adapted to this setting. 
Parametric \MDP{}s~\cite{HahnHZ11nfm}, to the contrary, allow such dependencies as every probability is described as a rational function of a finite set of global parameters.  
\IMDP{}s extend \IMC{}s by inclusion of non-determinism and are a subset of uncertain \MDP{}s and parametric \MDP{}s.

Secondly, from the side of algorithmic developments, several verification methods for uncertain Markov models have been proposed. 
The problems of computing reachability probabilities and expected total reward for \IMC{}s and \IMDP{}s were first investigated in~\cite{CHK13,WK08}. 
Then, several of their PCTL and LTL model checking algorithms were introduced in~\cite{BenediktLW13,ChatterjeeSH08,CHK13} and~\cite{WolffTM12,PuggelliLSS13,Lahijanian:TAC:2015}, respectively.  
As regards to strategy synthesis algorithms, the works in~\cite{HahnHZ11nfm,NilimG05} considered synthesis for parametric \MDP{}s and \MDP{}s with ellipsoidal uncertainty in the verification community.
In the control community, such synthesis problems were mostly studied for uncertain Markov models in~\cite{GivanLD00,NilimG05,WK08} with the aim to maximise expected finite-horizon (un)discounted rewards.  
All these works, however, consider solely single objective properties, and their extension to multi-objective synthesis is not trivial.

Multi-objective model checking of probabilistic models with respect to various quantitative objectives has been recently investigated in a few works. 
The works in~\cite{FKNPQ11,FKP12,KNPQ13,etessami2007} focused on multi-objective verification of ordinary \MDP{}s. 
In~\cite{chen2013stochastic}, these algorithms were extended to the more general models of 2-player stochastic games. 
These models, however, cannot capture the continuous uncertainty in the transition probabilities as \IMDP{}s do.  
For the purposes of synthesis though, it is possible to transform an \IMDP{} into a 2-player stochastic game; 
nevertheless, such a transformation raises an extra exponential factor to the complexity of the decision problem.  
This exponential blowup has been avoided in our setting.

\paragraph*{Structure of the paper.} 
We start with necessary preliminaries in Section~\ref{sec:Preliminaries}. 
In Section~\ref{sec:multiObRobustSyn}, we introduce multi-objective robust strategy synthesis for \IMDP{}s and present our solution 
approach. 
In Section~\ref{sec:case-studies}, we demonstrate our approach on two case studies and present promising experimental results.
Finally, in Section~\ref{sec:conclusion} we conclude the paper.

%% file: preliminaries.tex
\section{Mathematical Preliminaries}
\label{sec:Preliminaries}

For a set $X$, denote by $\Disc{X}$ the sets of discrete probability distributions over $X$. 
A discrete probability distribution $\gd$ is a function $\gd \colon X \to \posreals$ such that $\sum_{x \in X} \gd(x) = 1$; 
for $X' \subseteq X$, we write $\gd(X')$ for $\sum_{x \in X'} \gd(x)$. 
Given $\gd \in \Disc{X}$, we denote by $\Supp{\gd}$ the set $\setcond{x \in X}{\gd(x) >0}$, and by $\dirac{x}$, where $x \in X$, the \emph{Dirac} distribution such that $\gd(y) = 1$ for $y = x$, $0$ otherwise. 
For a probability distribution $\gd$, we also write $\gd = \setcond{(x, p_x)}{x \in X}$ where $p_{x}$ is the probability of $x$.

For a vector $\vct{x} \in \reals^{n}$ we denote by $x_{i}$, its $i$-th component, and we call $\vct{x}$ a \emph{weight vector} if $x_{i} \geq 0$ for all $i$ and $\sum_{i=1}^{n} x_{i} = 1$.
The Euclidean inner product $\vct{x} \cdot \vct{y}$ of two vectors $\vct{x}, \vct{y} \in \reals^{n}$ is defined as $\sum_{i=1}^{n} x_{i} \cdot y_{i}$.
For a set of vectors $S = \setnocond{\vct{s}_1, \dotsc, \vct{s}_{t}} \subseteq \reals^{n}$, we say that $\vct{s}$ is a \emph{convex combination} of elements of $S$, if $\vct{s} = \sum_{i=1}^{t} w_{i} \cdot \vct{s}_{i}$ for some weight vector $\vct{w} \in \reals^{t}$. 
Furthermore, we denote by $\dwc{S}$ the \emph{downward closure} of the convex hull of $S$ which is defined as $\dwc{S} = \setcond{\vct{y} \in \reals^{n}}{\text{$\vct{y} \leq \vct{z}$ for some convex combination $\vct{z}$ of $S$}}$. 
For a given convex set $X$, we say that a point $\vct{x} \in X$ is on the boundary of $X$, denoted by $\vct{x} \in \boundary{X}$, if
for every $\epsilon > 0$ there is a point $\vct{y} \notin X$ such that the Euclidean distance between $\vct{x}$ and $\vct{y}$ is at most $\epsilon$.
Given a downward closed set $X \in \reals^{n}$, for any $\vct{z} \in \reals^{n}$ such that $\vct{z} \in \boundary{X}$ or $\vct{z} \notin X$, there is a weight vector $\vct{w} \in \reals^{n}$ such that $\vct{w} \cdot \vct{z} \geq \vct{w} \cdot \vct{x}$ for all $\vct{x} \in X$~\cite{boyd2004}. 
We say that $\vct{w}$ separates $\vct{z}$ from $\dwc{X}$.

Given a set $Y \subseteq \reals^{k}$, we call a vector $\vct{y} \in Y$ \emph{Pareto optimal} in $Y$ if there does not exist a vector $\vct{z} \in Y$ such that $\vct{y} \leq \vct{z}$ and $\vct{y} \neq \vct{z}$.  We define the \emph{Pareto set} or \emph{Pareto curve} of $Y$ to be the set of all Pareto optimal vectors in $Y$, i.e., 
Pareto set $\mathcal{Y} = \setcond{\vct{y} \in Y}{\text{$\vct{y}$ is Pareto optimal}}$.

\subsection{Interval Markov Decision Processes}
\label{subsec:intervalMDPs}

We now define \emph{Interval Markov Decision Processes}~(\IMDP{}s) as an extension of \MDP{}s, which allows for the inclusion of transition probability uncertainties as \emph{intervals}. 
\IMDP{}s belong to the family of uncertain \MDP{}s and allow to describe a set of \MDP{}s with identical (graph) structures that differ in distributions associated with transitions. 
Formally, 
\begin{definition}[\IMDP{}s]
\label{def:imdp}
	An \emph{Interval Markov Decision Process} (\IMDP{}) $\imdp$ is a tuple $(\States,\InitState, \ActionSet, \intTransitionProbability)$, where 
	$\States$ is a finite set of \emph{states}, 
	$\InitState \in \States$ is the initial state, 
	$\ActionSet$ is a finite set of \emph{actions}, 
	and 
	$\intTransitionProbability \colon \States \times \ActionSet \times \States \to \intervalSet \cup \setnocond{0}$ is an \emph{interval transition probability function with $\intervalSet = \setcond{ \interval{a}{b}}{ \interval{a}{b} \subseteq (0,1]}$}. 
\end{definition}
We denote the set of available actions at state $s \in \States$ by $\StateActionSet{s}$. 
Furthermore, for each state $s$ and action $a \in \StateActionSet{s}$, we write $s \tra{a} \fixeduncertainty{a}{s}$ if $\fixeduncertainty{a}{s} \in \Disc{\States}$ is a \emph{feasible distribution}, i.e. for each state $s' \in S$ we have $\elementfixeduncertainty{a}{s}{s'} = \fixeduncertainty{a}{s}(s') \in \intTransitionProbability(s,a,s')$.
We denote by $\uncertainty{a}{s}=\setcond{\fixeduncertainty{a}{s}}{s \tra{a} \fixeduncertainty{a}{s}}$ the set of feasible distributions for state $s$ and action $a$ and we require that $\uncertainty{a}{s}$ is non-empty for each state $s$ and action $a \in \StateActionSet{s}$.

\begin{remark}\label{remark:sizeIMDP}
The size of a given $\imdp$ is determined as follows. 
Let $\size{\States}$ denote the number of states in $\imdp$. 
Then each state has $\bigO{\size{\ActionSet}}$ actions and at most $\bigO{\size{\ActionSet} \size{\States}}$ transitions, each of which is associated with a probability interval. 
Therefore, the overall size of $\imdp$ denoted by $\size{\imdp}$ is in {$\bigO{\size{\ActionSet} \size{\States}^2}$}. 
\end{remark}

The formal semantics of an \IMDP{} is as follows. A \emph{path} in $\imdp$ is a finite or infinite sequence of states in the form
 $\Pat = s_{0} \xrightarrow{\elementfixeduncertainty{a_0}{s_{0}}{s_1}} s_{1} \xrightarrow{\elementfixeduncertainty{a_1}{s_{1}}{s_2}} s_2 \cdots$, 
where $s_{0} = \InitState$ and for each $i \geq 0$, $s_{i} \in \States$, $a_{i} \in \StateActionSet{s_{i}}$, the transition probability $\elementfixeduncertainty{a_{i}}{s_{i}}{s_{i+1}} > 0$. 
Path $\Pat$ can be finite or infinite. 
The sets of all finite and infinite paths in $\imdp$ are denoted by $\Fpat$ and $\Infpat$, respectively. 
The $i$-th state and action along the path $\Pat$ are denoted by $\Pat[i]$ and $\Pat(i)$, respectively. 
For a finite path $\Pat \in \Fpat$, let $\last{\Pat}$ indicate its last state. 
Moreover, let $\Ifpat{\Pat} = \setcond{\Pat\Pat'}{\Pat' \in \Infpat}$ denote the set of infinite paths with the prefix $\Pat \in \Fpat$ which is also as known the \emph{cylinder} set of $\Pat$.

The nondeterministic choices between available actions and feasible distributions present in an \IMDP{} are resolved by strategies and natures, respectively. 
Formally, 
\begin{definition}[Strategy and Nature in \IMDP{}s]
\label{def:strategy}
Given an \IMDP{} $\imdp$, a \emph{strategy}
is a function $\str \colon \Fpat \to \Disc{\ActionSet}$ that to each finite path $\Pat$ assigns a distribution over the set of actions enabled by the last state of $\Pat$, that is, $\str(\Pat) \in \Disc{\ActionSet(\last{\Pat}}$. 
A \emph{nature} is a function $\env \colon \Fpat \times \ActionSet \to \Disc{\States}$ that to each finite path $\Pat$ and action $a \in \ActionSet(\last{\Pat})$ assigns a feasible distribution, i.e. an element of $\uncertainty{a}{s}$ where $s = \last{\Pat}$.
The sets of all strategies and all natures are denoted by $\Str$ and $\Env$, respectively. 
\end{definition}
A strategy $\str$ is said to be \emph{deterministic} if $\str(\Pat) = \dirac{a}$ for all finite paths $\Pat$ and some $a \in \ActionSet(\last{\Pat})$. 
Similarly, a nature is said to be \emph{deterministic} if $\env(\Pat,a) = \dirac{\fixeduncertainty{a}{\last{\Pat}}}$ for all finite paths $\Pat$, all $a \in \ActionSet(\last{\Pat})$, and some $\fixeduncertainty{a}{\last{\Pat}} \in \uncertainty{a}{\last{\Pat}}$. 
Furthermore, a strategy $\str$ is \emph{Markovian} if it depends only on $\last{\Pat}$, e.g., for each $\Pat, \Pat' \in \Fpat$, if $\last{\Pat} = \last{\Pat'}$, then $\str(\Pat) = \str(\Pat')$, and similarly for a nature $\env$. 
Given a finite path $\Pat$ of an \IMDP{}, a strategy $\str$, and a nature $\env$, the system evolution proceeds as follows. 
First, an action $a \in \StateActionSet{s_{i}}$, where $s_{i} = \last{\Pat}$, is chosen according to $\str(\Pat)$. 
Then, $\env$ resolves the uncertainties and chooses one feasible distribution $\fixeduncertainty{a}{s_{i}} \in \uncertainty{a}{s_{i}}$. 
Finally, the next state $s_{i+1}$ is chosen randomly according to $\fixeduncertainty{a}{s_{i}}$, and path $\Pat$ is appended by $s_{i+1}$.

For a strategy $\str$ and a nature $\env$, let $\Prob^{\str,\env}$ denote the unique probability measure over $(\Infpat, \sigmaalgebra)$\footnote{Here, $\sigmaalgebra$ is the standard $\sigma$-algebra over $\Infpat$ generated from the set of all cylinder sets $\setcond{\Ifpat{\Pat}}{\Pat \in \Fpat}$. 
The unique probability measure is obtained by the application of the extension theorem (see, e.g.~\cite{Billingsley1979}).
} 
such that $\Prob^{\str,\env}[\Ifpat{s'}] = \dirac{\InitState}(s')$ and the probability $\Prob^{\str,\env}[\Ifpat{\Pat s'}]$ of traversing a finite path $\Pat s'$ equals $\Prob^{\str,\env}[\Ifpat{\Pat}] \cdot \sum_{a \in \ActionSet(\last{\Pat})} \str(\Pat)(a) \cdot \env(\Pat,a)(s')$.

In order to model additional quantitative measures of an \IMDP{}, we associate a reward to actions available in each state. This is done by introducing a \emph{reward structure}:
\begin{definition}[Reward Structure]
\label{def:reward-structure}
A \emph{reward structure} for an \IMDP{} is a function $\rew \colon \States \times \ActionSet \to\reals$ that assigns to each state-action pair $(s,a)$, where $s \in \States$ and $a \in \ActionSet(s)$, a reward $\rew(s, a) \in \reals$. Given a (possibly infinite) path $\Pat$ and a step number $k \in \extNaturals$, the total accumulated reward in $k$ steps for $\Pat$ over $\rew$ is $\accumPathRew{\Pat}{k} = \sum_{i=0}^{k-1} \rew(\Pat[i],\Pat(i))$. 
\end{definition}
Note that we allow negative rewards in this definition, but that due to later assumptions their use is restricted.

% \begin{figure}[t]
\begin{wrapfigure}[13]{right}{50mm}
	\centering
	\vskip-7mm
	\begin{tikzpicture}[->, auto, >=stealth', semithick, baseline, 
		state/.style={draw, circle, inner sep=0, text centered, minimum size=6mm},
		trannode/.style={draw, circle, fill, minimum size=1mm, inner sep=0mm},
		prob/.style={font=\scriptsize,sloped, above}
		]
		\path[use as bounding box] (0.75,1.5) rectangle (4.8,5.8);
		\node[state, initial above, initial text={}] (0) at (3,5) {\vphantom{$t$}$s$};
		\node[trannode] (01) at ($(0) + (-0.75,-1)$) {};
		\node[trannode] (02) at ($(0) + (0.75,-1)$) {};
		\node[state] (2) at ($(01) + (-0.75,-1.5)$) {$t$};
		\node[state] (3) at ($(02) + (0.75,-1.5)$) {\vphantom{$t$}$u$};
		\draw
		(0) edge [-] node[left=3pt] {\scriptsize $a,\underline{3}$} (01)
		(0) edge [-] node[right=3pt] {\scriptsize $b,\underline{1}$} (02)
		(01) edge [bend right=10] node[prob] {$\interval{\frac{1}{3}}{\frac{2}{3}}$} (2)
		(01) edge [bend left=10] node[prob, below, near end] {$\interval{\frac{1}{10}}{1}$} (3)
		(02) edge [bend right=10] node[prob, below, near end] {$\interval{\frac{2}{5}}{\frac{3}{5}}$} (2)
		(02) edge [bend left=10] node[prob] {$\interval{\frac{1}{4}}{\frac{2}{3}}$} (3)
		(2) edge [loop below] node[below] {\scriptsize $a,\underline{0}$} node[left, near end] {\scriptsize $\interval{1}{1}$}(2)
		(3) edge [loop below] node[below] {\scriptsize $b,\underline{0}$} node[left, near end] {\scriptsize $\interval{1}{1}$} (3)
		;
	\end{tikzpicture}
	\vspace*{-1.5cm}
	\caption{An example of \IMDP{}.}
	\label{fig:uncertainMDP}
% \end{figure}
\end{wrapfigure}
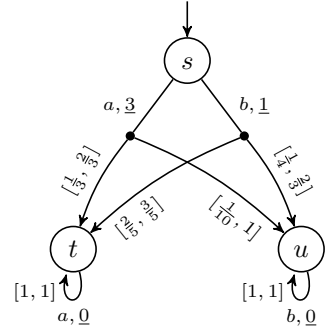
% \begin{myexample}
As an example of \IMDP{} with a reward structure, consider the \IMDP{} $\imdp$ depicted in Fig.~\ref{fig:uncertainMDP}. 
The set of states is $\States = \setnocond{s, t, u}$ with $s$ being the initial one. 
The set of actions is $\ActionSet = \setnocond{a,b}$, and the non-zero transition probability intervals are $\intTransitionProbability(s,a,t)= \interval{\frac{1}{3}}{\frac{2}{3}}$, $\intTransitionProbability(s,a,u)= \interval{\frac{1}{10}}{1}$, $\intTransitionProbability(s,b,t)= \interval{\frac{2}{5}}{\frac{3}{5}}$, $\intTransitionProbability(s,b,u)= \interval{\frac{1}{4}}{\frac{2}{3}}$, and $\intTransitionProbability(t,a,t) = \intTransitionProbability(u,b,u) = \interval{1}{1}$.
The underlined numbers indicate the reward structure $\rew$ with $\rew(s,a)=3$, $\rew(s,b)=1$, and $\rew(t,a) = \rew(u,b) = 0$.
Among the uncountable many distributions belonging to $\uncertainty{a}{s}$, two possible choices for nature $\env$ on $s$ and $a$ are $\env(s,a) = \setnocond{(t,\frac{3}{5}), (u,\frac{2}{5})}$ and $\env(s,a) = \setnocond{(t,\frac{1}{3}), (u,\frac{2}{3})}$.

%% file: multi-objective-queries.tex
\section{Multi-objective Robust Strategy Synthesis for IMDP\lowercase{s}}
\label{sec:multiObRobustSyn}

In this paper, we consider two main classes of properties for \IMDP{}s; 
the \emph{probability of reaching a target} and the \emph{expected total reward}. 
The reason that we focus on these properties is that their algorithms usually serve as the basis for more complex properties.  For instance, they can be easily extended to answer queries with linear temporal logic properties as shown in \cite{etessami2007}. 
To this aim, we lift the satisfaction definitions of these two classes of properties from \MDP{}s in~\cite{FKP12,FKNPQ11} to \IMDP{}s by encoding the notion of robustness for strategies.

\begin{center}
  \centering
  Please note that all proofs for this section are contained in Appendix~\ref{apx:proofs}.
\end{center}
\begin{definition}[Reachability Predicate \& its Robust Satisfaction]
	A \emph{reachability predicate} $\reachPred{k}{\GoalSet}{\sim p}$ consists of a set of target states $\GoalSet \subseteq \States$, a relational operator $\mathord{\sim} \in \setnocond{\leq, \geq}$, a rational probability bound $p \in \unitRationals$ and a time bound $k\in\extNaturals$. 
	It indicates that the probability of reaching $\GoalSet$ within $k$ time steps satisfies $\sim p$.  

	Robust satisfaction of $\reachPred{k}{\GoalSet}{\sim p}$ by \IMDP{} $\imdp$ under strategy $\str \in \Str$ is denoted by $\inducedMC \models_{\Env} \reachPred{k}{\GoalSet}{\sim p}$ and indicates that the probability of the set of all paths that reach $T$ under $\str$ satisfies the bound $\sim p$ for every choice of nature $\env \in \Env$.  
	Formally,
	$\inducedMC \models_{\Env} \reachPred{k}{\GoalSet}{\sim p}$ 
	iff
	$\Prob^{\str}(\boundReach{k}{\GoalSet}) \sim p$ 
	where 
	$
	\Prob^{\str}(\boundReach{k}{\GoalSet}) = \opt_{\env \in \Env}\Prob^{\str,\env}\setcond{\Pat \in \Infpat}{\exists i\leq k: \Pat[i] \in \GoalSet}
	$
	and
	$\opt = \min$ if $\mathord{\sim} = \mathord{\geq}$ and $\opt = \max$ if $\mathord{\sim} = \mathord{\leq}$.
	Furthermore, $\str$ is referred to as a robust strategy.
\end{definition}

\begin{definition}[Reward Predicate \& its Robust Satisfaction]
	A \emph{reward predicate} $\rewPred{\rew}{k}{r}$ consists of a reward structure $\rew$, a time bound $k \in \naturals \cup \setnocond{\infty}$, a relational operator $\mathord{\sim} \in \setnocond{\leq, \geq}$ and a reward bound $r \in \rationals$. 
	It indicates that the expected total accumulated reward within $k$ steps satisfies $\sim r$.  

	Robust satisfaction of $\rewPred{\rew}{k}{r}$ by \IMDP{} $\imdp$ under strategy $\str \in \Str$ is denoted by $\inducedMC \models_{\Env} \rewPred{\rew}{k}{r}$ and indicates that the expected total reward over the set of all paths under $\str$ satisfies the bound $\sim r$ for every choice of nature $\env \in \Env$.  
	Formally,
	$\inducedMC \models_{\Env} \rewPred{\rew}{k}{r}$ iff $\ExpRew{\rew}{k} \sim r$
	where
	$
	\ExpRew{\rew}{k} = \opt_{\env\in\Env} \int_{\Pat} \accumPathRew{\Pat}{k} \,\mathrm{d}\Prob^{\str,\env}
	$
	and
	$\opt =\min$ if $\mathord{\sim} = \mathord{\geq}$ and $\opt =\max$ if $\mathord{\sim} = \mathord{\leq}$. 
	Furthermore, $\str$ is referred to as the robust strategy.
\end{definition}
For the purpose of algorithm design, we also consider weighted sum of rewards. Formally,

\begin{definition}[Weighted Reward Sum]
	Given a weight vector $\vct{w} \in \reals^{n}$, vector of time bounds $\vct{k} = (k_{1}, \dotsc, k_{n})\in (\extNaturals)^{n}$ and reward structures $\rew = (\rew_{1}, \dotsc, \rew_{n})$ for \IMDP{} $\imdp$,
	the \emph{weighted reward sum} $\vct{w} \cdot \rew[\vct{k}]$ over a path $\Pat$ is defined as $\weightedAccumPathRew{\vct{w}}{\rew}{\vct{k}}{\Pat} = \sum_{i=1}^{n} \weightedAccumPathRew{w_{i}}{\rew_{i}}{{k}}{\Pat}$. 
	The \emph{expected total weighted sum} is defined as 
	$\ExpRew{\vct{w} \cdot \rew}{\vct{k}} = \max_{\env \in \Env} \int_{\Pat} \weightedAccumPathRew{\vct{w}}{\rew}{\vct{k}}{\Pat}\,\mathrm{d}\Prob^{\str,\env}$ for bounds $\leq$ and accordingly minimises over natures for $\geq$; 
	for a given strategy $\str$, we have: 
	$\ExpRew{\vct{w} \cdot \rew}{\vct{k}} = \sum_{i=1}^{n} w_{i} \cdot \ExpRew{\rew_{i}}{k_{i}}$.     
\end{definition}

\subsection{Multi-objective Queries}
\label{sec:multi-objective-queries}

Multi-objective properties for \IMDP{}s essentially require multiple predicates to be satisfied at the same time under the same strategy for every choice of the nature. 
We now explain how to formalise multi-objective queries for \IMDP{}s. 

\begin{definition}[Multi-objective Predicate]
\label{def:multi-objectivePredicate}
	A \emph{multi-objective predicate} is a vector $\phi = (\phi_{1},\dotsc,\phi_{n})$ of reachability or reward predicates. 
	We say that $\phi$ is satisfied by \IMDP{} $\imdp$ under strategy $\str$ for every choice of nature $\env \in \Env$, denoted by $\inducedMC \models_{\Env} \phi$ if, for each $1 \leq i\leq n$, it is $\inducedMC \models_{\Env} \phi_{i}$. 
	We refer to $\str$ as a robust strategy. 
	Furthermore, we call $\phi$ a basic multi-objective predicate if it is of the form $(\rewPredMin{\rew_{1}}{k_{1}}{r_{1}},\dotsc,\rewPredMin{\rew_{n}}{k_{n}}{r_{n}})$, i.e., it includes only lower-bounded reward predicates.  
\end{definition}

We formulate multi-objective queries for \IMDP{}s in three ways, namely \emph{synthesis queries}, \emph{quantitative queries} and \emph{Pareto queries}. 
Due to lack of space, we only focus on the synthesis queries and discuss the other types of queries in Appendix~\ref{app:other-queries}. 
We formulate multi-objective synthesis queries for \IMDP{}s as follows. 
\begin{definition}[Synthesis Query]
\label{def:synthesisQueries}
	Given an \IMDP{} $\imdp$ and a multi-objective predicate $\phi$, the \emph{synthesis query} asks if there exists a robust strategy $\str \in \Str$ such that $\inducedMC \models_{\Env} \phi$. 
\end{definition}
Note that the synthesis queries check for the existence of a robust strategy that satisfies a multi-objective predicate $\phi$ for every resolution of nature. 

In order to avoid unusual behaviours in strategy synthesis such as infinite total expected reward,
we need to limit the usage of rewards by assuming reward-finiteness for the strategies that satisfy the
reachability predicates in the given multi-objective query $\phi$. 
\begin{assumption}[Reward-finiteness]
\label{ASSUMPTION:REWARDFINITENESS}
	Suppose that an \IMDP{} $\imdp$ and a synthesis query $\phi$ are given. 
	Let $\phi =(\reachPred{k_{1}}{\GoalSet_{1}}{\sim p_{1}},\dotsc,\reachPred{k_{n}}{\GoalSet_{n}}{\sim p_{n}}, \rewPred{\rew_{n+1}}{k_{n+1}}{r_{n+1}},\dotsc,\rewPred{\rew_m}{k_m}{r_m})$. 
	We say that $\phi$ is reward-finite if for each $n+1\leq i \leq m$ such that $k_{i}=\infty$,
	$\sup\setcond{\ExpRew{\rew_{i}}{k_{i}}}{\inducedMC\models_{\Env}(\reachPred{k_{1}}{\GoalSet_{1}}{\sim p_{1}},\dotsc,\reachPred{k_{n}}{\GoalSet_{n}}{\sim p_{n}})}<\infty$. 
\end{assumption}
Due to lack of space, we provide in Appendix~\ref{app:rewardFinitenessAssumption} a method to check for reward-finiteness assumption of a given \IMDP{} $\imdp$ and a synthesis query $\phi$, a preprocessing procedure that removes actions with non-zero rewards from the end components of $\imdp$, and a proof for the correctness of this procedure with respect to $\phi$.
Therefore, in the rest of the paper, we assume that all queries are reward-finite. 
Furthermore, for the soundness of our analysis we also require that for any \IMDP{} $\imdp$ and $\phi$ given as in Assumption~\ref{ASSUMPTION:REWARDFINITENESS}:
\begin{inparaenum}[\itshape (i)]
\item
each reward structure $\rew_{i}$ assigns only non-negative values; 
\item 
$\phi$ is reward-finite; and 
\item
for indices $n+1\leq i\leq m$ such that $k_{i}=\infty$, either all $\sim_{i}$s are $\leq$ or all are $\geq$.
\end{inparaenum}

%% file: strategy-synthesis.tex
\subsection{Robust Strategy Synthesis}
\label{sec:strategy-syntheis}
We first study the computational complexity of multi-objective robust strategy synthesis problem for \IMDP{}s. 
Formally,
\begin{restatable}{theorem}{synthesisComplexityIMDPs}
\label{thm:synthesisComplexityIMDPs}
Given an \IMDP{} $\imdp$ and a multi-objective predicate $\phi$, the problem of synthesising a strategy $\str \in \Str$ such that $\inducedMC\models_{\Env} \phi$ is \pspace-hard. 
\end{restatable}

As the first step towards derivation of a solution approach for the robust strategy synthesis problem, we need to convert all reachability predicates to reward predicates and therefore, to transform an arbitrarily given query to a query over a basic predicate on a modified \IMDP{}. 
This can be simply done by adding, once for all, a reward of one at the time of reaching the target set and also negating the objective of predicates with upper-bounded relational operators.
We correct and extend the procedure in~\cite{FKP12} to reduce a general multi-objective predicate on an \IMDP{} model to a basic form on a modified \IMDP{}. 
\begin{restatable}{proposition}{reductionToBasicForm}
	\label{appLemma:reductionToBasicForm}
	\sloppy
	Given an \IMDP{} $\imdp = (\States,\InitState, \ActionSet, \intTransitionProbability)$ and a multi-objective predicate $\phi =(\reachPred{k_{1}}{\GoalSet_{1}}{\sim_{1} p_{1}}, \dotsc, \reachPred{k_{n}}{\GoalSet_{n}}{\sim_{n} p_{n}}, \rewPred[n+1]{\rew_{n+1}}{k_{n+1}}{r_{n+1}}, \dotsc, \rewPred[m]{\rew_m}{k_m}{r_m})$, let $\imdp' = (\States', \InitState', \ActionSet', \intTransitionProbability')$ be the \IMDP{} whose components are defined as follows:
	\begin{itemize}
		\item $\States' = \States \times 2^{\setnocond{1, \dotsc, n}}$;
		\item $\InitState' = (\InitState,\emptyset)$;
		\item $\ActionSet' = \ActionSet \times 2^{\setnocond{1, \dotsc, n}}$;
		and
		\item for all $s, s' \in S$, $a \in \ActionSet$, and $v, v', v'' \subseteq \setnocond{1, \dotsc, n}$,
		\[\kern-3mm
		\intTransitionProbability'((s,v),(a,v'),(s', v'')) = 
		\begin{cases}
		\intTransitionProbability(s,a,s') & \text{if $v' = \setcond{i}{s \in \GoalSet_{i}} \setminus v$ and $v'' = v \cup v'$,}\\
		0 & \text{otherwise.}
		\end{cases}
		\]
	\end{itemize}
	Now, let $\phi' = (\rewPredMin{\rew_{\GoalSet_{1}}}{k_{1}+1}{p'_{1}}, \dotsc, \rewPredMin{\rew_{\GoalSet_{n}}}{{k_{n}+1}}{p'_{n}}, \rewPredMin{\bar{\rew}_{n+1}}{k_{n+1}}{r'_{n+1}}, \dotsc, \rewPredMin{\bar{\rew}_{m}}{k_{m}}{r'_{m}})$ where, for each $i \in \setnocond{1, \dotsc, n}$,
	\[
	\begin{array}{ccc}
	p'_{i} = 
	\begin{cases}
	p_{i} & \text{if $\mathord{\sim_{i}} = \mathord{\geq}$,}\\
	-p_{i} & \text{if $\mathord{\sim_{i}} = \mathord{\leq}$;}
	\end{cases}
	&
	\text{~~and~~}
	&
	\rew_{\GoalSet_{i}}((s,v), (a,v')) =
	\begin{cases}
	1  & \text{if $i \in v'$ and $\mathord{\sim_{i}} = \mathord{\geq}$,}\\
	-1 & \text{if $i \in v'$ and $\mathord{\sim_{i}} = \mathord{\leq}$,}\\
	0  & \text{otherwise;}
	\end{cases}
	\end{array}
	\]
	and, for each $j \in \setnocond{n+1, \dotsc, m}$,
	\[
	\begin{array}{ccc}
	r'_{j} = 
	\begin{cases}
	r_{j} & \text{if $\mathord{\sim_{j}} = \mathord{\geq}$,}\\
	-r_{j} & \text{if $\mathord{\sim_{j}} = \mathord{\leq}$;}
	\end{cases}
	&
	\text{~~and~~}
	&
	\bar{\rew}_{j}((s,v), (a,v'))= 
	\begin{cases}
	\rew_{j}(s,a)  & \text{if $\mathord{\sim_{j}} = \mathord{\geq}$,}\\
	-\rew_{j}(s,a) & \text{if $\mathord{\sim_{j}} = \mathord{\leq}$.}
	\end{cases}
	\end{array}
	\]
	Then $\phi$ is satisfiable in $\imdp$ if and only if $\phi'$ is satisfiable in $\imdp'$.	
\end{restatable}

We therefore need to only consider the basic multi-objective predicates of the form $(\rewPredMin{\rew_{1}}{k_{1}}{r_{1}}, \dotsc, \rewPredMin{\rew_{n}}{k_{n}}{r_{n}})$ for the purpose of robust strategy synthesis. 
For a basic multi-objective predicate, we define its Pareto curve as follows.
\begin{definition}[Pareto Curve of a Multi-objective Predicate]
Given an \IMDP{} $\imdp$ and a basic multi-objective predicate $\phi = (\rewPredMin{\rew_{1}}{k_{1}}{r_{1}}, \dotsc, \rewPredMin{\rew_{n}}{k_{n}}{r_{n}})$, we define the set of achievable values with respect to $\phi$ as \sloppy$A_{\imdp,\phi} = \setcond{(r_1,\ldots,r_n) \in \reals^{n}}{\text{$(\rewPredMin{\rew_{1}}{k_{1}}{r_{1}}, \dotsc, \rewPredMin{\rew_{n}}{k_{n}}{r_{n}})$ is satisfiable}}$.
We define the Pareto curve of $\phi$ to be the Pareto curve of $A_{\imdp,\phi}$ and denote it by $\paretoCurve{\imdp,\phi}$.
\end{definition}

\begin{figure}[tb]
	\centering
		\begin{subfigure}[c]{0.65\textwidth}
			\centering
			\resizebox{\textwidth}{!}{
				\begin{tikzpicture}[->, auto, >=stealth', semithick, baseline, 
				state/.style={draw, ellipse, inner sep=0.2mm,text centered},
				trannode/.style={draw, circle, fill, minimum size=1mm, inner sep=0mm},
				prob/.style={font=\scriptsize,sloped, above}
				]
				\path[use as bounding box] (0.9,0.8) rectangle (10.35,5.75);
				\node[state, initial above, initial text={}] (0) at (3,5) {$(s,\emptyset)$};
				\node[state] (1) at (8,5) {$(s,\{1\})$};
				\node[trannode] (01) at ($(0) + (-0.75,-1)$) {};
				\node[trannode] (02) at ($(0) + (0.75,-1)$) {};
				\node[trannode] (11) at ($(1) + (-0.75,-1)$) {};
				\node[trannode] (12) at ($(1) + (0.75,-1)$) {};
				\node[state] (2) at ($(01) + (-0.75,-2)$) {$(t,\emptyset)$};
				\node[state] (3) at ($(02) + (0.75,-2)$) {$(u,\emptyset)$};
				\node[state] (4) at ($(11) + (-0.75,-2)$) {$(t,\{1\})$};
				\node[state] (5) at ($(12) + (0.75,-2)$) {$(u,\{1\})$};
				\draw
				(0) edge [-] node[left=3pt] {\scriptsize $(a,\emptyset),\underline{3},\underline{0}$} (01)
				(0) edge [-] node[right=3pt] {\scriptsize $(b,\emptyset),\underline{1},\underline{0}$} (02)
				(1) edge [-] node[left=3pt] {\scriptsize $(a,\emptyset),\underline{3},\underline{0}$} (11)
				(1) edge [-] node[right=3pt] {\scriptsize $(b,\emptyset),\underline{1},\underline{0}$} (12)
				(01) edge [bend right=10] node[prob] {$\interval{\frac{1}{3}}{\frac{2}{3}}$} (2)
				(01) edge [bend left=10] node[prob, above=-2pt, near end] {$\interval{\frac{1}{10}}{1}$} (3)
				(02) edge [bend left=20] node[prob] {$\interval{\frac{2}{5}}{\frac{3}{5}}$} (2)
				(02) edge [bend left=10] node[prob] {$\interval{\frac{1}{4}}{\frac{2}{3}}$} (3)
				(11) edge [bend right=10] node[prob] {$\interval{\frac{1}{3}}{\frac{2}{3}}$} (4)
				(11) edge [bend left=10] node[prob, above=-2pt, near end] {$\interval{\frac{1}{10}}{1}$} (5)
				(12) edge [bend left=20] node[prob] {$\interval{\frac{2}{5}}{\frac{3}{5}}$} (4)
				(12) edge [bend left=10] node[prob] {$\interval{\frac{1}{4}}{\frac{2}{3}}$} (5)
				(2) edge [bend right=30] node[below] {\scriptsize $(a,\{1\}),\underline{1},\underline{1}$} (4)
				(3) edge [loop below] node[left, near end] {\scriptsize $(b,\emptyset),\underline{1},\underline{0}$} (3)
				(4) edge [loop below] node[below] {\scriptsize $(a,\emptyset),\underline{1},\underline{0}$} (4)
				(5) edge [loop below] node[below] {\scriptsize $(b,\emptyset),\underline{1},\underline{0}$} (5)
				;
				\end{tikzpicture}
			}
			\caption{The transformed \IMDP{} $\imdp'$}
			\label{fig:modifiedUncertainMDP}
	\end{subfigure}
	\hfill
	\begin{subfigure}[c]{.25\textwidth}
		\centering
		\resizebox{\textwidth}{!}{\includegraphics{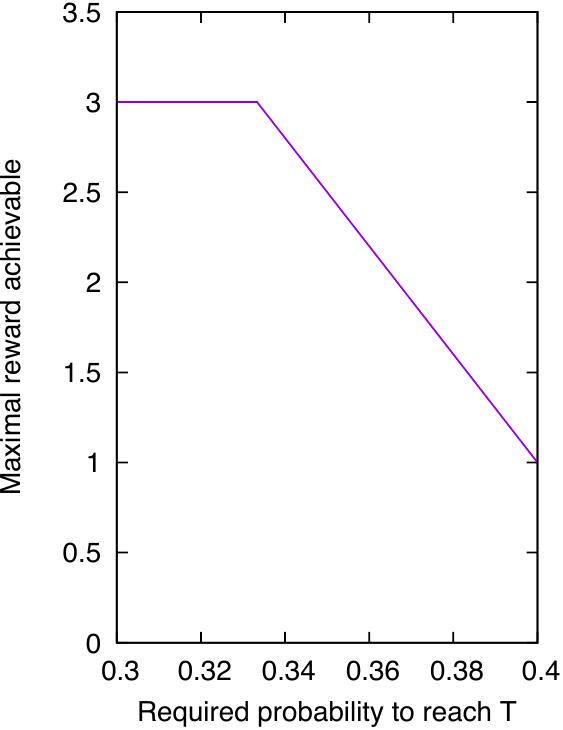}}
		\vskip5mm
		\caption{Pareto curve.\label{fig:pareto-running-example}}
	\end{subfigure}
	\caption{Example of \IMDP{} transformation. (a) The \IMDP{} $\imdp'$ generated from $\imdp$ shown in Fig.~\ref{fig:uncertainMDP}. (b) Pareto curve for the property $([\rew_{\GoalSet}]^{\leq 2}_{\max}, [\rew]^{\leq 1}_{\max})$.}
	% \label{fig:robot_env}
\end{figure}
\begin{myexample}
	To illustrate the transformation presented in Proposition~\ref{appLemma:reductionToBasicForm}, consider again the \IMDP{} depicted in Fig.~\ref{fig:uncertainMDP}. 
	Assume that the target set is $\GoalSet = \setnocond{t}$ and consider the property $\phi = (\reachPred{1}{\GoalSet}{\geq\frac{1}{3}}, \rewPredMin{\rew}{1}{\frac{1}{4}})$. 
	The reduction converts $\phi$ to the property 
	$\phi'=(\rewPredMin{\rew_{\GoalSet}}{2}{\frac{1}{3}}, \rewPredMin{\rew}{1}{\frac{1}{4}})$ on the modified $\imdp'$ depicted in Fig.~\ref{fig:modifiedUncertainMDP}. 
	We show two different reward structures $\bar{\rew}$ and $\rew_{\GoalSet}$ besides each action, respectively.

	In Fig.~\ref{fig:pareto-running-example} we show the Pareto curve for this property.
	As we see, until required probability $\frac{1}{3}$ to reach $\GoalSet$, the maximal reward value is $3$.
	Afterwards, the reward obtainable linearly decreases, until at required probability $\frac{2}{5}$ it is just $1$.
	For higher required probabilities, the problem becomes infeasible.
	The reason for this behaviour is that, up to minimal probability $\frac{1}{3}$, action $a$ can be chosen in state $s$, because the lower interval bound to reach $t$ is $\frac{1}{3}$, which in turn leads to a reward of $3$ being obtained.
	For higher reachability probabilities required, choosing action $b$ with a certain probability is required, which however provides a lower reward.
	There is no strategy with which $t$ is reached with a probability larger than $\frac{2}{5}$.
\end{myexample}

It is not difficult to see that the Pareto curve is in general an infinite set, and therefore, it is usually not possible to derive an exact representation of it in polynomial time. 
However, it can be shown that an $\epsilon$-approximation of it can be computed efficiently~\cite{etessami2007}.

In the rest of this section, we describe an algorithm to solve the synthesis query. 
We follow the well-known \emph{normalisation} approach in order to solve the multi-objective predicate which is essentially based on
normalising multiple objectives into one single objective. 
It is known that the optimal solution of the normalised (single-objective) predicate, if it exists, is the Pareto optimal solution of the multi-objective predicate~\cite{ehrgott2006}.

\begin{algorithm}[!t]
	\small 
	\setstretch{0.9}
	    \caption{Algorithm for solving robust synthesis queries\label{alg:synthesis}}
		\SetKwInOut{Input}{Input}\SetKwInOut{Output}{Output}
		\KwIn{An \IMDP{} $\imdp$, multi-objective predicate $\phi= (\rewPredMin{\rew_{1}}{k_{1}}{r_{1}}, \dotsc, \rewPredMin{\rew_{n}}{k_{n}}{r_{n}})$}
		\KwOut{true if there exists a strategy $\str \in \Str$ such that $\inducedMC\models_{\Env} \phi$, false if not.}
		\Begin{
			$X := \emptyset$; $\rew := (\rew_{1}, \dotsc, \rew_{n})$;\\
			$\vct{k} := (k_{1}, \dotsc, k_{n})$; $\vct{r} := (r_{1}, \dotsc, r_{n})$; \\
			\While{$\vct{r} \notin \dwc{X}$}{
				Find $\vct{w}$ separating $\vct{r}$ from $\dwc{X}$;\\
				Find strategy $\str$ maximising $\ExpRew{\vct{w} \cdot \rew}{\vct{k}}$;\label{Alg1:eq1Optimization}\\
				$\vct{g} := (\ExpRew{\rew_{i}}{k_{i}})_{1\leq i\leq n}$;\label{Alg1:eq2Optimization}\\
				\If {$\vct{w} \cdot \vct{g} < \vct{w} \cdot\vct{r}$}{ \Return false;}
				$X := X \cup \setnocond{\vct{g}}$;
			}
			\Return true;
		}
\end{algorithm}

The robust synthesis procedure is detailed in Algorithm~\ref{alg:synthesis}.  This algorithm basically aims to construct a sequential approximation to the Pareto curve $\paretoCurve{\imdp,\phi}$ while the quality of approximations gets better and more precise along the iterations. 
In other words, along the course of Algorithm~\ref{alg:synthesis} a sequence of weight vectors $\vct{w}$ are generated and corresponding to each of them, a $\vct{w}$-weighted sum of $n$ objectives is optimised through lines~\ref{Alg1:eq1Optimization}-\ref{Alg1:eq2Optimization}. 
The optimal strategy $\str$ is then used in order to generate a point $\vct{g}$ on the Pareto curve $\paretoCurve{\imdp,\phi}$.
We collect all these points in the set $X$. 
The multi-objective predicate $\phi$ is satisfiable once we realise that $\vct{r}$ belongs to $\dwc{X}$.

The optimal strategies for the multi-objective robust synthesis queries are constructed following the approach of~\cite{FKP12} and as a result of termination of Algorithm~\ref{alg:synthesis}. 
In particular, when Algorithm~\ref{alg:synthesis} terminates, a sequence of points $\vct{g}^{1}, \dotsc, \vct{g}^{t}$ on the Pareto curve $\paretoCurve{\imdp,\phi}$ are generated each of which corresponds to a deterministic strategy $\str_{\vct{g}^{j}}$ for the current point $\vct{g}^{j}$. 
The resulting optimal strategy $\str_{opt}$ is subsequently constructed from these using a randomised weight vector $\vct{\alpha} \in \reals^{t}$ satisfying $r_{i} \leq \sum_{j=1}^{t} {\alpha}_{i} \cdot {g_{i}}^{j}$, cf. Appendix~\ref{app:memoryless-strategies-generation} of the extended version of this paper.

\begin{remark}
\label{transformationIMDPsStochaticGames} 
It is worthwhile to mention that the synthesis query for \IMDP{}s cannot be solved on the \MDP{}s generated from \IMDP{}s by computing all feasible extreme transition probabilities and then applying the algorithm in~\cite{FKP12}. 
The latter is a valid approach provided the cooperative semantics is applied for resolving the two sources of nondeterminism in \IMDP{}s. 
With respect to the competitive semantics needed here, one can instead transform \IMDP{}s to $2\frac{1}{2}$-player games~\cite{basset2014compositional} and then  along the lines of the previous approach apply the algorithm in~\cite{chen2013stochastic}. 
Unfortunately, the transformation to (\MDP{}s or) $2\frac{1}{2}$-player games induces an exponential blowup, adding an exponential factor to the worst case time complexity of the decision problem.             
Our algorithm avoids this by solving the robust synthesis problem directly on the \IMDP{}  so that the core part, i.e., lines~\ref{Alg1:eq1Optimization}-~\ref{Alg1:eq2Optimization} of Algorithm~\ref{alg:synthesis} can be solved with time complexity polynomial in~$\size{\imdp}$.
\end{remark}
\begin{algorithm}[!t]
	\small
	\setstretch{0.9}
	\caption{Value iteration-based algorithm to solve lines 6-7 of Algorithm~\ref{alg:synthesis}\label{alg:valueAlg}}
	\SetKwInOut{Input}{Input}\SetKwInOut{Output}{Output}
	\KwIn{An \IMDP{} $\imdp$, weight vector $w$, reward structures $\rew=(\rew_{1}, \dotsc, \rew_{n})$, time-bound vector $\vct{k}\in(\extNaturals)^{n}$, threshold $\epsilon$}
	\KwOut{strategy $\str$ maximising $\ExpRew{\vct{w} \cdot \rew}{\vct{k}}$, $\vct{g} := (\ExpRew{\rew_{i}}{k_{i}})_{1 \leq i \leq n}$}
	\Begin{
		$\vct{x} := 0$; $\vct{x}^{1} := 0$; \ldots; $\vct{x}^{n} := 0$;\\
		$\vct{y} := 0$; $\vct{y}^{1} := 0$; \ldots; $\vct{y}^{n} := 0$;\\
		$\str^{\infty}(s) := \bot$ for all $s \in \States$\\
		\While{$\delta > \epsilon$}{
			\ForEach{$s \in \States$}{$y_{s} := \max_{a \in \StateActionSet{s}}(\sum_{\setcond{i}{k_{i}=\infty}} w_{i} \cdot \rew_{i}(s,a) + \min_{\fixeduncertainty{a}{s} \in \uncertainty{a}{s}}{\sum_{s' \in \States} \fixeduncertainty{a}{s}(s') \cdot x_{s'}})$;\label{robustEq1}\\
			$\str^{\infty}(s) := \arg\max_{a \in \StateActionSet{s}}(\sum_{\setcond{i}{k_{i} = \infty}} w_{i} \cdot \rew_{i}(s,a) + \min_{\fixeduncertainty{a}{s} \in \uncertainty{a}{s}}{\sum_{s' \in \States} \fixeduncertainty{a}{s}(s') \cdot x_{s'}})$\\
			$\optimalfixeduncertainty{\str^{\infty}(s)}{s}(s') := \arg\min_{\fixeduncertainty{a}{s} \in \uncertainty{a}{s}}{\sum_{s' \in \States} \fixeduncertainty{a}{s}(s') \cdot x_{s'}}$
			}
			$\delta := \max_{s \in \States} (y_{s} - x_{s})$; $\vct{x} := \vct{y}$;
		}
		\While{$\delta >\epsilon$}{
			\ForEach{$s \in \States$ and $i \in \setnocond{1, \dotsc, n}$ \Where $k_{i} = \infty$}
			{$y^{i}_{s} := \rew_{i}(s,\str^{\infty}(s)) + \sum_{s' \in \States} \optimalfixeduncertainty{\str^{\infty}(s)}{s}(s') \cdot x^{i}_{s'}$;
			}
			$\delta := \max_{i=1}^{n} \max_{s \in \States} (y^{i}_{s} - x^{i}_{s})$; $\vct{x}^{1} := \vct{y}^{1}$; \ldots;  $\vct{x}^{n} := \vct{y}^{n}$;
		}
		\For {$j = \max\setcond{k_{b} < \infty}{b \in \setnocond{1, \dotsc, n}}$ \DownTo $1$}
		{
		\ForEach{$s \in \States $}{$y_{s} := \max_{a \in \StateActionSet{s}}(\sum_{\setcond{i}{k_{i} \geq j}} w_{i} \cdot \rew_{i}(s,a) + \min_{\fixeduncertainty{a}{s} \in \uncertainty{a}{s}}{\sum_{s' \in \States} \fixeduncertainty{a}{s}(s') \cdot x_{s'}})$;\label{robustEq2}\\
			$\str^{j}(s) := \arg\max_{a \in \StateActionSet{s}}(\sum_{\setcond{i}{k_{i}\geq j}} w_{i} \cdot \rew_{i}(s,a) + \min_{\fixeduncertainty{a}{s} \in \uncertainty{a}{s}}{\sum_{s' \in \States} \fixeduncertainty{a}{s}(s') \cdot x_{s'}})$;\\
			$\optimalfixeduncertainty{\str^{j}(s)}{s}(s') := \arg\min_{\fixeduncertainty{a}{s} \in \uncertainty{a}{s}}{\sum_{s' \in \States} \fixeduncertainty{a}{s}(s') \cdot x_{s'}}$;\\
		    \ForEach{$i \in \setnocond{1, \dotsc, n}$ \Where $k_{i}\geq j$}{$y^{i}_{s} := \rew_{i}(s,\str^{j}(s)) + \sum_{s' \in \States} \optimalfixeduncertainty{\str^{j}(s)}{s}(s') \cdot x^{i}_{s'}$;}
			$\vct{x} := \vct{y}$; $\vct{x}^{1} := \vct{y}^{1}$; \ldots; $\vct{x}^{n} := \vct{y}^{n}$;}		
			}
			\For{$i=1$ \KwTo $n$}{$g_{i} := y_{\InitState}^{i}$;}	
			$\str$ acts as $\str^{j}$ in $j^{th}$ step when $j < \max_{i \in \setnocond{1, \dotsc, n}} k_{i}$ and as $\str^{\infty}$ afterwards;\\
			\Return $\str, \vct{g}$
	}
\end{algorithm}
Algorithm~\ref{alg:valueAlg} represents a value iteration-based algorithm which extends the value iteration-based algorithm in~\cite{FKP12} and adjusts it for \IMDP{} models by encoding the notion of robustness. 
More precisely, the core difference is indicated in lines~\ref{robustEq1} and~\ref{robustEq2} where the optimal strategy is computed so as to be robust against any choice of nature. 
\begin{restatable}{theorem}{SynthesisAlgorithmComplexity}
\label{thm:complexityAnalysis}
Algorithm~\ref{alg:synthesis} is sound, complete and has runtime exponential in $\size{\imdp}$, $\vct{k}$, and $n$.
\end{restatable}
\begin{remark}
It is worthwhile to mention that our robust strategy synthesis approach can also be applied to \MDP{}s with richer formalisms for uncertainties such as likelihood or ellipsoidal uncertainties while preserving the computational complexity. 
In particular, in every inner optimisation problem in Algorithm~\ref{alg:synthesis}, the optimality of a Markovian deterministic strategy and nature is guaranteed as long as the uncertainty set is convex, the set of actions is finite and the inner optimisation problem which minimises/maximises the objective function over the choices of nature achieves its optimum~(cf.~\cite[Proposition~4.1]{PuggelliThesis2014}).
Furthermore, due to the convexity of the generated optimisation problems, the computational complexity of our approach remains intact.
\end{remark}

%% file: case-studies.tex
\section{Case Studies}
\label{sec:case-studies}
We implemented the proposed multi-objective robust strategy synthesis algorithm and applied them to two case studies: 
\begin{inparaenum}[(1)]
\item
	motion planning for a robot with noisy continuous dynamics and 
\item
	autonomous nondeterministic tour guides drawn from~\cite{CRI07,hashemi2016reward}.
\end{inparaenum}
All experiments took a few seconds to complete on a standard laptop PC.

\subsection{Robot Motion Planning under Uncertainty}
\label{sec:robot-planning}

\begin{figure}[tb]
	\centering
		\begin{subfigure}[c]{0.3\textwidth}
		\includegraphics[width=\textwidth]{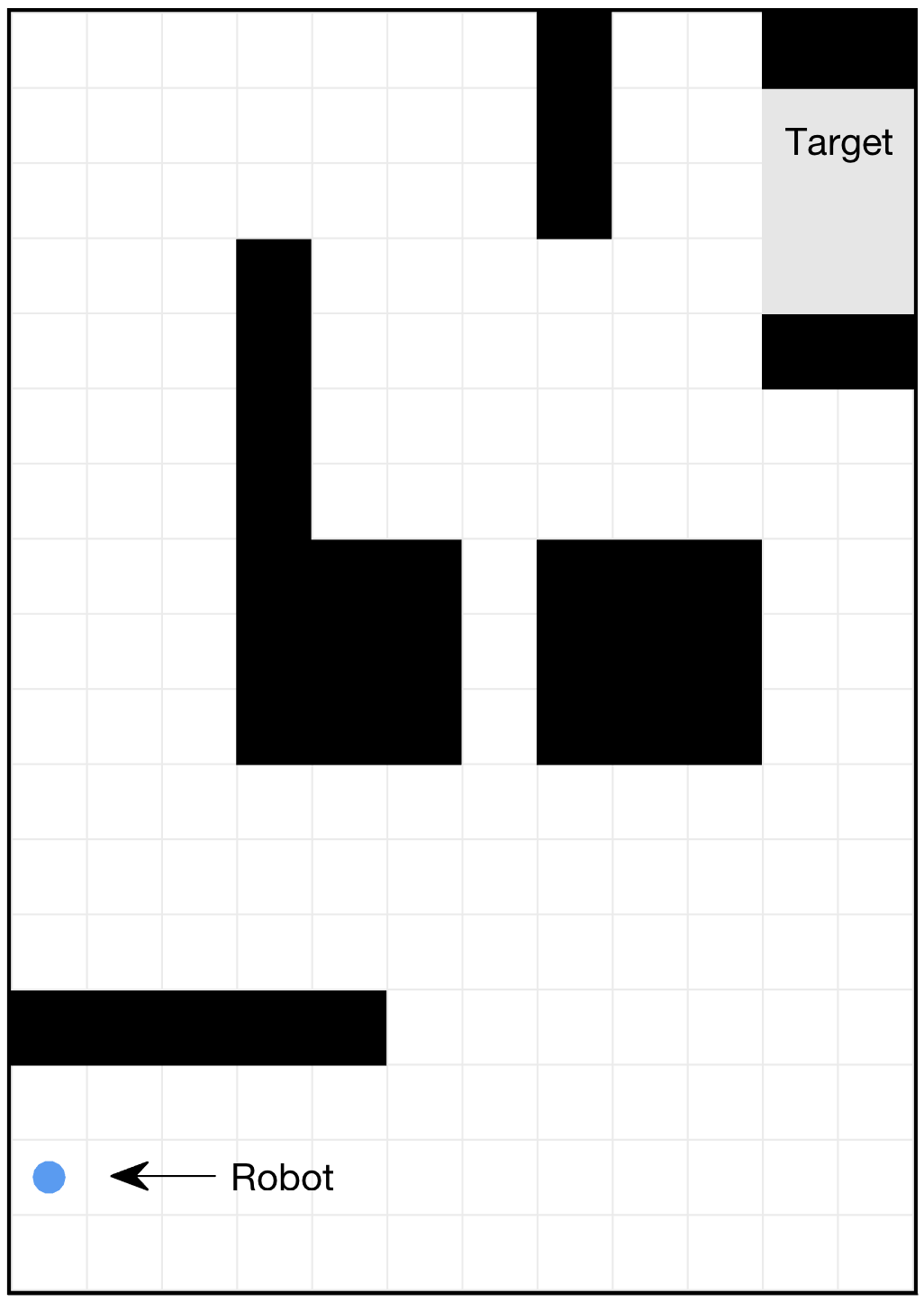}
		\caption{Robot Environment}
		\label{fig:robot_env}
	\end{subfigure}
	\quad
	\begin{subfigure}[c]{.54\textwidth}
		\includegraphics[width=1\textwidth]{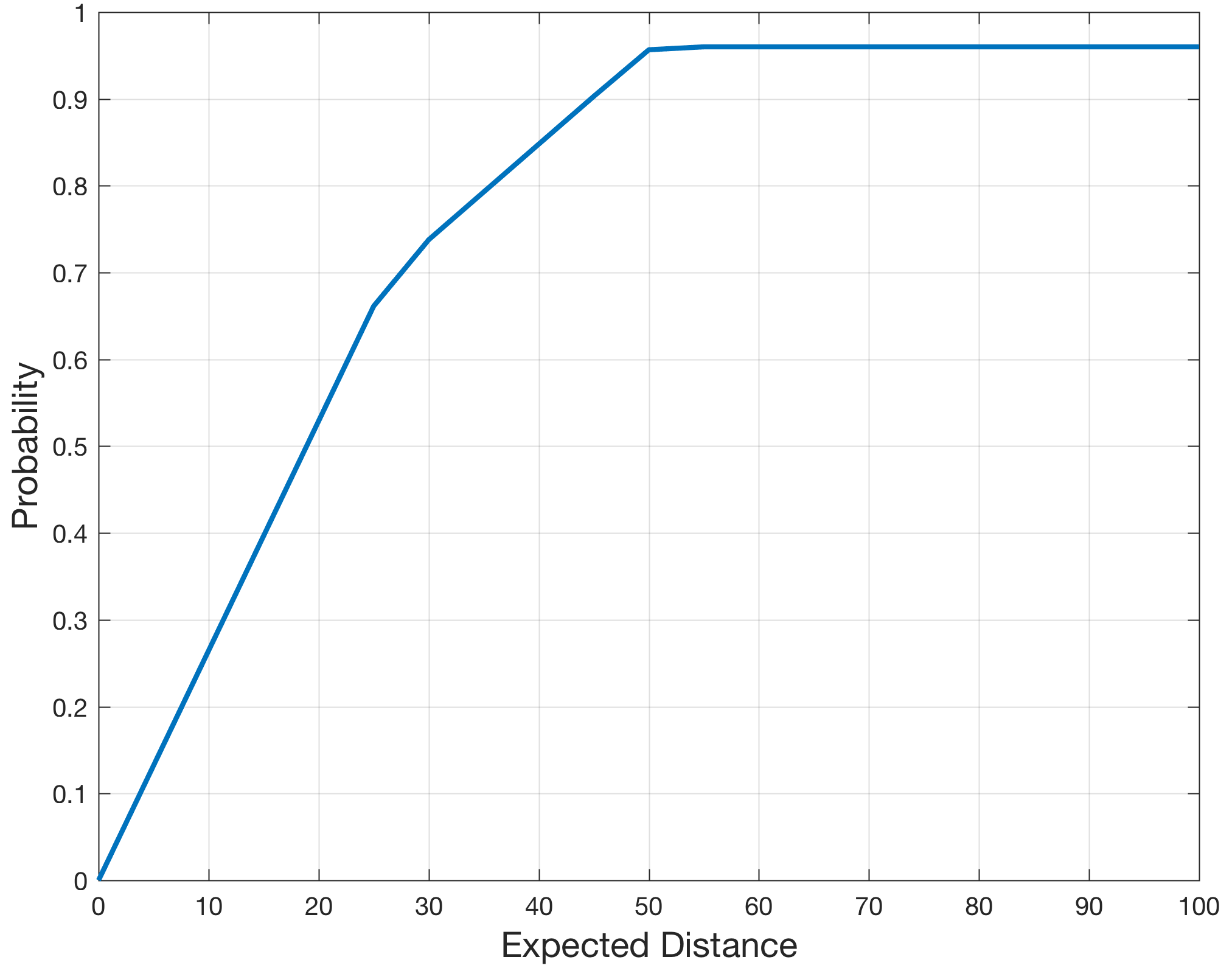}
		\caption{Pareto Curve}
		\label{fig:pareto}
	\end{subfigure}
	\caption{Robotic Scenario. (a) Environment map, where obstacles and target are shown in black and gray, respectively. (b) Pareto curve for the property $([\rew_p]^{\leq \infty}_{\max}, [\rew_d]^{\leq \infty}_{\min})$.}
	% \label{fig:robot_env}
\end{figure}
In robot motion planning, designers often seek a plan that simultaneously satisfies multiple objectives \cite{Lahijanian:CDC:2016}, e.g., \emph{maximising the chances of reaching the target while minimising the energy consumption}.  
These objectives are usually in conflict with each other; 
hence, presenting the Pareto curve, i.e., the set of achievable points with optimal trade-off between the objectives, is helpful to the designers.  
They can then choose a point on the curve according to their desired guarantees and obtain the corresponding plan (strategy) for the robot.  
In this case study, we considered such a motion planning problem for a noisy robot with continuous dynamics in an environment with obstacles and a target region, as depicted in Fig.~\ref{fig:robot_env}.  
The robot's motion model was a single integrator with additive Gaussian noise. 
The initial state of the robot was on the bottom-left of the environment.  
The objectives were to reach the target safely while reducing the energy consumption, which is proportional to the travelled distance.  

We approached this problem by first abstracting the motion of the noisy robot in the environment as an \IMDP{} $\imdp$ and then computing strategies on $\imdp$ as in \cite{luna:wafr:2014,luna:aaai:2014,luna:icra:2014}. 
The abstraction was achieved by partitioning the environment into a grid and computing local (continuous) controllers to allow transitions from every cell to each of its neighbours. 
The cells and the local controllers were then associated to the states and actions of the \IMDP{}, respectively, resulting in 204 states (cells) and 4 actions per state.  
The boundaries of the environment were also associated with a state. 
Note that the transition probabilities between cells were raised by the noise in the dynamics and their ranges were due to variation of the possible initial robot (continuous) state within each cell. 

The \IMDP{} states corresponding to obstacles (including boundaries) were given deterministic self-transitions, modelling robot termination as the result of a collision.  
To allow for the computation of the probability of reaching target, we included an extra state in the \IMDP{} with a deterministic self-transition and then added incoming deterministic transitions to this state from the target states.  
A reward structure $\rew_p$, which assigns a reward of $1$ to these transitions and $0$ to all the others, in fact, computes the probability of reaching the target.  
To capture the travelled distance, we defined a reward structure $\rew_d$ assigning a reward of $0$ to the state-action pairs with self-transitions and $1$ to the rest.

\begin{figure*}[t!]
	% \vspace{-15mm}
	\centering
	% \begin{subfigure}[c]{.44\textwidth}
	% 	\includegraphics[width=1\textwidth]{pareto}
	% 	\caption{$\phi_1$}
	% \end{subfigure}
	\begin{subfigure}[c]{.25\textwidth}
		\includegraphics[width=1\textwidth]{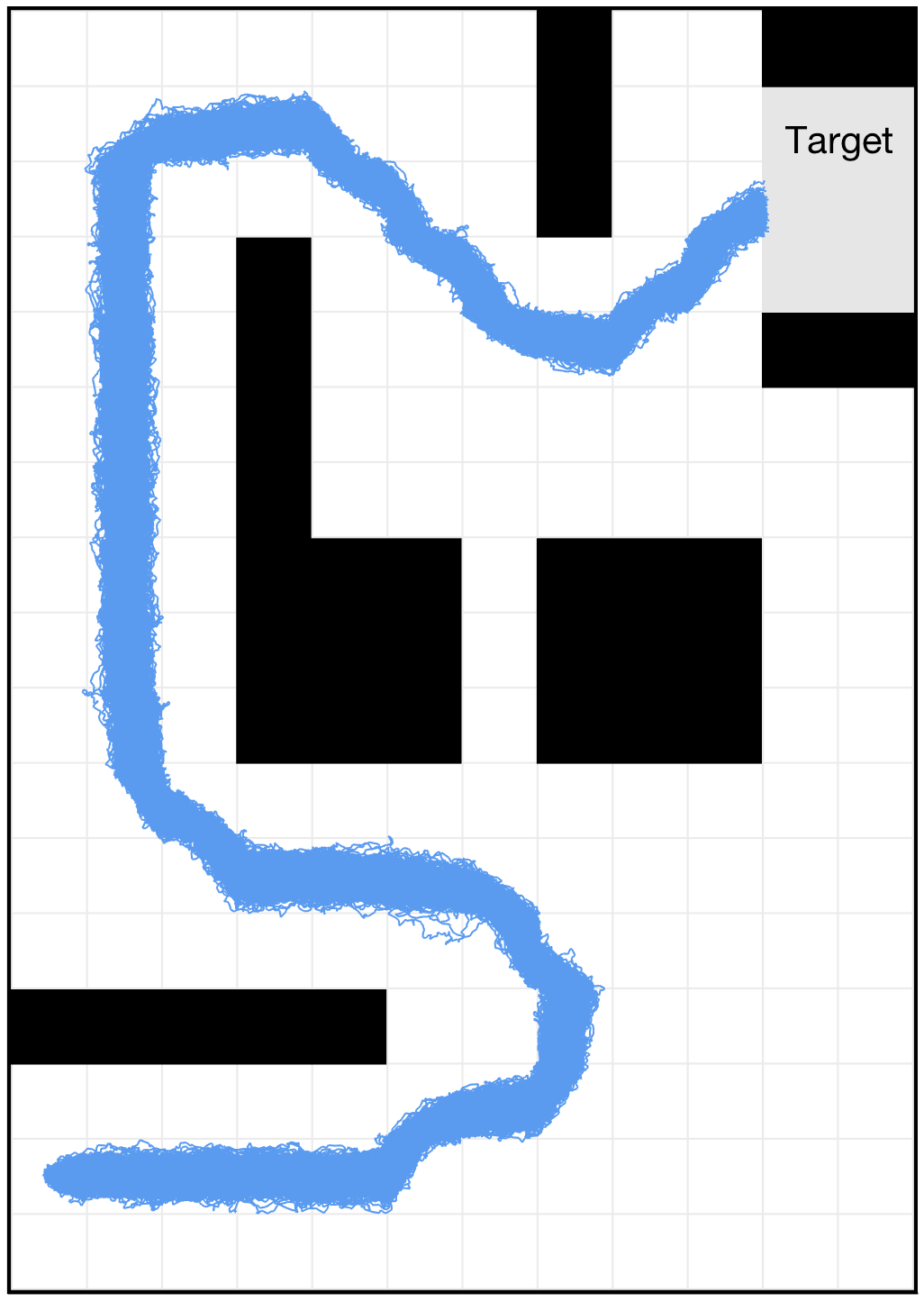}
		\caption{$\phi_1$}
		\label{fig:samplePath1}
	\end{subfigure}
	\quad
	\begin{subfigure}[c]{.25\textwidth}
		\includegraphics[width=1\textwidth]{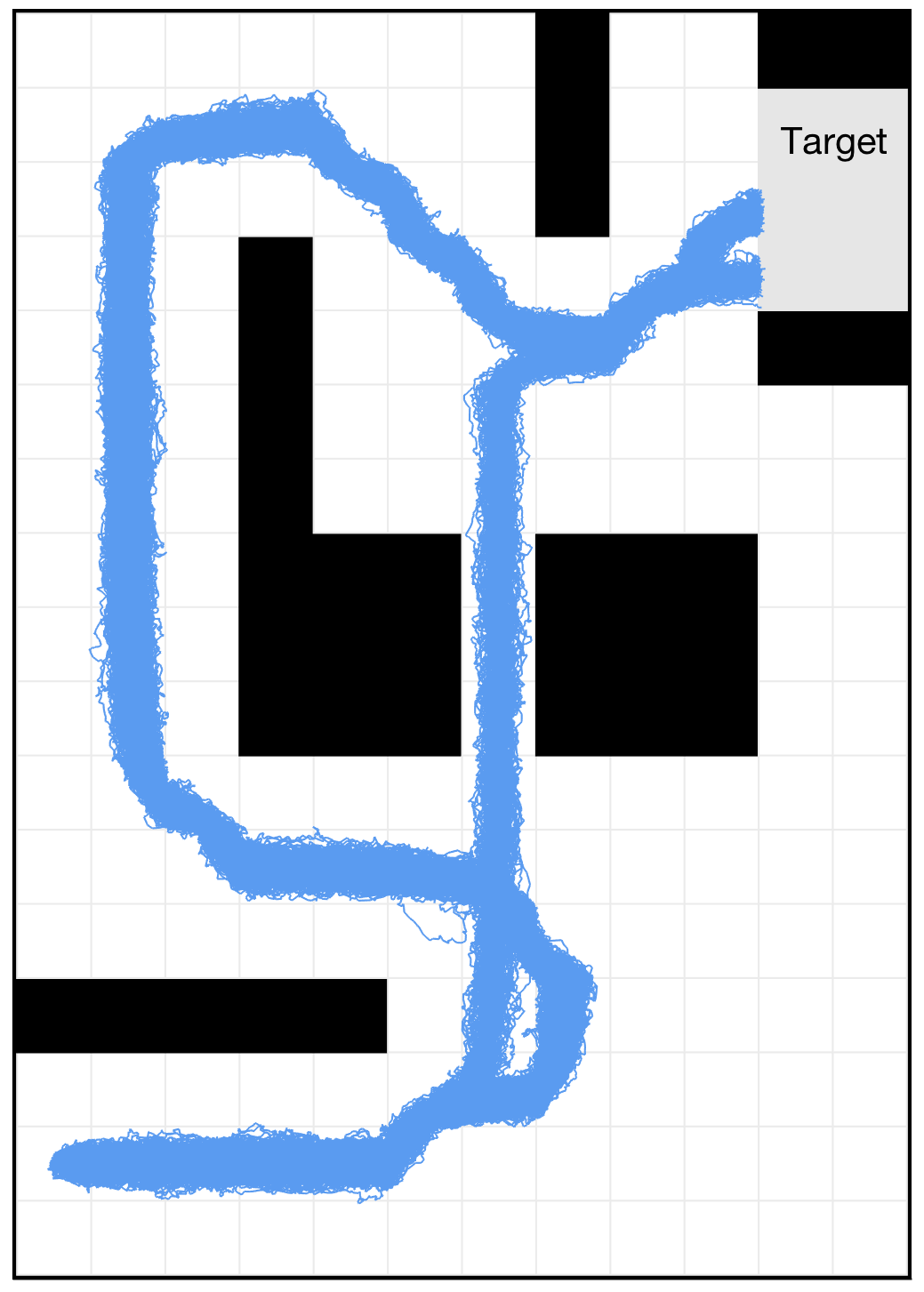}
		\caption{$\phi_2$}
		\label{fig:samplePath2}
	\end{subfigure}
	\quad
	\begin{subfigure}[c]{.25\textwidth}
		\includegraphics[width=1\textwidth]{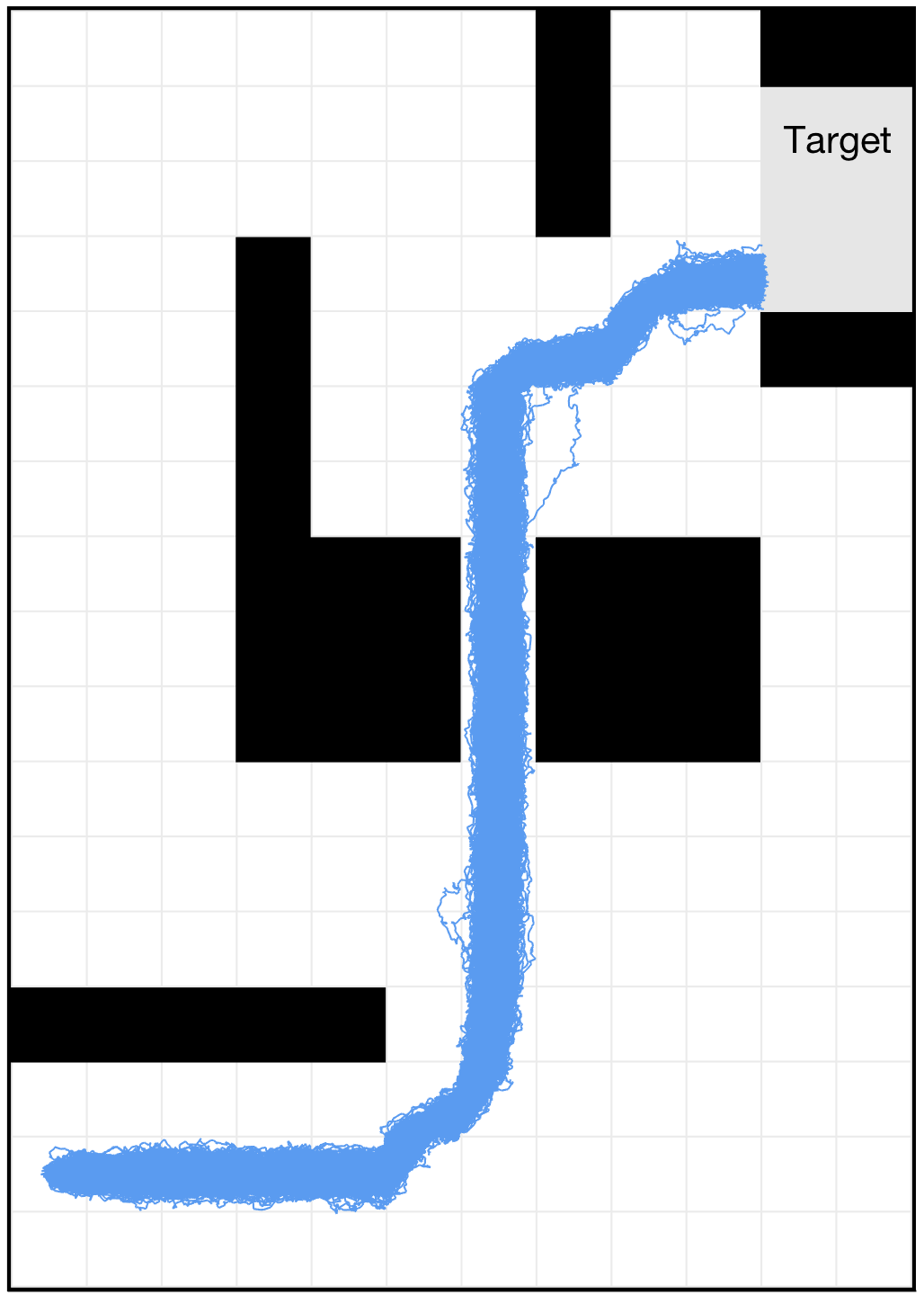}
		\caption{$\phi_3$}
		\label{fig:samplePath3}
	\end{subfigure}
	\caption{Robot sample paths under strategies for $\phi_1$, $\phi_2$, and $\phi_3$}
	\label{fig:robot_samplepaths}
\end{figure*}

The two robot objectives then can be expressed as: $([\rew_p]^{\leq \infty}_{\max}, [\rew_d]^{\leq \infty}_{\min})$ - see Appendix \ref{app:other-queries} for Pareto queries. 
We first computed the Pareto curve for the property, which is shown in Fig.~\ref{fig:pareto}, to find the set of all achievable values (optimal trade-offs) for the reachability probability and expected travelled distance.  
The Pareto curve shows that there is clearly a trade-off between the two objectives.  
To achieve high probability of reaching target safely, the robot needs to travel a longer distance, i.e., spend more energy, and vice versa.  
We chose three points on the curve and computed the corresponding robust strategies for 
\[
\begin{array}{cp{2mm}cp{2mm}c}
	\phi_1 = ([\rew_p]^{\leq \infty}_{\geq {0.95}}, [\rew_d]^{\leq \infty}_{\leq {50}}),
	&&
	\phi_2 = ([\rew_p]^{\leq \infty}_{\geq {0.90}}, [\rew_d]^{\leq \infty}_{\leq {45}}),
	&&
	\phi_3 = ([\rew_p]^{\leq \infty}_{\geq {0.66}}, [\rew_d]^{\leq \infty}_{\leq {25}}).
\end{array}
\]
We then simulated the robot under each strategy 500 times.  
The statistical results of these simulations are consistent with the bounds in $\phi_1$, $\phi_2$, and $\phi_3$.  
The collision-free robot trajectories are shown in Fig.~\ref{fig:robot_samplepaths}.  
These trajectories illustrate that the robot is conservative under $\phi_1$ and takes a longer route with open spaces around it to go to target in order to be safe (Fig.~\ref{fig:samplePath1}), while it becomes reckless under $\phi_3$ and tries to go through a narrow passage with the knowledge that its motion is noisy and could collide with the obstacles (Fig.~\ref{fig:samplePath3}).  
This risky behaviour, however, is required in order to meet the bound on the expected travelled distance in $\phi_3$. 
The sample trajectories for $\phi_2$ (Fig.~\ref{fig:samplePath2}) demonstrate the stochastic nature of the strategy.  
That is, the robot probabilistically chooses between being safe and reckless in order to satisfy the bounds in $\phi_2$.

\subsection{The Model of Autonomous Nondeterministic Tour Guides}
Our second case study is inspired by ``Autonomous Nondeterministic Tour
Guides'' (ANTG) in~\cite{CRI07,hashemi2016reward}, which models a complex
museum with a variety of collections. 
We note that the model introduced in~\cite{CRI07} is an \MDP{}.  In this case study, we use an \IMDP{} model by inserting uncertainties into the \MDP{}.

Due to the popularity of the museum, there are many visitors at the same time. 
Different visitors may have different preferences of arts. 
We assume the museum divides all collections into different categories so that visitors can choose what they would like to visit and pay tickets according to their preferences. 
In order to obtain the best experience, a visitor can first assign certain weights to all categories denoting their preferences to the museum, and then design the best strategy for a target. 
However, the preference of a sort of arts to a visitor may depend on many factors like price, weather, or the length of queue at that moment etc., hence it is hard to assign fixed values to these preferences. 
In our model we allow uncertainties of preferences such that their values may lie in an interval. 

\begin{figure}[!t]
\centering
\begin{subfigure}{0.4\linewidth}
\resizebox{\linewidth}{!}{
  \begin{tikzpicture}[every node/.style={minimum size=.35cm-\pgflinewidth, outer sep=0pt}, scale=0.7]
    \input{museum-arena.tex}
    \input{museum-strategy-main.tex}
	\end{tikzpicture}
	}
\caption{\scriptsize{The ANTG model for $n=14$.
    The yellow, black and green cells represent the entrance, closed and exit parts of the museum, respectively.
The red arrows indicate an example strategy.}}
	\label{fig:ANTG-caseStudy}
\end{subfigure}
\hfill
\begin{subfigure}{0.58\linewidth}
\includegraphics[width=1\textwidth]{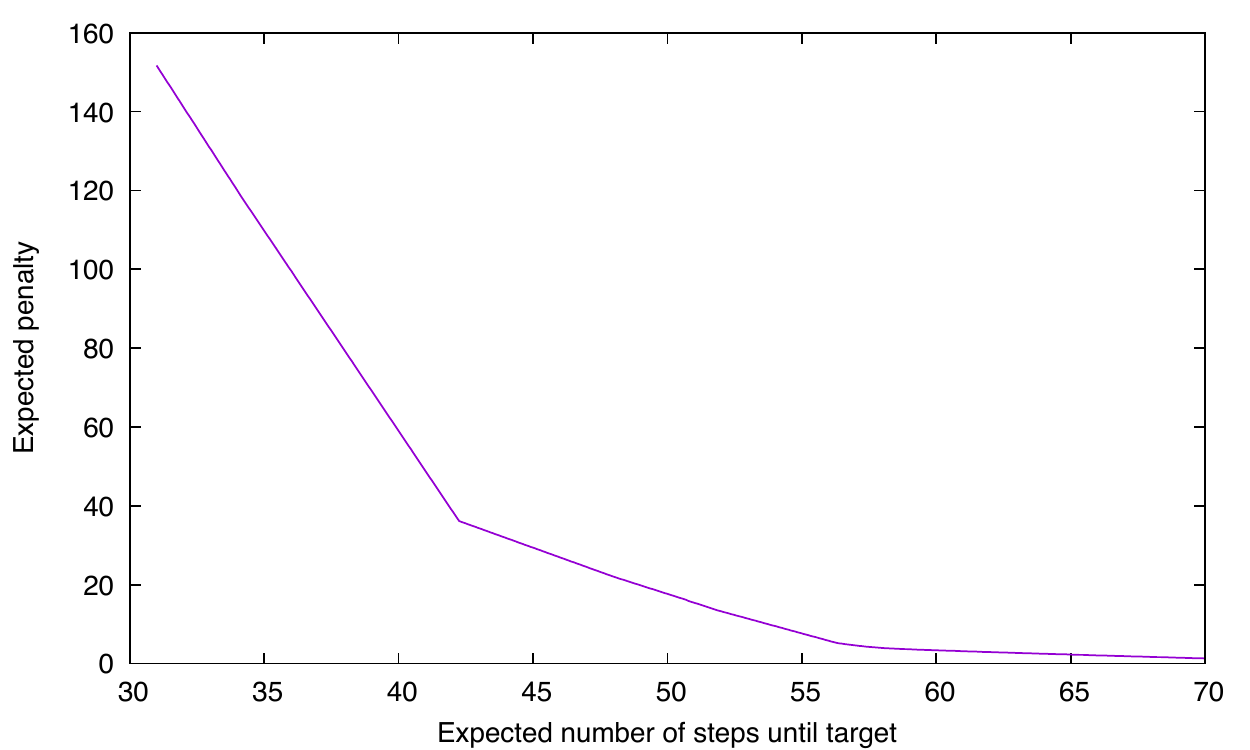} 
\caption{\scriptsize{The Pareto Curve}}
	\label{fig:ANTG-paretoCurve}
\end{subfigure}
\caption{The ANTG case study: model and analysis}
\end{figure}

For simplicity we assume all collections are organised in an $n \times n$ square with $n \ge 10$, with $(0,0)$ being the south-west corner of the museum and $(n-1,n-1)$ the north-east one. 
Let $c = \frac{n-1}{2}$; 
note that $(c,c)$ is at the center of the museum. 
We assume all collections at $(x,y)$ are assigned with a weight interval $[3,4]$ if $\max\setnocond{\abs{x-c}, \abs{y-c}} \leq \frac{n}{10}$, with a weight $2$ if $\frac{n}{10} < \max\setnocond{\abs{x-c}, \abs{y-c}} \leq \frac{n}{5}$, and a weight $1$ if $\max\setnocond{\abs{x-c}, \abs{y-c}} > \frac{n}{5}$. 
In other words, we expect collections in the center to be more popular and subject to more uncertainties than others.
Furthermore, we assume that people at each location $(x,y)$ have four nondeterministic choices of moving to $(x',y')$ in the north east, south east, north west, and south west of $(x,y)$ (limited to the boundaries of the museum). The outcome of these choices, however, is not deterministic.  That is, deciding to go to $(x',y')$ takes the visitor to either $(x,y')$ or $(x',y)$ depending on the weight intervals of $(x,y')$ and $(x',y)$. Thus, the actual outcome of the move is probabilistic to north, south, east or west.
To obtain an \IMDP{}, weights are normalised.
For instance, if the visitor chooses to go to the north east and on $(x,y+1)$ there is a weight interval of $[3,4]$ and on $(x+1,y)$ there is a weight interval of $[2,2]$, it will go to $(x,y+1)$ with probability interval $[3/(3+2),4/(4+2)]$ and to $(x+1,y)$ with probability interval $[2/(2+4),2/(2+3)]$.

Therefore a model with parameter $n$ has $n^2$ states in total and roughly $4n^2$ transitions, a few of which are associated with uncertain transition probabilities. 
An instance of the museum model for $n=14$ is depicted in Fig.~\ref{fig:ANTG-caseStudy}.
In this instantiation, we assume that the visitor starts in the lower left corner (marked yellow) and wants to move to the upper right corner (marked green) with as few steps as possible.
On the other hand, it wants to avoid moving to the black cells, because they correspond to exhibitions which are closed.
For closed exhibitions located at $x=2$, the visitor receive a penalty of $2$,
for those at $x=5$ it receives a penalty of $4$,
for $x=8$ one of $16$ and
for $x=11$ one of $64$.
Therefore, there is a tradeoff between leaving the museum as fast as possible and minimising the penalty received.
With $\rew_s$ being the reward structure for the number of steps and $\rew_p$ denoting the penalty accumulated, $([\rew_s]^{\leq \infty}_{\leq {40}}, [\rew_p]^{\leq \infty}_{\leq {70}})$ requires that we leave the museum within $40$ steps but with a penalty of no more than $70$.
The red arrows indicate a strategy which has been used when computing the Pareto curve by our tool.
Here, the tourist mostly ignores closed exhibitions at $x=2$ but avoids them later.
In Appendix~\ref{app:ANTGother-strategies}, we provide a few more strategies occurring during the computation.
We provide the Pareto curve for this situation in Fig.~\ref{fig:ANTG-paretoCurve}.
With an increasing step bound considered acceptable, the optimal accumulated penalty decreases.
This is expected, because with an increasing step bound, the visitor has more time to walk around more of the closed exhibitions, thus facing a lower penalty.

%% file: museum-arena.tex
\draw[step=0.5cm,color=black] (-3,-3) grid (4,4);
\node[fill=green!70!black] at (3.75,3.75) {};
\node[fill=yellow] at (-2.75,-2.75) {};
\node[fill=black] at (-1.75,-2.75) {};
\node[fill=black] at (1.25,-2.75) {};
\node[fill=black] at (-1.75,-2.25) {};
\node[fill=black] at (-1.75,-1.75) {};
\node[fill=black] at (-1.75,-1.25) {};
\node[fill=black] at (-1.75,-0.75) {};
\node[fill=black] at (-1.75,-0.25) {};
\node[fill=black] at (-1.75,0.25) {};
\node[fill=black] at (1.25,-2.25) {};
\node[fill=black] at (1.25,-1.75) {};
\node[fill=black] at (1.25,-1.25) {};
\node[fill=black] at (1.25,-0.75) {};
\node[fill=black] at (1.25,-0.25) {};
\node[fill=black] at (1.25,0.25) {};
\node[fill=black] at (-0.25,0.25) {};
\node[fill=black] at (1.25,0.25) {};
\node[fill=black] at (2.75,0.25) {};
\node[fill=black] at (-1.75,0.75) {};
\node[fill=black] at (-0.25,0.75) {};
\node[fill=black] at (1.25,0.75) {};
\node[fill=black] at (2.75,0.75) {};
\node[fill=black] at (-0.25,1.25) {};
\node[fill=black] at (2.75,1.25) {};
\node[fill=black] at (-0.25,1.75) {};
\node[fill=black] at (2.75,1.75) {};
\node[fill=black] at (-0.25,2.25) {};
\node[fill=black] at (2.75,2.25) {};
\node[fill=black] at (-0.25,2.75) {};
\node[fill=black] at (2.75,2.75) {};
\node[fill=black] at (-0.25,3.25) {};
\node[fill=black] at (2.75,3.25) {};
\node[fill=black] at (-0.25,3.75) {};
\node[fill=black] at (2.75,3.75) {};
%	\node[fill=orange] at (+0.25,+0.75) {};
%	\node[fill=yellow] at (+0.75,+0.75) {};
%	\node[fill=purple!70] at (-0.75,+0.25) {};

%% file: museum-strategy-main.tex
\placeArrowne{0}{0}
\placeArrowne{0}{1}
\placeArrowne{1}{0}
\placeArrowne{0}{2}
\placeArrowne{1}{1}
\placeArrowne{2}{0}
\placeArrowne{0}{3}
\placeArrowne{1}{2}
\placeArrowne{2}{1}
\placeArrowne{3}{0}
\placeArrowne{0}{4}
\placeArrowne{1}{3}
\placeArrowne{2}{2}
\placeArrowne{3}{1}
\placeArrowne{4}{0}
\placeArrowne{0}{5}
\placeArrowne{1}{4}
\placeArrowne{2}{3}
\placeArrowne{3}{2}
\placeArrowne{4}{1}
\placeArrowne{5}{0}
\placeArrowne{0}{6}
\placeArrownw{1}{5}
\placeArrowne{2}{4}
\placeArrowne{3}{3}
\placeArrowne{4}{2}
\placeArrowne{5}{1}
\placeArrowne{6}{0}
\placeArrowne{0}{7}
\placeArrownw{1}{6}
\placeArrowne{2}{5}
\placeArrowne{3}{4}
\placeArrowne{4}{3}
\placeArrowne{5}{2}
\placeArrowne{6}{1}
\placeArrownw{7}{0}
\placeArrowne{0}{8}
\placeArrownw{1}{7}
\placeArrowne{2}{6}
\placeArrowse{3}{5}
\placeArrowne{4}{4}
\placeArrowne{5}{3}
\placeArrowne{6}{2}
\placeArrownw{7}{1}
\placeArrowne{8}{0}
\placeArrowse{0}{9}
\placeArrowne{1}{8}
\placeArrowne{2}{7}
\placeArrowse{3}{6}
\placeArrowse{4}{5}
\placeArrowne{5}{4}
\placeArrowne{6}{3}
\placeArrownw{7}{2}
\placeArrowne{8}{1}
\placeArrowne{9}{0}
\placeArrowse{0}{10}
\placeArrowse{1}{9}
\placeArrowne{2}{8}
\placeArrowse{3}{7}
\placeArrowsw{4}{6}
\placeArrowse{5}{5}
\placeArrowne{6}{4}
\placeArrownw{7}{3}
\placeArrowne{8}{2}
\placeArrowne{9}{1}
\placeArrowne{10}{0}
\placeArrowse{0}{11}
\placeArrowse{1}{10}
\placeArrowse{2}{9}
\placeArrowse{3}{8}
\placeArrowsw{4}{7}
\placeArrowse{5}{6}
\placeArrowne{6}{5}
\placeArrownw{7}{4}
\placeArrowse{8}{3}
\placeArrowne{9}{2}
\placeArrowne{10}{1}
\placeArrowne{11}{0}
\placeArrowse{0}{12}
\placeArrowse{1}{11}
\placeArrowse{2}{10}
\placeArrowse{3}{9}
\placeArrowsw{4}{8}
\placeArrowne{5}{7}
\placeArrowne{6}{6}
\placeArrownw{7}{5}
\placeArrowsw{8}{4}
\placeArrowne{9}{3}
\placeArrowne{10}{2}
\placeArrowne{11}{1}
\placeArrowne{12}{0}
\placeArrowse{0}{13}
\placeArrowse{1}{12}
\placeArrowse{2}{11}
\placeArrowse{3}{10}
\placeArrowsw{4}{9}
\placeArrowne{5}{8}
\placeArrowne{6}{7}
\placeArrownw{7}{6}
\placeArrownw{8}{5}
\placeArrowne{9}{4}
\placeArrowne{10}{3}
\placeArrowne{11}{2}
\placeArrowne{12}{1}
\placeArrownw{13}{0}
\placeArrowse{1}{13}
\placeArrowse{2}{12}
\placeArrowse{3}{11}
\placeArrowsw{4}{10}
\placeArrowne{5}{9}
\placeArrowne{6}{8}
\placeArrownw{7}{7}
\placeArrownw{8}{6}
\placeArrowse{9}{5}
\placeArrowne{10}{4}
\placeArrowne{11}{3}
\placeArrowne{12}{2}
\placeArrownw{13}{1}
\placeArrowse{2}{13}
\placeArrowse{3}{12}
\placeArrowsw{4}{11}
\placeArrowse{5}{10}
\placeArrowse{6}{9}
\placeArrowne{7}{8}
\placeArrowne{8}{7}
\placeArrowse{9}{6}
\placeArrowse{10}{5}
\placeArrowne{11}{4}
\placeArrowne{12}{3}
\placeArrownw{13}{2}
\placeArrowse{3}{13}
\placeArrowsw{4}{12}
\placeArrowse{5}{11}
\placeArrowse{6}{10}
\placeArrowse{7}{9}
\placeArrowne{8}{8}
\placeArrowse{9}{7}
\placeArrowsw{10}{6}
\placeArrowse{11}{5}
\placeArrowne{12}{4}
\placeArrownw{13}{3}
\placeArrowsw{4}{13}
\placeArrowse{5}{12}
\placeArrowse{6}{11}
\placeArrowse{7}{10}
\placeArrowse{8}{9}
\placeArrowse{9}{8}
\placeArrowsw{10}{7}
\placeArrowse{11}{6}
\placeArrowne{12}{5}
\placeArrownw{13}{4}
\placeArrowse{5}{13}
\placeArrowse{6}{12}
\placeArrowse{7}{11}
\placeArrowse{8}{10}
\placeArrowse{9}{9}
\placeArrowsw{10}{8}
\placeArrowse{11}{7}
\placeArrowne{12}{6}
\placeArrownw{13}{5}
\placeArrowse{6}{13}
\placeArrowse{7}{12}
\placeArrowse{8}{11}
\placeArrowse{9}{10}
\placeArrowsw{10}{9}
\placeArrowse{11}{8}
\placeArrowne{12}{7}
\placeArrownw{13}{6}
\placeArrowse{7}{13}
\placeArrowse{8}{12}
\placeArrowse{9}{11}
\placeArrowsw{10}{10}
\placeArrowse{11}{9}
\placeArrowne{12}{8}
\placeArrownw{13}{7}
\placeArrowse{8}{13}
\placeArrowse{9}{12}
\placeArrowsw{10}{11}
\placeArrowse{11}{10}
\placeArrowne{12}{9}
\placeArrownw{13}{8}
\placeArrowse{9}{13}
\placeArrowsw{10}{12}
\placeArrowse{11}{11}
\placeArrowne{12}{10}
\placeArrownw{13}{9}
\placeArrowsw{10}{13}
\placeArrowse{11}{12}
\placeArrowne{12}{11}
\placeArrownw{13}{10}
\placeArrowse{11}{13}
\placeArrowne{12}{12}
\placeArrownw{13}{11}
\placeArrowse{12}{13}
\placeArrownw{13}{12}

%% file: conclusion.tex
\section{Concluding Remarks}
\label{sec:conclusion}
In this paper, we have analysed interval Markov decision processes under controller synthesis semantics in a dynamic setting. 
In particular, we discussed the problem of multi-objective robust strategy synthesis for \IMDP{}s, aiming for strategies that satisfy a given multi-objective predicate under all resolutions of the uncertainty in the transition probabilities. 
We first showed that this problem is \pspace-hard and then introduced a value iteration-based decision algorithm to approximate the Pareto set. 
Finally, we presented results obtained with a prototype tool on several real-world case studies to show the effectiveness of the developed algorithms.

Even though we focused on \IMDP{}s with multi-objective reachability and reward properties in this paper, the proposed robust synthesis algorithm can also handle \MDP{}s with convex uncertain sets and any $\omega$-regular properties such as LTL. 
For future work, we aim to explore the upper bound of the time complexity of the multi-objective robust strategy synthesis problem for \IMDP{}s which is left open in this paper.

%% file: appendix.tex
\section*{Appendix} 
\appendix 

This appendix contains the supplementary materials and also proofs of the results enunciated in the main part of the paper.
It is available for the reviewers in case they want to verify the correctness of the presented results;
it is not meant to be included in the final version of the paper.

\section{Proofs of the Results Enunciated in the Paper} % in Section~\ref{Sec:compositionality}}
\label{apx:proofs}
%=============================================
\synthesisComplexityIMDPs*
In order to prove the theorem, we need to define the multiple reachability problem for \MDP{}s. Formally,  
\begin{definition}
\label{app-def:multiReachabilityMDPs}
Given an \MDP{} $\mdp$ and a reachability predicate described as a vector $\phi=(\phi_{1}, \dotsc, \phi_{n})$ where $\phi_{j}= \reachPred{k_{j}}{\GoalSet_{j}}{\sim p_{j}}$ for $j \in \setnocond{1,\dotsc,n}$, the multiple reachability problem asks to check if there exists a strategy $\str$ of $\mdp$ such that $\mdp, \str \models \phi$. 
The almost-sure multiple reachability problem restricts to $\mathord{\sim} = \mathord{\geq}$ and $p_{j} = 1$ for all $j \in \setnocond{1, \dotsc, n}$.
\end{definition}
The proof makes use of the following lemma:
\begin{lemma}[Complexity of the multi-objective reachability problem for \MDP{}s~\cite{RRS15}]
	\label{lem:synthesizComplexityMDPs}
	Given an \MDP{} $\mdp$, the almost-sure multiple reachability problem is \pspace-complete and strategies need exponential memory in the query size.
\end{lemma}
\begin{myproof}[of Theorem~\ref{thm:synthesisComplexityIMDPs}]
	We reduce the problem in Lemma~\ref{lem:synthesizComplexityMDPs} to the one under our analysis.
	In fact, any instance of the multiple reachability problem for \MDP{} $\mdp$ can be seen as an instance of 
	the multi-objective robust strategy synthesis problem for an \IMDP{} $\imdp$ generated from $\mdp$ by replacing 
	all probability values with point intervals. Since the multiple reachability problem for \MDP{}s is \pspace-complete
	and the reduction is performed in polynomial time therefore, solving the robust strategy synthesis problem for \IMDP{}s
	is at least \pspace-hard. 
\end{myproof}
	
\SynthesisAlgorithmComplexity*	
\begin{myproof}
	The proof follows closely the one in~\cite{FKP12}. In every iteration of the loop in Algorithm~\ref{alg:synthesis}, a point $\vct{g}$ on a unique face of the Pareto curve is identified. 
	The number of faces of the Pareto curve $\mathcal{P}_{\imdp,\phi}$ is, in the worst case, exponential in $\size{\imdp}$, $\vct{k}$, and $n$~\cite{etessami2007}.  
	Therefore, termination of Algorithm~\ref{alg:synthesis} is guaranteed and the correctness is ensured as a result of the correctness of Algorithm~1 in~\cite{FKP12}.
	The soundness and completeness of the Algorithm~\ref{alg:synthesis} is followed by the fact that in every iteration of the algorithm through lines~\ref{Alg1:eq1Optimization}-\ref{Alg1:eq2Optimization}, the individual model checking problems can be solved in polynomial time in $\size{\imdp}$ by formulating the weighted sum of $n$ objectives as a linear programming problem. 
	To see this, without loss of generality, assume that $k_{i}=\infty$ for all $i \in \setnocond{1,\dotsc,n}$. 
	Therefore, following the approach in~\cite{PuggelliThesis2014}, the problem of maximising the $\ExpRew{\vct{w} \cdot \rew}{\vct{k}}$ across the range of strategies $\str\in \Str$ can be formulated as the following optimisation problem:
	\[
		\begin{array}{lp{10mm}l} 
			\min\limits_x \quad \vct{x}^T\vct{1} \\
			\text{subject to:} \\
			x_s \geq \sum_{i=1}^n w_{i} \cdot \rew_{i}(s,a) + \min\limits_{\fixeduncertainty{a}{s} \in \uncertainty{a}{s}} \vct{x}^T \fixeduncertainty{a}{s} && \forall s\in \States, \forall a\in\StateActionSet{s}\\
		\end{array} 
	\]     
    We now modify the above optimisation problem to simplify derivation of the LP problem. To this aim, we transform the optimisation operator ``$\min$'' to ``$\max$''. 
    Therefore, we get the following optimisation problem:
	\[
	\begin{array}{lp{10mm}l} 
	\max\limits_x \quad -\vct{x}^T\vct{1} \\
	\text{subject to:} \\
	x_s \geq \sum_{i=1}^n w_{i} \cdot \rew_{i}(s,a) + \min\limits_{\fixeduncertainty{a}{s} \in \uncertainty{a}{s}} \vct{x}^T \fixeduncertainty{a}{s} && \forall s\in \States, \forall a\in\StateActionSet{s}\\
	\end{array} 
	\]
As it is clear from the set of constraints in the latter optimization problem, the inner optimisation problem is not linear. 
In order to overcome this difficulty and induce the LP formulation, we follow the techniques in~\cite{PuggelliThesis2014} and use dual of the inner optimisation problem. 
To this aim, consider the inner optimisation problem with fixed $\vct{x}$:
\[
\begin{array}{lp{10mm}l} 
P(\vct{x}) := \min\limits_{\fixeduncertainty{a}{s} \in \uncertainty{a}{s}} \vct{x}^T \fixeduncertainty{a}{s}
\end{array} 
\]
Based on the general description of the interval uncertainty set $\uncertainty{a}{s} = \setcond{\fixeduncertainty{a}{s}}{\vec{0} \leq \underline{\fixeduncertainty{a}{s}} \leq \fixeduncertainty{a}{s} \leq \overline{\fixeduncertainty{a}{s}}\leq \vec{1}, \mathbf{1}^T\fixeduncertainty{a}{s}=1}$, we can rewrite the latter inner optimisation problem as:
\[
\begin{array}{lp{10mm}lp{10mm}l} 
\multicolumn{5}{l}{P(\vct{x}) := \min\vct{x}^T \fixeduncertainty{a}{s}}\\
\text{subject to:} \\
\mathbf{1}^T \fixeduncertainty{a}{s} = 1\\
\underline{\fixeduncertainty{a}{s}} \leq \fixeduncertainty{a}{s} \leq \overline{\fixeduncertainty{a}{s}}
\end{array} 
\] 
The dual of the above problem is formulated as follows:
\[
\begin{array}{lp{10mm}lp{10mm}l} 
\multicolumn{5}{l}{D(\vct{x}) := \max\limits_{\gamma_{j,1}^{s,a}, \gamma_{j,2}^{s,a}, \gamma_{j,3}^{s,a}} \gamma_{j,1}^{s,a} + \underline{\fixeduncertainty{a}{s}}^T \gamma_{j,2}^{s,a} - \overline{\fixeduncertainty{a}{s}}^T \gamma_{j,3}^{s,a}} \\
\text{subject to:} \\
\vct{x} - \gamma_{j,2}^{s,a} + \gamma_{j,3}^{s,a} - \gamma_{j,1}^{s,a} \vct{1} = \vct{0} &&  \\
\gamma_{j,2}^{s,a} \geq 0, \gamma_{j,3}^{s,a} \geq 0 &&
\end{array} 
\] 
Since the latter inner optimisation problem with fixed $\vct{x}$ is an LP, therefore due to the strong duality theorem~\cite{BT97}, we have $P^*(\vct{x})=D^*(\vct{x})$ where $P^*(\vct{x})$ and $D^*(\vct{x})$ are the primal and dual optimal values, respectively.
Therefore, we can replace the original inner optimisation problem with its dual LP to derive the ultimate LP formulation. 
Note that the inner optimisation operator is removed as the outer optimisation operator will find the least underestimate to maximise its objective function. 
Hence, maximising the expected total reward for \IMDP{} $\imdp$ with respect to the reward structure $\vct{w}\cdot\rew$ is formulated as the following LP which can in turn be solved in polynomial time.
\[
\begin{array}{lp{10mm}l} 
\multicolumn{3}{l}{\max\limits_{x,\gamma} \quad  -\vct{x}^T\vct{1} } \\
\text{subject to:} \\
x_s \geq \sum_{i=1}^n w_{i} \cdot \rew_{i}(s,a)+ \gamma_{j,1}^{s,a} + \underline{\fixeduncertainty{a}{s}}^T \gamma_{j,2}^{s,a} - \overline{\fixeduncertainty{a}{s}}^T \gamma_{j,3}^{s,a}  && \forall s \in \States, \forall a \in \StateActionSet{s} \\
\vct{x} - \gamma_{j,2}^{s,a} + \gamma_{j,3}^{s,a} - \gamma_{j,1}^{s,a} \vct{1} = \vct{0} && \forall s \in \States, \forall a \in \StateActionSet{s} \\
\gamma_{j,2}^{s,a}, \gamma_{j,3}^{s,a} \geq 0 && \forall s \in \States, \forall a \in \StateActionSet{s} 
\end{array} 
\]
\end{myproof}
\reductionToBasicForm*
\begin{myproof}
	Given a state $(s,v) \in \States'$, let $v_{e} = \setcond{i \in \setnocond{1, \dotsc, n}}{s \in \GoalSet_{i}} \setminus v$.
	By definition of the transition probability function, it follows that the only successors $(s',v')$ that can be reached from $(s,v)$ must have $v' = v \cup v_{e}$; 
	moreover, the action performed for such a transition must be of the form $(a,v_{e})$.
	This means that the sets $v_{e}$ and $v'$ are uniquely determined by the current state $(s,v)$;
	let $\nu \colon \States' \to 2^{\setnocond{1, \dotsc, n}}$ be the function such that $\nu(s,v) = \setcond{i \in \setnocond{1, \dotsc, n}}{s \in \GoalSet_{i}} \setminus v$ for each $(s,v) \in \States'$,
	$\nu_{\ActionSet} \colon \States' \times \ActionSet \to \ActionSet'$ be the function such that $\nu_{\ActionSet}((s,v),a) = (a, \nu(s,v))$ for each $(s,v) \in \States'$ and $a \in \ActionSet$, 
	and 
	$\nu_{\States} \colon \States' \times \States \to \States'$ be the function such that $\nu_{\States}((s,v), s') = (s', v \cup \nu(s,v))$ for each $(s,v) \in \States'$ and $s' \in \States$.
	
	It is immediate to see that every path $\Pat'$ of $\imdp'$,
	\[
		\Pat' = (s_{0},v_{0}) \xrightarrow{\elementfixeduncertainty{(a_{0},v'_{0})}{(s_{0},v_{0})}{(s_{1},v_{1})}} (s_{1},v_{1}) \xrightarrow{\elementfixeduncertainty{(a_{1},v'_{1})}{(s_{1},v_{1})}{(s_2,v_{2})}} (s_{2},v_{2}) \dots\text{,}
	\]
	is actually of the form 
	\[
		\Pat' = (s_{0},v_{0}) \xrightarrow{\elementfixeduncertainty{\nu_{\ActionSet}((s_{0},v_{0}),a_{0})}{(s_{0},v_{0})}{\nu_{\States}((s_{0},v_{0}),s_{1})}} (s_{1},v_{1}) \xrightarrow{\elementfixeduncertainty{\nu_{\ActionSet}((s_{1},v_{1}),a_{1})}{(s_{1},v_{1})}{\nu_{\States}((s_{1},v_{1}),s_{2})}} (s_{2},v_{2}) \dots
	\]
	where $(s_{j+1},v_{j+1}) = \nu_{\States}((s_{j},v_{j}), s_{j+1})$ for each $j \in \naturals$, i.e., $v_{j+1} = v_{j} \cup \nu(s_{j},v_{j})$.
	This means that we can define a bijection $\sharp \colon \paths \to \paths'$ as follows: 
	given a path $\Pat = s_{0} \xrightarrow{\elementfixeduncertainty{a_{0}}{s_{0}}{s_{1}}} s_{1} \xrightarrow{\elementfixeduncertainty{a_{1}}{s_{1}}{s_{2}}} s_{2} \dots$ of $\imdp$, $\sharp(\Pat)$ is defined as
	\[
		\sharp(\Pat) = (s_{0},v_{0}) \xrightarrow{\elementfixeduncertainty{(a_{0},v'_{0})}{(s_{0},v_{0})}{(s_{1},v_{1})}} (s_{1},v_{1}) \xrightarrow{\elementfixeduncertainty{(a_{1},v'_{1})}{(s_{1},v_{1})}{(s_{2},v_{2})}} (s_{2},v_{2}) \dots
	\]
	where $v_{0} = \emptyset$ and for each $j \in \naturals$, $(a_{j},v'_{j}) = \nu_{\ActionSet}((s_{j},v_{j}),a_{j})$ and $(s_{j+1},v_{j+1}) = \nu_{\States}((s_{j},v_{j}),s_{j})$.
	
	The inverse $\flat \colon \paths' \to \paths$ of $\sharp$ is just the projection on $\imdp$: 
	given a path $\Pat' = (s_{0},v_{0}) \xrightarrow{\elementfixeduncertainty{(a_{0},v'_{0})}{(s_{0},v_{0})}{(s_{1},v_{1})}} (s_{1},v_{1}) \xrightarrow{\elementfixeduncertainty{(a_{1},v'_{1})}{(s_{1},v_{1})}{(s_{2},v_{2})}} (s_{2},v_{2}) \dots$ of $\imdp'$, $\flat(\Pat')$ is defined as 
	\[
		\flat(\Pat') = s_{0} \xrightarrow{\elementfixeduncertainty{a_{0}}{s_{0}}{s_{1}}} s_{1} \xrightarrow{\elementfixeduncertainty{a_{1}}{s_{1}}{s_{2}}} s_{2} \dots\text{.}
	\]

	Moreover, since the sequence of sets $v_{0} v_{1} v_{2} \dots$ is monotonic non-decreasing with respect to the subset inclusion partial order, we have that, for a given $i \in \setnocond{1, \dotsc, n}$, if $i \in v_{N}$ for some $N \in \naturals$, then there exists exacly one $l \in \naturals$ such that $i \notin v_{j}$ for each $0 \leq j < l$ and $i \in v_{j}$ for each $j \geq l$, i.e., $s_{l}$ is the first time a state $s \in \GoalSet_{i}$ occurs along $\flat(\Pat')$.
	Therefore, it follows that $i \in \nu(s_{l},v_{l})$ while $i \notin \nu(s_{j},v_{j})$ for each $j \in \naturals \setminus \setnocond{l}$.
	This implies that $\rew_{\GoalSet_{i}}(\Pat'[l],\Pat'(l)) = 1$ if $\mathord{\sim_{i}} = \mathord{\geq}$ or $\rew_{\GoalSet_{i}}(\Pat'[l],\Pat'(l)) = -1$ if $\mathord{\sim_{i}} = \mathord{\leq}$ while $\rew_{\GoalSet_{i}}(\Pat'[j],\Pat'(j)) = 0$ for each $j \in \naturals \setminus \setnocond{l}$, thus
	\[
		\accumPathRew[\rew_{\GoalSet_{i}}]{\Pat'}{k} = 
		\begin{cases}
			1 & \text{if $l < k$ and $\mathord{\sim_{i}} = \mathord{\geq}$,} \\
			-1 & \text{if $l < k$ and $\mathord{\sim_{i}} = \mathord{\leq}$,} \\
			0 & \text{otherwise.}
		\end{cases}
	\]
	Note that, if $i \notin v_{j}$ for each $j \in \naturals$, then this means that $i \notin \nu(s_{j},v_{j})$ for each $j \in \naturals$, thus $\rew_{\GoalSet_{i}}(\Pat'[j],\Pat'(j)) = 0$ for each $j \in \naturals$ and $\accumPathRew[\rew_{\GoalSet_{i}}]{\Pat'}{k} = 0$.

	Similarly, for each $h \in \setnocond{n+1, \dotsc, m}$, we get that $\accumPathRew[\bar{\rew}_{h}]{\Pat'}{k} = \accumPathRew[\rew_{h}]{\Pat}{k}$ if $\mathord{\sim_{h}} = \mathord{\geq}$ and $\accumPathRew[\bar{\rew}_{h}]{\Pat'}{k} = -\accumPathRew[\rew_{h}]{\Pat}{k}$ if $\mathord{\sim_{h}} = \mathord{\leq}$.
	
	We are now ready to prove the statement of the proposition, by considering the two implications separately.
	
	Suppose that $\phi$ is satisfiable in $\imdp$:
	by definition, it follows that there exists a strategy $\str$ of $\imdp$ such that $\inducedMC \models_{\Env} \phi$, that is, $\inducedMC \models_{\Env} \reachPred{k_{i}}{\GoalSet_{i}}{\sim_{i} p_{i}}$ for each $i \in \setnocond{1, \dotsc, n}$ and $\inducedMC \models_{\Env} \rewPred[h]{\rew_{h}}{k_{h}}{r_{h}}$ for each $h \in \setnocond{n+1, \dotsc, m}$.
	Let $\str'$ be the strategy of $\imdp'$ such that, for each finite path $\Pat' \in \Fpat'$ and action $a \in \ActionSet$, $\str(\Pat')(\nu_{\ActionSet}(\last{\Pat'},a)) = \str(\flat(\Pat'))(a)$, $0$ otherwise.
	Intuitively, $\str'$ chooses the next action $(a,v)$ exactly as $\str$ chooses $a$ since $v$ is uniquely determined by $\Pat'$.
	We claim that $\str'$ is such that $\inducedMC['] \models_{\Env} \phi'$.
	
	Let $i \in \setnocond{1, \dotsc, n}$ and consider $\phi'_{i} = \rewPredMin{\rew_{\GoalSet_{i}}}{k_{i}+1}{p'_{i}}$:
	there are two cases depending on the original bound $\mathord{\sim_{i}}$.
	
	If $\mathord{\sim_{i}} = \mathord{\geq}$, then $\rewPredMin{\rew_{\GoalSet_{i}}}{k_{i}+1}{p'_{i}} = \rewPredMin{\rew_{\GoalSet_{i}}}{k_{i}+1}{p_{i}}$;
	$\inducedMC['] \models_{\Env'} \rewPredMin{\rew_{\GoalSet_{i}}}{k_{i}+1}{p_{i}}$ if and only if $\min_{\env'\in\Env'} \int_{\Pat'} \accumPathRew[\rew_{\GoalSet_{i}}]{\Pat'}{k_{i}+1} \,\mathrm{d}\Prob[\imdp']^{\str',\env'} \geq p_{i}$.
	Since for each path $\Pat' \in \paths'$, $\accumPathRew[\rew_{\GoalSet_{i}}]{\Pat'}{k_{i}+1} = 1$ if there exists $l < k_{i}+1$ such that $\flat(\Pat')[l] \in \GoalSet_{i}$, $\accumPathRew[\rew_{\GoalSet_{i}}]{\Pat'}{k_{i}+1} = 0$ otherwise, by the way $\intTransitionProbability'$ and $\str'$ are defined it follows that $\min_{\env'\in\Env'} \int_{\Pat'} \accumPathRew[\rew_{\GoalSet_{i}}]{\Pat'}{k_{i}+1} \,\mathrm{d}\Prob[\imdp']^{\str',\env'} = \min_{\env\in\Env} \Prob^{\str,\env} \setcond{\Pat \in \Infpat}{\exists l \leq k: \Pat[l] \in \GoalSet_{i}}$.
	Since by hypothesis $\phi$ is satisfiable in $\imdp$, then it follows that $\min_{\env\in\Env} \Prob^{\str,\env} \setcond{\Pat \in \Infpat}{\exists l \leq k: \Pat[l] \in \GoalSet_{i}} \geq p_{i}$, thus $\min_{\env'\in\Env'} \int_{\Pat'} \accumPathRew[\rew_{\GoalSet_{i}}]{\Pat'}{k_{i}+1} \,\mathrm{d}\Prob[\imdp']^{\str',\env'} \geq p_{i}$ holds as well, hence $\inducedMC['] \models_{\Env'} \rewPredMin{\rew_{\GoalSet_{i}}}{k_{i}+1}{p_{i}} = \rewPredMin{\rew_{\GoalSet_{i}}}{k_{i}+1}{p'_{i}}$ is satisfied, as required.
	
	Consider now the second case: 
	if $\mathord{\sim_{i}} = \mathord{\leq}$, then $\rewPredMin{\rew_{\GoalSet_{i}}}{k_{i}+1}{p'_{i}} = \rewPredMin{\rew_{\GoalSet_{i}}}{k_{i}+1}{-p_{i}}$;
	$\inducedMC['] \models_{\Env'} \rewPredMin{\rew_{\GoalSet_{i}}}{k_{i}+1}{-p_{i}}$ if and only if $\min_{\env'\in\Env'} \int_{\Pat'} \accumPathRew[\rew_{\GoalSet_{i}}]{\Pat'}{k_{i}+1} \,\mathrm{d}\Prob[\imdp']^{\str',\env'} \geq -p_{i}$.
	Since for each path $\Pat' \in \paths'$, $\accumPathRew[\rew_{\GoalSet_{i}}]{\Pat'}{k_{i}+1} = -1$ if there exists $l < k_{i}+1$ such that $\flat(\Pat')[l] \in \GoalSet_{i}$, $\accumPathRew[\rew_{\GoalSet_{i}}]{\Pat'}{k_{i}+1} = 0$ otherwise, by the way $\intTransitionProbability'$ and $\str'$ are defined it follows that $\min_{\env'\in\Env'} \int_{\Pat'} \accumPathRew[\rew_{\GoalSet_{i}}]{\Pat'}{k_{i}+1} \,\mathrm{d}\Prob[\imdp']^{\str',\env'} = -\max_{\env\in\Env} \Prob^{\str,\env} \setcond{\Pat \in \Infpat}{\exists l \leq k: \Pat[l] \in \GoalSet_{i}}$.
	Since by hypothesis we have that $\phi$ is satisfiable in $\imdp$, then it follows that $\max_{\env\in\Env} \Prob^{\str,\env} \setcond{\Pat \in \Infpat}{\exists l \leq k: \Pat[l] \in \GoalSet_{i}} \leq p_{i}$, thus $\min_{\env'\in\Env'} \int_{\Pat'} \accumPathRew[\rew_{\GoalSet_{i}}]{\Pat'}{k_{i}+1} \,\mathrm{d}\Prob[\imdp']^{\str',\env'} \geq -p_{i}$ holds as well, hence $\inducedMC['] \models_{\Env'} \rewPredMin{\rew_{\GoalSet_{i}}}{k_{i}+1}{-p_{i}} = \rewPredMin{\rew_{\GoalSet_{i}}}{k_{i}+1}{p'_{i}}$ is satisfied, as required.
	
	This completes the analysis of the case $\phi'_{i} = \rewPredMin{\rew_{\GoalSet_{i}}}{k_{i}+1}{p'_{i}}$ for each $i \in \setnocond{1, \dotsc, n}$.
	
	Let $h \in \setnocond{n+1, \dotsc, m}$ and consider $\phi'_{h} = \rewPredMin{\bar{\rew}_{h}}{k_{h}}{r'_{h}}$:
	there are two cases depending on the original bound $\mathord{\sim_{h}}$.
	
	If $\mathord{\sim_{h}} = \mathord{\geq}$, then $\rewPredMin{\bar{\rew}_{h}}{k_{h}}{r'_{h}} = \rewPredMin{\bar{\rew}_{h}}{k_{h}}{r_{h}}$;
	$\inducedMC['] \models_{\Env'} \rewPredMin{\bar{\rew}_{h}}{k_{h}}{r_{h}}$ holds if and only if $\min_{\env'\in\Env'} \int_{\Pat'} \accumPathRew[\bar{\rew}_{h}]{\Pat'}{k_{h}} \,\mathrm{d}\Prob[\imdp']^{\str',\env'} \geq r_{h}$ holds.
	Since for each path $\Pat' \in \paths'$, $\accumPathRew[\bar{\rew}_{h}]{\Pat'}{k} = \accumPathRew[\rew_{h}]{\flat(\Pat')}{k}$, by the way the components $\intTransitionProbability'$, $\bar{\rew}_{h}$, and $\str'$ are defined it follows that $\min_{\env'\in\Env'} \int_{\Pat'} \accumPathRew[\bar{\rew}_{h}]{\Pat'}{k_{h}} \,\mathrm{d}\Prob[\imdp']^{\str',\env'} = \min_{\env\in\Env} \int_{\Pat} \accumPathRew[\rew_{h}]{\Pat}{k_{h}} \,\mathrm{d}\Prob[\imdp]^{\str,\env}$.
	Since by hypothesis $\phi$ is satisfiable in $\imdp$, then it follows that $\min_{\env\in\Env} \int_{\Pat} \accumPathRew[\rew_{h}]{\Pat}{k_{h}} \,\mathrm{d}\Prob[\imdp]^{\str,\env} \geq r_{h}$, thus $\min_{\env'\in\Env'} \int_{\Pat'} \accumPathRew[\bar{\rew}_{h}]{\Pat'}{k_{h}} \,\mathrm{d}\Prob[\imdp']^{\str',\env'} \geq r_{h}$ holds as well, hence $\inducedMC['] \models_{\Env'} \rewPredMin{\bar{\rew}_{h}}{k_{h}}{r_{h}} = \rewPredMin{\bar{\rew}_{h}}{k_{h}}{r'_{h}}$ is satisfied, as required.
	
	Consider now the second case: 
	if $\mathord{\sim_{h}} = \mathord{\leq}$, then $\rewPredMin{\bar{\rew}_{h}}{k_{h}}{r'_{h}} = \rewPredMin{\bar{\rew}_{h}}{k_{h}}{-r_{h}}$;
	$\inducedMC['] \models_{\Env'} \rewPredMin{\bar{\rew}_{h}}{k_{h}}{-r_{h}}$ if and only if $\min_{\env'\in\Env'} \int_{\Pat'} \accumPathRew[\bar{\rew}_{h}]{\Pat'}{k_{h}} \,\mathrm{d}\Prob[\imdp']^{\str',\env'} \geq -r_{h}$.
	Since for each path $\Pat' \in \paths'$, $\accumPathRew[\bar{\rew}_{h}]{\Pat'}{k} = -\accumPathRew[\rew_{h}]{\flat(\Pat')}{k}$, by the way $\intTransitionProbability'$, $\bar{\rew}_{h}$, and $\str'$ are defined it follows that $\min_{\env'\in\Env'} \int_{\Pat'} \accumPathRew[\bar{\rew}_{h}]{\Pat'}{k_{h}} \,\mathrm{d}\Prob[\imdp']^{\str',\env'} = -\max_{\env\in\Env} \int_{\Pat} \accumPathRew[\rew_{h}]{\Pat}{k_{h}} \,\mathrm{d}\Prob[\imdp]^{\str,\env}$.
	Since by hypothesis $\phi$ is satisfiable in $\imdp$, then it follows that $\max_{\env\in\Env} \int_{\Pat} \accumPathRew[\rew_{h}]{\Pat}{k_{h}} \,\mathrm{d}\Prob[\imdp]^{\str,\env} \leq r_{h}$, thus $\min_{\env'\in\Env'} \int_{\Pat'} \accumPathRew[\bar{\rew}_{h}]{\Pat'}{k_{h}} \,\mathrm{d}\Prob[\imdp']^{\str',\env'} \geq -r_{h}$ holds as well, hence $\inducedMC['] \models_{\Env'} \rewPredMin{\bar{\rew}_{h}}{k_{h}}{-r_{h}} = \rewPredMin{\bar{\rew}_{h}}{k_{h}}{r'_{h}}$ is satisfied, as required.

	This completes the analysis of the case $\phi'_{h} = \rewPredMin{\bar{\rew}_{h}}{k_{h}}{r'_{h}}$ for each $h \in \setnocond{n+1, \dotsc, m}$; 
	since $\inducedMC['] \models_{\Env'} \phi'_{j}$ for each $j \in \setnocond{1, \dotsc, m}$, it follows that $\phi$ is satisfiable in $\imdp'$, as required to prove that ``if $\phi$ is satisfiable in $\imdp$, then $\phi'$ is satisfiable in $\imdp'$''.
	
	Suppose now the other implication, namely ``if $\phi'$ is satisfiable in $\imdp'$, then $\phi$ is satisfiable in $\imdp$'' and assume that $\phi'$ is satisfiable in $\imdp'$:
	by definition, it follows that there exists a strategy $\str'$ of $\imdp'$ such that $\inducedMC['] \models_{\Env'} \phi'$, that is, $\inducedMC['] \models_{\Env'} \rewPredMin{\rew_{\GoalSet_{i}}}{k_{i}+1}{p'_{i}}$ for each $i \in \setnocond{1, \dotsc, n}$ and $\inducedMC['] \models_{\Env'} \rewPredMin{{\bar{\rew}_{h}}}{k_{h}}{r'_{h}}$ for each $h \in \setnocond{n+1, \dotsc, m}$.
	Let $\str$ be the strategy of $\imdp$ such that, for each finite path $\Pat \in \Fpat$ and action $a \in \ActionSet$, $\str(\Pat)(a) = \str'(\sharp(\Pat))(a,v)$, $0$ otherwise, where $(a,v) = \nu_{\ActionSet}(\last{\sharp(\Pat)},a)$.
	Intuitively, $\str$ chooses the next action $a$ exactly as $\str'$ chooses $(a,v)$ since $v$ is uniquely determined by $\Pat'$.
	We claim that $\str$ is such that $\inducedMC \models_{\Env} \phi$.

	Let $i \in \setnocond{1, \dotsc, n}$ and consider $\phi_{i} = \reachPred{k_{i}}{\GoalSet_{i}}{\sim_{i} p_{i}}$:
	there are two cases depending on the bound $\mathord{\sim_{i}}$.
	
	If $\mathord{\sim_{i}} = \mathord{\geq}$, then $\inducedMC \models_{\Env} \reachPred{k_{i}}{\GoalSet_{i}}{\geq p_{i}}$ if and only if $\min_{\env\in\Env} \Prob^{\str,\env} \setcond{\Pat \in \Infpat}{\exists l \leq k: \Pat[l] \in \GoalSet_{i}} \geq p_{i}$.
	Since for each path $\Pat \in \paths$, $\accumPathRew[\rew_{\GoalSet_{i}}]{\sharp(\Pat)}{k_{i}+1} = 1$ if there exists $l < k_{i}+1$ such that $\Pat[l] \in \GoalSet_{i}$, $\accumPathRew[\rew_{\GoalSet_{i}}]{\sharp(\Pat)}{k_{i}+1} = 0$ otherwise, by the way $\intTransitionProbability'$ and $\str$ are defined it follows that $\min_{\env\in\Env} \Prob^{\str,\env} \setcond{\Pat \in \Infpat}{\exists l \leq k: \Pat[l] \in \GoalSet_{i}} = \min_{\env'\in\Env'} \int_{\Pat'} \accumPathRew[\rew_{\GoalSet_{i}}]{\Pat'}{k_{i}+1} \,\mathrm{d}\Prob[\imdp']^{\str',\env'}$.
	Since by hypothesis $\phi'$ is satisfiable in $\imdp'$, then it follows that $\min_{\env'\in\Env'} \int_{\Pat'} \accumPathRew[\rew_{\GoalSet_{i}}]{\Pat'}{k_{i}+1} \,\mathrm{d}\Prob[\imdp']^{\str',\env'} \geq p_{i}$, thus $\min_{\env\in\Env} \Prob^{\str,\env} \setcond{\Pat \in \Infpat}{\exists l \leq k: \Pat[l] \in \GoalSet_{i}} \geq p_{i}$ holds as well, hence $\inducedMC \models_{\Env} \reachPred{k_{i}}{\GoalSet_{i}}{\geq p_{i}} = \reachPred{k_{i}}{\GoalSet_{i}}{\sim_{i} p_{i}}$ is satisfied, as required.
	
	Consider now the second case: 
	If $\mathord{\sim_{i}} = \mathord{\leq}$, then $\inducedMC \models_{\Env} \reachPred{k_{i}}{\GoalSet_{i}}{\leq p_{i}}$ if and only if $\max_{\env\in\Env} \Prob^{\str,\env} \setcond{\Pat \in \Infpat}{\exists l \leq k: \Pat[l] \in \GoalSet_{i}} \leq p_{i}$.
	Since for each path $\Pat \in \paths$, $\accumPathRew[\rew_{\GoalSet_{i}}]{\sharp(\Pat)}{k_{i}+1} = -1$ if there exists $l < k_{i}+1$ such that $\Pat[l] \in \GoalSet_{i}$, $\accumPathRew[\rew_{\GoalSet_{i}}]{\sharp(\Pat)}{k_{i}+1} = 0$ otherwise, by the way $\intTransitionProbability'$ and $\str$ are defined it follows that $\max_{\env\in\Env} \Prob^{\str,\env} \setcond{\Pat \in \Infpat}{\exists l \leq k: \Pat[l] \in \GoalSet_{i}} = -\min_{\env'\in\Env'} \int_{\Pat'} \accumPathRew[\rew_{\GoalSet_{i}}]{\Pat'}{k_{i}+1} \,\mathrm{d}\Prob[\imdp']^{\str',\env'}$.
	Since by hypothesis $\phi'$ is satisfiable in $\imdp'$, then it follows that $\min_{\env'\in\Env'} \int_{\Pat'} \accumPathRew[\rew_{\GoalSet_{i}}]{\Pat'}{k_{i}+1} \,\mathrm{d}\Prob[\imdp']^{\str',\env'} \geq -p_{i}$, thus $\max_{\env\in\Env} \Prob^{\str,\env} \setcond{\Pat \in \Infpat}{\exists l \leq k: \Pat[l] \in \GoalSet_{i}} \leq p_{i}$ holds as well, hence $\inducedMC \models_{\Env} \reachPred{k_{i}}{\GoalSet_{i}}{\leq p_{i}} = \reachPred{k_{i}}{\GoalSet_{i}}{\sim_{i} p_{i}}$ is satisfied, as required.
	
	This completes the analysis of the case $\phi_{i} = \reachPred{k_{i}}{\GoalSet_{i}}{\sim_{i} p_{i}}$ for each $i \in \setnocond{1, \dotsc, n}$.
	
	Let $h \in \setnocond{n+1, \dotsc, m}$ and consider $\phi_{h} = \rewPred[h]{\rew_{h}}{k_{h}}{r_{h}}$:
	there are two cases depending on the original bound $\mathord{\sim_{h}}$.
	
	If $\mathord{\sim_{h}} = \mathord{\geq}$, then $\inducedMC \models_{\Env} \rewPredMin{\rew_{h}}{k_{h}}{r_{h}}$ if and only if $\min_{\env\in\Env} \int_{\Pat} \accumPathRew[\rew_{h}]{\Pat}{k_{h}} \,\mathrm{d}\Prob^{\str,\env} \geq r_{h}$.
	Since for each path $\Pat \in \paths$, $\accumPathRew[\bar{\rew}_{h}]{\sharp(\Pat)}{k} = \accumPathRew[\rew_{h}]{\Pat}{k}$, by the way $\intTransitionProbability'$, $\bar{\rew}_{h}$, and $\str$ are defined it follows that $\min_{\env\in\Env} \int_{\Pat} \accumPathRew[\rew_{h}]{\Pat}{k_{h}} \,\mathrm{d}\Prob[\imdp]^{\str,\env} = \min_{\env'\in\Env'} \int_{\Pat'} \accumPathRew[\bar{\rew}_{h}]{\Pat'}{k_{h}} \,\mathrm{d}\Prob[\imdp']^{\str',\env'}$.
	Since by hypothesis $\phi'$ is satisfiable in $\imdp'$, then $\min_{\env'\in\Env'} \int_{\Pat'} \accumPathRew[\bar{\rew}_{h}]{\Pat'}{k_{h}} \,\mathrm{d}\Prob[\imdp']^{\str',\env'} \geq r_{h}$, thus $\min_{\env\in\Env} \int_{\Pat} \accumPathRew[\rew_{h}]{\Pat}{k_{h}} \,\mathrm{d}\Prob^{\str,\env} \geq r_{h}$ holds as well, hence $\inducedMC \models_{\Env} \rewPredMin{\rew_{h}}{k_{h}}{r_{h}} = \rewPred[h]{\rew_{h}}{k_{h}}{r_{h}}$ is satisfied, as required.
	
	Consider now the second case: 
	\sloppy
	if $\mathord{\sim_{h}} = \mathord{\leq}$, then $\inducedMC \models_{\Env} \rewPredMax{\rew_{h}}{k_{h}}{r_{h}}$ if and only if $\max_{\env\in\Env} \int_{\Pat} \accumPathRew[\rew_{h}]{\Pat}{k_{h}} \,\mathrm{d}\Prob^{\str,\env} \leq r_{h}$.
	Since for each path $\Pat \in \paths$, $-\accumPathRew[\bar{\rew}_{h}]{\sharp(\Pat)}{k} = \accumPathRew[\rew_{h}]{\Pat}{k}$, by the definition of the components $\intTransitionProbability'$, $\bar{\rew}_{h}$, and $\str$ it is the case that $\max_{\env\in\Env} \int_{\Pat} \accumPathRew[\rew_{h}]{\Pat}{k_{h}} \,\mathrm{d}\Prob[\imdp]^{\str,\env} = -\min_{\env'\in\Env'} \int_{\Pat'} \accumPathRew[\bar{\rew}_{h}]{\Pat'}{k_{h}} \,\mathrm{d}\Prob[\imdp']^{\str',\env'}$.
	Since by hypothesis $\phi'$ is satisfiable in $\imdp'$, then $\min_{\env'\in\Env'} \int_{\Pat'} \accumPathRew[\bar{\rew}_{h}]{\Pat'}{k_{h}} \,\mathrm{d}\Prob[\imdp']^{\str',\env'} \geq -r_{h}$, thus $\max_{\env\in\Env} \int_{\Pat} \accumPathRew[\rew_{h}]{\Pat}{k_{h}} \,\mathrm{d}\Prob^{\str,\env} \leq r_{h}$ holds as well, hence $\inducedMC \models_{\Env} \rewPredMax{\rew_{h}}{k_{h}}{r_{h}} = \rewPred[h]{\rew_{h}}{k_{h}}{r_{h}}$ is satisfied, as required.
	
	This completes the analysis of the case $\phi_{h} = \rewPred[h]{\rew_{h}}{k_{h}}{r_{h}}$ for each $h \in \setnocond{n+1, \dotsc, m}$; 
	since $\inducedMC \models_{\Env} \phi_{j}$ for each $j \in \setnocond{1, \dotsc, m}$, it follows that $\phi$ is satisfiable in $\imdp$, as required to prove that ``if $\phi'$ is satisfiable in $\imdp'$, then $\phi$ is satisfiable in $\imdp$''.
	
	Having proved both implications, the statement of the proposition ``$\phi$ is satisfiable in $\imdp$ if and only if $\phi'$ is satisfiable in $\imdp'$'' holds, as required.
\end{myproof}
%=============================================
\section{A Procedure to Check the Reward-Finiteness Assumption~\ref{ASSUMPTION:REWARDFINITENESS}}% in Section~\ref{Sec:compositionality}}
\label{app:rewardFinitenessAssumption}
%=============================================
In this section, we discuss in detail how reward-finiteness assumption for a given \IMDP{} $\imdp$ and a synthesis query $\phi$ can be ensured. 

In order to describe the procedure that checks Assumption~\ref{ASSUMPTION:REWARDFINITENESS}, first we need to define a counterpart of end components of \MDP{}s for \IMDP{}s, to which we refer as a \textit{strong end-component} (SEC).  Intuitively, a SEC of an \IMDP{} is a sub-\IMDP{} for which there exists a strategy that forces the sub-\IMDP{} to remain in the end component and visit all its states infinitely often under any nature.  It is referred to as strong because it is independent of the choice of nature.  Formally,
\begin{definition}[Strong End-Component]
	\label{def:strong_endcomp}
	A \emph{strong end-component} (SEC) of an \IMDP{} $\imdp$ is $\SEC = (\States',\ActionSet')$, where $\States' \subseteq \States$ and $\ActionSet' \subseteq \bigcup_{s \in \States'} \ActionSet(s)$ such that 
	\begin{inparaenum}[(1)]
		\item
		$\sum_{s' \in \States'} \fixeduncertainty{a}{ss'} = 1$ for every $s \in \States'$ and $a \in \ActionSet'(s)$, and all $\fixeduncertainty{a}{s} \in \uncertainty{a}{s}$, and
		\item
		for all $s, s' \in \States'$ there is a finite path $\Pat = \Pat[0] \cdots \Pat[n]$ such that $\Pat[0] = s$, $\Pat[n] = s'$ and for all $0 \leq i \leq n-1$ we have $\Pat[i] \in \States'$ and $\Pat(i) \in \ActionSet'$.
	\end{inparaenum}
\end{definition}

\begin{remark} 
	\label{remark:computeSEC}
	The SECs of an \IMDP{} $\imdp$ can be identified by using any end-component-search algorithm of \MDP{}s on its underlying graph structure.  That is, since the lower transition probability bounds of $\imdp$ are strictly greater than zero for the transitions whose upper probability bounds are non-zero, the underlying graph structure of $\imdp$ is identical to the graph structure of every \MDP{} it contains.  Therefore, a SEC of $\imdp$ is an end-component of every contained \MDP{}, and vice versa.
\end{remark}

\begin{lemma}
	\label{lem:finiteOccurance}
	If state-action pair $(s,a)$ is not contained in a SEC, then 
	\begin{eqnarray*}
		\sup_{\str \in \Str} \inf_{\env \in \Env} \mathit{occ}^{\str}_\env ((s,a)) < \infty,
	\end{eqnarray*}
	where $\mathit{occ}^{\str}_\env((s,a))$ denotes the expected total number of occurrences of $(s,a)$ under $\str$ and $\env$.
\end{lemma}
\begin{myproof}
	If $(s,a)$ is not contained in a SEC of $\imdp$, then starting from $s$ and under action $a$, the probability of returning to $s$ is less than one, independent of the choice of strategy and nature. %, i.e., there always exists a path with non-zero probability starting from $(s,a)$ that never returns to $s$.
	Then, the proof follows from basic results of probability theory.
\end{myproof}

\begin{proposition}
	\label{appLemma:rewardfiniteness}
	Let $\SEC = (\States',\ActionSet')$ denote a SEC of \IMDP{} $\imdp$.  
	Then, we have \sloppy$\sup\setcond{\ExpRew{\rew}{\infty}}{\inducedMC\models_{\Env}(\reachPred{k_{1}}{\GoalSet_{1}}{\sim p_{1}},\dotsc,\reachPred{k_{n}}{\GoalSet_{n}}{\sim p_{n}})} = \infty$ for a reward structure $\rew$ of $\imdp$ iff there is a strategy $\str$ of $\imdp$ that $\inducedMC\models_{\Env}(\reachPred{k_{1}}{\GoalSet_{1}}{\sim p_{1}},\dotsc,\reachPred{k_{n}}{\GoalSet_{n}}{\sim p_{n}})$, $\SEC$ is reachable under $\str$, and $\rew(\Pat[i],\Pat(i))>0$, where $\Pat$ is a path under $\str$ with $\Pat[i] \in \States'$ and $\Pat(i) \in \ActionSet'(\Pat[i])$ for some  $i \geq 0$.
	
\end{proposition}	
\begin{myproof}
	We prove this proposition by adapting the proof from~\cite[Proposition~1]{FKNPQ11}.
	
	Direction $\Rightarrow$. Assume that, for a reward structure $\rew$, \sloppy$\sup\setcond{\ExpRew{\rew}{\infty}}{\inducedMC\models_{\Env}(\reachPred{k_{1}}{\GoalSet_{1}}{\sim p_{1}},\dotsc,\reachPred{k_{n}}{\GoalSet_{n}}{\sim p_{n}})} = \infty$.  From Lemma~\ref{lem:finiteOccurance}, it follows that if state-action pair $(s,a)$ occurs infinitely often, $s$ and $a$ are contained in a SEC $\SEC$.  Therefore, to satisfy the assumed condition, there must exist some strategy $\str$ such that $\inducedMC\models_{\Env}(\reachPred{k_{1}}{\GoalSet_{1}}{\sim p_{1}},\dotsc,\reachPred{k_{n}}{\GoalSet_{n}}{\sim p_{n}})$ and a SEC is reachable, in which $\str$ picks action $a$ at reachable state $s$ with positive probability, and $\rew(s,a) > 0$.
	
	Direction $\Leftarrow$. Assume that there is a strategy $\str$ such that $\inducedMC\models_{\Env}(\reachPred{k_{1}}{\GoalSet_{1}}{\sim p_{1}},\dotsc,\reachPred{k_{n}}{\GoalSet_{n}}{\sim p_{n}})$, a SEC $\SEC = (\States',\ActionSet')$ is reachable, and $\rew(\Pat[n],\Pat(n))>0$, where $\Pat$ is a finite path of length $n+1$ under $\str$ with $\Pat[n] \in \States'$ and $\Pat(n) \in \ActionSet'(\Pat[n])$ for some  $n \geq 0$.  To complete the proof, it is enough to show that there is a sequence of strategies $\{\str_k\}_{k\in \naturals}$ under which (i) the probabilistic predicates $\reachPred{k_{1}}{\GoalSet_{1}}{\sim p_{1}},\dotsc,\reachPred{k_{n}}{\GoalSet_{n}}{\sim p_{n}}$ are satisfied and (ii) $\lim_{k \rightarrow \infty} \ExpRewK{\rew}{k} = \infty$.\\
	(i) Let $\Pat[n] = s$ and $\Pat(n) = a$. For $k \in \naturals$ consider $\str_k$ that
	\begin{itemize}
		\item for the paths that do not have the prefix $\Pat$, $\str_k$ emulates $\str$.
		\item when the path $\Pat$ is performed, $\str_k$ forces the system to stay in $\SEC$ containing $(s,a)$.  After $k$ occurrences of $(s,a)$, the next time $s$ is visited, the strategy $\str_k$ emulates $\str$ again as if the performed path segment after $\Pat[n]$ was never executed.
	\end{itemize}
	Under $\str_k$, the reachability predicates are satisfied for any $k \in \naturals$.  To see this, consider $\theta_k$ that maps each path $\Pat$ of $\str$ to the paths of $\str_k$.  We now have $\theta(\Pat) \cap \theta(\Pat') = \emptyset$ for all $\Pat \neq \Pat'$, and for all sets $\Omega$ and two natures $\env$ and $\env_k$, where $\env_k$ emulates $\env$ the same way $\str_k$ emulates $\str$, we have $\Prob^{\str,\env}(\Omega) = \Prob^{\str_k,{\env_k}}(\theta(\Omega))$, independent of the choice of $\env_k$ during the execution of the path segment that $\str_k$ forces the stay in $\SEC$.  The satisfaction of the reachability predicates under each $\str_k$ follows from the fact that, for any path $\Pat$ of $\str$, $\Pat$ satisfies a reachability predicate iff each path in $\theta(\Omega)$ satisfies the reachability predicate.\\
	(ii) To show that $\lim_{k \rightarrow \infty} \ExpRewK{\rew}{k} = \infty$, recall that the probability of reaching $(s,a)$ under $\str_k$ for the first time is some positive value $p_{1}$.  From the properties of SEC, the probability of returning to $s$ within $l$ steps, where $l=|\States|$, is also some positive value $p_2$.  By construction, $(s,a)$ is picked $k$ times, therefore, $\ExpRewK{\rew}{k} \geq p_{1} p_2 \frac{k}{l} \rew(s,a)$, and hence, $\lim_{k \rightarrow \infty} \ExpRewK{\rew}{k} = \infty$.
\end{myproof}
We can now construct, from $\imdp$, an \IMDP{} $\bar{\imdp}$ that is equivalent to $\imdp$ in terms of satisfaction of $\phi$ but does not include actions with positive rewards in its SEC.  
The algorithm is similar to the one introduced in~\cite{FKNPQ11} for \MDP{}s and is as follows. 
First, remove action $a$ from $\StateActionSet{s}$ if $(s,a)$ is contained in a SEC and $\rew(s,a) > 0$ for some maximizing reward structure $\rew$.  
Second, recursively remove states with no outgoing transitions and transitions that lead to non-existent states until a fixpoint is reached.
\begin{proposition}
	\label{appLemma:rewardfiniteness_AlgCorr}
	There is a strategy $\str$ of $\imdp$ such that $\ExpRew{\rew}{\infty} = x < \infty$ and $\inducedMC\models_{\Env} \phi$ iff there is a strategy $\bar{\str}$ of $\bar{\imdp}$ such that $\ExpRewBar{\rew}{\infty} = x$ and $\inducedMCBar\models_{\Env} \phi$.
\end{proposition}
\begin{myproof}
	The proof follows straightforwardly from Proposition~\ref{appLemma:rewardfiniteness}.
\end{myproof}

%=============================================
\section{Multi-objective Robust Strategy Synthesis: Other Queries} 
\label{app:other-queries}
%=============================================
For the sake of completeness of our approach, in this section we discuss other types of multi-objective queries
and present algorithms to solve them. In particular, we follow the same direction as~\cite{FKP12} and show how   
Algorithm~\ref{alg:synthesis} can be adapted to solve these types of queries.

We first start with the definition of quantitative and Pareto queries. Formally,

\begin{definition}[Quantitative Queries~\cite{FKP12}]
	\label{def:quantitativeQueries}
	Given an \IMDP{} $\imdp$ and a multi-objective predicate $\phi$, a quantitative query is of the form $\quantity(\objective{k_{1}}{o}{\star}, (\phi_2,\dots,\phi_{n}))$, consisting of a multi-objective predicate $(\phi_2, \dotsc, \phi_{n})$ of size $n-1$ and an objective $\objective{k_{1}}{o}{\star}$ where $o$ is a target set $\GoalSet$ or a reward structure $\rew$, $k_{1} \in \extNaturals$ and $\star \in \setnocond{\min,\max}$. 
	We define:
	\begin{align*}
		\quantity(\objective{k_{1}}{o}{\min}, (\phi_2, \dotsc, \phi_{n})) & = \inf\setcond{x \in \reals}{\text{$(\objective{k_{1}}{o}{\leq x}, \phi_2, \dotsc, \phi_{n})$  is satisfiable}} \\
		\quantity(\objective{k_{1}}{o}{\max}, (\phi_2, \dotsc, \phi_{n})) &= \sup \setcond{x \in \reals}{\text{$(\objective{k_{1}}{o}{\geq x}, \phi_2, \dotsc,\phi_{n})$ is satisfiable}}\text{.}
	\end{align*}
\end{definition}

\begin{definition}[Pareto Queries~\cite{FKP12}]
	\label{def:paretoQueries}
	Given an \IMDP{} $\imdp$ and a multi-objective predicate $\phi$, a Pareto query is of the form $\mathit{Pareto}(\objective{k_{1}}{o_{1}}{\star_{1}}, \dotsc, \objective{k_{n}}{o_{n}}{\star_{n}})$,
	where each $\objective{k_{i}}{o_{i}}{\star_{i}}$ is an objective in which $o_{i}$ is either a target set $\GoalSet$ or a reward structure $\rew$, $k_{i} \in \extNaturals$ and $\star_{i} \in \setnocond{\min,\max}$. 
	We define the set of achievable values as $A = \setcond{\vct{x} \in \reals^{n}}{\text{$(\objective{k_{1}}{o_{1}}{\sim_{1} x_{1}}, \dotsc, \objective{k_{n}}{o_{n}}{\sim_{n} x_{n}})$ is satisfiable}}$ where
	\[	
		\mathord{\sim}_{i} = 
		\begin{cases}
		\geq & \text{if $\star_{i}=\max$} \\
		\leq & \text{if $\star_{i}=\min$}	
		\end{cases}
	\]
	Then,  	
	\[
		\mathit{Pareto}(\objective{k_{1}}{o_{1}}{\star_{1}}, \dotsc, \objective{k_{n}}{o_{n}}{\star_{n}}) = \setcond{\vct{x} \in A}{\text{$\vct{x}$ is Pareto optimal}}.
	\]	
\end{definition}

Note that the quantitative queries asks to maximise or minimise the reachability/reward
objective over the set of strategies satisfying $\phi$. 
The Pareto queries ask to determine the Pareto set for a given set of objectives. 

\subsection{Algorithms for Robust Synthesis of Multi-objective Queries}

We now discuss algorithmic solutions to solve quantitative and Pareto queries. These algorithms are in fact designed as an adaption of Algorithm~\ref{alg:synthesis}
as detailed below and can also be considered as an extension of their counterparts in~\cite{FKP12} under presence of model uncertainty. 

\paragraph{Quantitative queries.} 
Let us first focus on the quantitative queries. 
To this end, without loss of generality, consider the quantitative query 
$\quantity(\objective{k_{1}}{\rew_{1}}{\max}, (\rewPredMin{\rew_{2}}{k_{2}}{r_{2}}, \dotsc, \rewPredMin{\rew_{n}}{k_{n}}{r_{n}})$. 
Algorithm~\ref{alg:quantitative}, similarly to Algorithm~\ref{alg:synthesis}, generates a sequence of points $\vct{g}$ on the Pareto curve from a sequence of weight vectors $\vct{w}$.
In order to optimise the objective $\rew_{1}$ as detailed in~\cite{FKP12}, a sequence of lower bounds $r_{1}$ is generated which are used in the same manner as Algorithm~\ref{alg:synthesis}. 
In particular, in the initial step we let $r_{1}$ be the minimum value for $\rew_{1}$ that can be computed with an instance of value iteration~\cite{PuggelliThesis2014}. 
The sequence of non-decreasing values for $r_{1}$ are generated at the next steps based on the set of points $X$ specified so far. 
In each step, the computation in the lines~\ref{QuantitativeAlg:eq1Optimization}-\ref{QuantitativeAlg:eq2Optimization} of Algorithm~\ref{alg:quantitative} can again be done using Algorithm~\ref{alg:valueAlg}.   
\begin{algorithm}[!t]
	\small
	\caption{Algorithm for solving robust quantitative queries\label{alg:quantitative}}
	\SetKwInOut{Input}{Input}\SetKwInOut{Output}{Output}
	\KwIn{An \IMDP{} $\imdp$, objective $\objective{k_{1}}{\rew_{1}}{\max}$, multi-objective predicate $(\rewPredMin{\rew_{2}}{k_{2}}{r_{2}}, \dotsc, \rewPredMin{\rew_{n}}{k_{n}}{r_{n}})$}
	\KwOut{value of $\quantity(\objective{k_{1}}{\rew_{1}}{\max}, (\rewPredMin{\rew_{2}}{k_{2}}{r_{2}}, \dotsc, \rewPredMin{\rew_{n}}{k_{n}}{r_{n}}))$}
	\Begin{
		$X = \emptyset$; $\rew = (\rew_{1}, \dotsc, \rew_{n})$;\\
		$\vct{k}=(k_{1}, \dotsc, k_{n})$; $\vct{r}=(\min_{\str\in\Str} \ExpRew{\rew_{1}}{\vct{k}}, r_{2}, \dotsc, r_{n})$;\label{QuantitativeAlg:eq3Optimization} \\
		\While{$\vct{r} \notin \dwc{X}$ or $\vct{w} \cdot \vct{g} > \vct{w} \cdot \vct{r}$}{
			Find $\vct{w}$ separating $\vct{r}$ from $\dwc{X}$ such that $w_{1} > 0$;\\
			Find strategy $\str$ maximising $\ExpRew{\vct{w} \cdot \rew}{\vct{k}}$;\label{QuantitativeAlg:eq1Optimization}\\
			$\vct{g} := (\ExpRew{\rew_{i}}{k_{i}})_{1\leq i\leq n}$;\label{QuantitativeAlg:eq2Optimization}\\
			\If {$\vct{w} \cdot \vct{g} < \vct{w} \cdot\vct{r}$}{ \Return $\perp$;}
			$X = X \cup \setnocond{\vct{g}}$; $r_{1}:= \max\setnocond{r_{1}, \max\setcond{r'}{(r',r_{2}, \dotsc, r_{n}) \in \dwc{X}}}$;
		}
		\Return $r_{1}$;
	}
\end{algorithm}

At this point it is worthwhile to mention that our extended Algorithm~\ref{alg:quantitative} is different from its counterpart in~\cite{FKP12} (cf. Algorithm~3) especially in lines~\ref{QuantitativeAlg:eq3Optimization},~\ref{QuantitativeAlg:eq1Optimization}-\ref{QuantitativeAlg:eq2Optimization}. 
In fact, all computations in these lines are performed while considering the behaviour of an adversarial nature as detailed in Algorithm~\ref{alg:valueAlg}. 
       
\paragraph{Pareto queries.} 
We next discuss the Pareto queries. 
Our algorithm is depicted as Algorithm~\ref{alg:pareto} which is in principle an extension of its counterpart in~\cite{FKP12} (cf. Algorithm~3). 
Likewise Algorithm~\ref{alg:quantitative}, the key differences of this algorithm with its counterpart
are in lines~\ref{ParetoAlg:eq1Optimization}-\ref{ParetoAlg:eq2Optimization} and~\ref{ParetoAlg:eq3Optimization}-\ref{ParetoAlg:eq4Optimization}.        
Following the same direction as in~\cite{FKP12}, we solely concentrate on two objectives while in theory this can be extended to an arbitrary number of objectives.
Since the number of faces of the Pareto curve is exponentially large and also the result of the value iteration algorithm to compute the individual points is an approximation, Algorithm~\ref{alg:pareto} only constructs an $\epsilon$-approximation of the Pareto curve.

\begin{algorithm}[!t]
	\small
	\caption{Algorithm for solving robust Pareto queries\label{alg:pareto}}
	\SetKwInOut{Input}{Input}\SetKwInOut{Output}{Output}
	\KwIn{An \IMDP{} $\imdp$, reward structures $\rew = (\rew_{1}, \rew_{2})$, time bounds $(k_{1}, k_{2})$, $\epsilon \in \posreals$}
	\KwOut{An $\epsilon$-approximation of the Pareto curve}
	\Begin{
		$X = \emptyset$; $Y \colon \reals^2 \to 2^{\reals^2}$ with initial $Y(x) = \emptyset$ for all $x$; $\vct{w} = (1,0)$;\\
		Find strategy $\str$ maximising $\ExpRew{\vct{w} \cdot \rew}{\vct{k}}$;\label{ParetoAlg:eq1Optimization}\\
		$\vct{g} := (\ExpRew{\rew_{1}}{k_{1}}, \ExpRew{\rew_{2}}{k_{2}})$;\label{ParetoAlg:eq2Optimization}\\	
		$X := X \cup \setnocond{\vct{g}}$; $Y(\vct{g}) := Y(\vct{g}) \cup \setnocond{\vct{w}}$; $\vct{w} := (0,1)$;\\
		\While{$\vct{w} \neq \bot$}{
			Find strategy $\str$ maximising $\ExpRew{\vct{w} \cdot \rew}{\vct{k}}$;\label{ParetoAlg:eq3Optimization}\\
			$\vct{g} := (\ExpRew{\rew_{1}}{k_{1}}, \ExpRew{\rew_{2}}{k_{2}})$;\label{ParetoAlg:eq4Optimization}\\
			$X := X \cup \setnocond{\vct{g}}$; $Y(\vct{g}) := Y(\vct{g}) \cup \setnocond{\vct{w}}$; $\vct{w} := \bot$;\\	
			Order $X$ to a sequence $\vct{x}^1, \dotsc, \vct{x}^m$ such that $\forall i: x^i_{1} \leq x^{i+1}_{1}$ and $x^i_2 \geq x^{i+1}_2$;\\
			\For{$i=1$ \KwTo $m$}{
				Let $\vct{u}$ be the element of $Y(\vct{x}^i)$ with maximal $u_{1}$;\\
				Let $\vct{u}'$ be the element of $Y(\vct{x}^{i+1})$ with minimal $u'_{1}$;\\
				Find a point $\vct{p}$ such that $\vct{u} \cdot \vct{p} = \vct{u} \cdot \vct{x}^i$ and $\vct{u}' \cdot \vct{p} = \vct{u}' \cdot \vct{x}^{i+1}$;\\	
				\If {distance of $\vct{p}$ from $\dwc{X}$ is $\geq \epsilon$}
				{Find $\vct{w}$ separating $\dwc{X}$ from $\vct{p}$, maximising $\vct{w} \cdot \vct{p} - \max_{\vct{x} \in \dwc{X}} \vct{w} \cdot \vct{x}$;\\
				\Break;}
			}}
			\Return $X$;
		}
\end{algorithm}
%=============================================
\section{The ANTG Case Study: Further Analysis on Strategies} 
\label{app:ANTGother-strategies}
%=============================================
In Fig.~\ref{fig:different-scheduling}, we provide strategies for different points on the Pareto curve in Fig.~\ref{fig:ANTG-paretoCurve}.
The lowest expected number of steps in which the museum can be left at all is 30.9665389.
To achive this number, there is a single optimal strategy sketched in Fig.~\ref{fig:different-scheduling-fewest}.
As we see, the tourist indeed leaves the museum as soon as possible, ignoring any closed exhibitions and receiving an expected penalty as high as 152.0609886.

In Fig.~\ref{fig:different-scheduling-31-1} and Fig.~\ref{fig:different-scheduling-31-2}, we give the tourist somewhat more time, namely 31 steps, so that the penalty of 151.7077821 is somewhat lower.
Here, with a high probability (0.9894174) the same strategy as for the previous case is chosen.
With a probability of 0.0105826 however, the less reckless strategy of Fig.~\ref{fig:different-scheduling-31-2} is used, which takes some efforts to avoid the last row of closed exhibitions at $x=11$.

If we further increase the time bound to 40, as in Fig.~\ref{fig:different-scheduling-40-1} and Fig.~\ref{fig:different-scheduling-40-2}, the strategies used become even less risky but more time consuming to execute.

For a step bound of 76.8658133 and larger, it is possible to avoid receiving any penalty by using the strategy of Fig.~\ref{fig:different-scheduling-maxsteps}, which circumvents all of the closed exhibitions.

\begin{figure}[!t]
  \centering
  30.9665389 steps, 152.0609886 penalty\\
  \begin{subfigure}{0.35\linewidth}
  \resizebox{\linewidth}{!}{
    \begin{tikzpicture}[every node/.style={minimum size=.35cm-\pgflinewidth, outer sep=0pt}, scale=0.7]
      \input{museum-arena.tex}
      \input{museum-strategy-fewest.tex}
  \end{tikzpicture}}
  \caption{Probability 1}
  \label{fig:different-scheduling-fewest}
\end{subfigure}%
\ \\\ \\\ \\
  31 steps, 151.7077821 penalty\\
\begin{subfigure}{0.35\linewidth}
  \resizebox{\linewidth}{!}{
    \begin{tikzpicture}[every node/.style={minimum size=.35cm-\pgflinewidth, outer sep=0pt}, scale=0.7]
      \input{museum-arena.tex}
      \input{museum-strategy-31-1.tex}
  \end{tikzpicture}}
  \caption{Probability 0.9894174}
  \label{fig:different-scheduling-31-1}
\end{subfigure}%
\hfill
\begin{subfigure}{0.35\linewidth}
  \resizebox{\linewidth}{!}{
    \begin{tikzpicture}[every node/.style={minimum size=.35cm-\pgflinewidth, outer sep=0pt}, scale=0.7]
      \input{museum-arena.tex}
      \input{museum-strategy-31-2.tex}
  \end{tikzpicture}}
  \caption{Probability 0.0105826}
  \label{fig:different-scheduling-31-2}
\end{subfigure}
\ \\\ \\\ \\
40 steps, 59.0123994 penalty\\
\begin{subfigure}{0.35\linewidth}
  \resizebox{\linewidth}{!}{
    \begin{tikzpicture}[every node/.style={minimum size=.35cm-\pgflinewidth, outer sep=0pt}, scale=0.7]
      \input{museum-arena.tex}
      \input{museum-strategy-40-1.tex}
  \end{tikzpicture}}
  \caption{Probability 0.7230247}
  \label{fig:different-scheduling-40-1}
\end{subfigure}%
\hfill
\begin{subfigure}{0.35\linewidth}
  \resizebox{\linewidth}{!}{
    \begin{tikzpicture}[every node/.style={minimum size=.35cm-\pgflinewidth, outer sep=0pt}, scale=0.7]
      \input{museum-arena.tex}
      \input{museum-strategy-40-2.tex}
  \end{tikzpicture}}
  \caption{Probability 0.2769753}
  \label{fig:different-scheduling-40-2}
\end{subfigure}
\ \\\ \\\ \\
76.8658133 steps, 0 penalty\\
\begin{subfigure}{0.35\linewidth}
  \resizebox{\linewidth}{!}{
    \begin{tikzpicture}[every node/.style={minimum size=.35cm-\pgflinewidth, outer sep=0pt}, scale=0.7]
      \input{museum-arena.tex}
      \input{museum-strategy-maxsteps.tex}
  \end{tikzpicture}}
  \caption{Probability 1}
  \label{fig:different-scheduling-maxsteps}
\end{subfigure}%
\caption{\label{fig:different-scheduling}Strategies for different points on the Pareto curve in Fig.~\ref{fig:ANTG-paretoCurve}.}
\end{figure}

%=====================================================================================================================================
\section{Generation of randomised strategies}
\label{app:memoryless-strategies-generation}
%=====================================================================================================================================
We consider a fixed \IMDP{} $\imdp = (\States, \InitState, \ActionSet, \intTransitionProbability)$ and a basic multi-objective predicate $(\rewPredMin{\rew_{1}}{k_{1}}{r_{1}},\dotsc,\rewPredMin{\rew_{n}}{k_{n}}{r_{n}})$.
For clarity, we assume that all $k_{i} = \infty$;
we discuss the extension to $k_{i} < \infty$ afterwards.
In the following, we will describe how we can obtain a randomised algorithm from the results computed by Algorithms~\ref{alg:synthesis},~\ref{alg:quantitative}, and~\ref{alg:pareto}.
These algorithms compute a set $X = \setnocond{\vct{g}_{1}, \dotsc, \vct{g}_{m}}$ of reward vectors $\vct{g}_{i} = (g_{i,1}, \dotsc, g_{i,n})$ and their corresponding set of strategies $\Sigma = \setnocond{ \str_{1}, \dotsc, \str_{m} }$, where strategy $\str_{i}$ achieves the reward vector $\vct{g}_{i}$.

In the descriptions of the given algorithms, the strategies $\str_{i}$ are not explicitly stored and mapped to the reward they achieve, but they can be easily adapted.
All used strategies are memoryless and deterministic; 
this means that we can treat them as functions of the form $\str_{i} \colon \States \to \ActionSet$ or, equivalently, as functions $\str_{i} \colon \States \times \ActionSet \to \setnocond{0, 1}$ where $\str_{i}(s, a) = 1$ if $\str_{i}(s) = a$ and $\str_{i}(s,\functionDot ) = 0$ otherwise. 

From the set $X$, we can compute a set $P = \setnocond{p_{1}, \dotsc, p_{m}}$ of the probabilities with which each of these strategies shall be executed.
If we execute each $\str_{i}$ with its according probability $p_{i}$, the vector of total expected rewards is $\vct{g} = \sum_{i=1}^{m} p_{i} \vct{g}_{i}$.
Let $\vct{r} = (r_{1}, \dotsc, r_{n})$ denote the vector of reward bounds of the multi-objective predicate.
To obtain $P$ after having executed Algorithm~\ref{alg:synthesis}, we can choose the values $p_{i}$ in $P$ such that they fulfill the constraints $\sum_{i=1}^{m} \vct{g}_{i} p_{i} \geq \vct{r}$, $\sum_{i=1}^{m} p_{i} = 1$ and $p_{i} \geq 0$ for each $1 \leq i \leq m$.
For the other algorithms, $P$ can be computed accordingly.

To obtain a stochastic process with expected values $\vct{g}$, we initially randomly choose one of the memoryless deterministic strategies $\str_{i}$ according to their probabilities in $P$.
Afterwards, we just keep executing the chosen $\str_{i}$.
The initial choice of the strategy to execute is the only randomised choice to be made.
We do \emph{not} perform a random choice after the initial choice of $\str_{i}$.

This process of obtaining the expected rewards $\vct{g}$ indeed uses memory, because we have to remember the deterministic strategy which was randomly chosen to be executed.
On the other hand, we only need a very limited way of randomisation.

We like to emphasise that indeed we cannot just construct a memoryless randomised strategy by choosing the strategy $\str_{i}$ with probability $p_{i}$ in each step anew.
\begin{wrapfigure}[15]{right}{33mm}
  \centering
%  \vskip-10mm
  \begin{tikzpicture}[->, auto, >=stealth', semithick, baseline, 
      state/.style={draw, circle, inner sep=0, text centered, minimum size=5mm},
      trannode/.style={draw, circle, fill, minimum size=1mm, inner sep=0mm},
      prob/.style={font=\scriptsize},
    ]
    \node[state, initial left, initial text={}] (s) at (0,0) {\vphantom{$t$}$s$};
    \node[state] (t) at (1.5,0) {$t$};
    \node[state] (u) at (0,-1.5) {\vphantom{$t$}$u$};
    \node[state] (v) at (1.5,-1.5) {\vphantom{$t$}$v$};
    \node[state] (w) at (0,-3) {\vphantom{$t$}$w$};

    \draw (s) to node[prob,above] {$a,\underline{1}$} (t);
    \draw (s) to node[prob,right] {$b,\underline{0}$} (u);
    \draw (u) to node[prob,above] {$a,\underline{0}$} (v);
    \draw (u) to node[prob,right] {$b,\underline{1}$} (w);
    \draw (t) edge[loop right] node[prob, right] {$a,\underline{0}$} (t);
    \draw (t) edge[loop below] node[prob, below] {$b,\underline{0}$} (t);
    \draw (v) edge[loop right] node[prob, right] {$a,\underline{0}$} (v);
    \draw (v) edge[loop below] node[prob, below] {$b,\underline{0}$} (v);
    \draw (w) edge[loop right] node[prob, right] {$a,\underline{0}$} (w);
    \draw (w) edge[loop below] node[prob, below] {$b,\underline{0}$} (w);
  \end{tikzpicture}
  \caption{Computing randomised strategies.}
  \label{fig:randomised-strategies}
\end{wrapfigure}
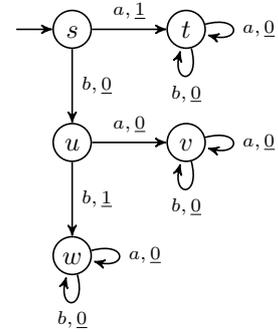
\begin{myexample}
  Consider the \IMDP{} in Fig.~\ref{fig:randomised-strategies}.
  We only have two possible actions, $a$ and $b$.
  The initial state is $s$ and all probability intervals are the interval $\interval{1}{1}$, which we omit for readability;
  thus, there is also only one possible nature $\env$.
  There is only a single reward structure, indicated by the underlined numbers.
  If we choose $a$ in state $s$, we end up in $t$ in the next step and obtain a reward of $1$ with certainty, while if we choose $b$, we will be in $u$ in the next step and obtain a reward of $0$, and accordingly for the other states.
  
  We consider the strategies $\str_a$ which chooses $a$ in each state and $\str_b$ which chooses $b$ in each state.
  With both strategies, we accumulate a reward of exactly $1$.
  Therefore, if we choose to execute $\str_a$ with probability $0.5$ and $\str_b$ with the same probability, this process will lead to a reward of $1$ as well.

  Now, consider a strategy which chooses the action selected by $\str_a$ \emph{in each} state with probability $0.5$, and with the same probability chooses the action selected by $\str_b$.
  It is easy to see that this strategy only obtains a reward of $0.5\cdot 1 + 0.5\cdot 0.5\cdot 1 = 0.75$.
  As we see, this naive way of combining the two deterministic strategies into a memoryless randomised strategy is not correct.
\end{myexample}

Thus, the way to construct a memoryless randomised strategy is somewhat more involved.
We will have to compute the \emph{state-action frequencies}, that is the average number of times a given state-action pair is seen.

At first, we fix an arbitrary memoryless nature $\env \colon \Fpat \times \ActionSet \to \Disc{\States}$, that is, $\env \colon \States \times \ActionSet \to \Disc{\States}$.
The particular choice of $\env$ is not important, which is due to the fact that our algorithms are robust against any choice of nature.
We then let $x_{i}^{\str}(s)$ denote the probability to be in state $s$ at step $i$ when strategy $\str$ is used (using nature $\env$ and under the condition that we have started in $\InitState$).

For any $\str \in \Sigma$, we have  
%$x_{i+1}^{\str}(s) = \sum_{s' \in \States} \env(s', \str(s'))(s) x_{i}^{\str}(s')$
$x_{i}^{\str}(s) = \sum_{\setcond{\Pat \in \Fpat}{\last{\Pat} = s, \size{\Pat} = i}} \Prob^{\str,\env}[\Ifpat{\Pat}]$, which can be shown to be equivalent to the inductive form $x^{\str}_{0}(\InitState) = 1$ and $x^{\str}_{0}(s) = 0$ for $s \neq \InitState$, and $x_{i+1}^{\str}(s) = \sum_{s' \in \States} \env(s', \str(s'))(s) \cdot x_{i}^{\str}(s')$.

The state-action frequency $y^{\str}(s,a)$ is the number of times action $a$ is chosen in state $s$ when using strategy $\str$.
We then have that $y^{\str}(s,a) = \sum_{i=0}^{\infty} x_{i}^{\str}(s) \str(s,a)$.
Thus, state-action frequencies can be approximated using a simple value iteration scheme.
The \emph{mixed state-action frequency} $y(s,a)$ is the average over all state action frequencies weighted by the probability with which a given strategy is executed.
Thus, $y(s,a) = \sum_{i=1}^{m} p_{i} y^{\str_{i}}(s,a)$ for all $s,a$.
To construct a memoryless randomised strategy $\str$, we normalise the probabilities to $\str(s,a) = \frac{y(s,a)}{\sum_{b \in \ActionSet} y(s,b)}$ for all $s,a$ (see also the description for the computation of strategies/adversaries below Proposition 4 of~\cite{FKNPQ11}).

\begin{myexample}
  In the model of Fig.~\ref{fig:randomised-strategies}, we have
  $y^{\str_a}(s,a) = 1$,
  $y^{\str_a}(s,b) = 0$,
  $y^{\str_a}(u,a) = 0$,
  $y^{\str_a}(u,b) = 0$,
  $y^{\str_b}(s,a) = 0$,
  $y^{\str_b}(s,b) = 1$,
  $y^{\str_b}(u,a) = 0$, and
  $y^{\str_b}(u,b) = 1$.
  If we choose both $\str_a$ and $\str_b$ with probability $0.5$, we obtain the mixed state-action frequencies
  $y(s,a) = 0.5$,
  $y(s,b) = 0.5$,
  $y(u,a) = 0$, and
  $y(u,b) = 0.5$.
  The memoryless randomised strategy $\str$ we can construct is then
  $\str(s,a) = 0.5$,
  $\str(s,b) = 0.5$,
  $\str(u,a) = 0$,
  $\str(u,b) = 1$,
  which indeed achieves a reward of $1$.
\end{myexample}

For the general case where $k_{i} < \infty$ for some $k$, we have to work with counting deterministic strategies and natures.
Let $k_{\max}$ be the largest non-infinite step bound.
The usage of memory is unavoidable here because it is required already in case of a single objective.
To achieve optimal values, the computed strategies have to be able to make their decision dependent on how many steps are left before the step bound is reached.
Thus, we have strategies of the form $\str_{i} \colon \States \times \setnocond{0, \dotsc, k_{\max}} \to \ActionSet$
or equivalently
$\str_{i} \colon \States \times \setnocond{0, \dotsc, k_{\max}} \times \ActionSet \to \setnocond{0,1}$
where $\str_{i}(s, j, a) = 1$ if $\str_{i}(s, j) = a$ and $\str_{i}(s, j, \functionDot) = 0$ otherwise.
For step $i$ with $i < k_\mathrm{max}$, a strategy $\str$ chooses action $\str(s, i)$ for state $s$ whereas for all $i \geq k_\mathrm{max}$ the decision $\str(s, k_\mathrm{max})$ is used.
Natures are of the form $\env \colon \States \times \ActionSet \times \setnocond{0, \dotsc, k_{\max}}  \to \Disc{\States}$.
The computation of the randomised strategy changes accordingly:
for any $\str \in \Sigma$, we have
$x^{\str}_{0}(\InitState) = 1$ and
$x^{\str}_{0}(s) = 0$ for $s \neq \InitState$, and
$x_{i+1}^{\str}(s) = \sum_{s' \in \States} \env(s', \str(s', i'), i')(s) x_{i'}^{\str}(s')$
where $i' = \min\setnocond{ i, k_{\max} }$.
Also the state-action frequencies are now defined as step-dependent.
For $i \in \setnocond{0, \dotsc, k_{\max} - 1}$ we define
$y^{\str}(s,i,a) = x_{i}^{\str}(s) \str(s,i,a)$
and
$y^{\str}(s,k_{\max},a) = \sum_{i\geq k_{\max}} x_{i}^{\str}(s) \str(s,a)$.

The mixed state-action frequency is then $y(s,i,a) = \sum_{j=1}^{m} p_{j} y^{\str_{j}}(s,i,a)$.
Again using normalisation we define the counting randomised strategy
$\str(s,i,a) = \frac{y(s,i,a)}{\sum_{b \in \ActionSet} y(s,i,b)}$.
Here, for step $i$ with $i < k_{\max}$ we use decisions from $\str(\functionDot, i, \functionDot)$ while for $i \geq k_{\max}$ we use decisions from $\str(\functionDot, k_{\max}, \functionDot)$.

The bounded step case can be derived from the unbounded step case in the following sense:
we can transform the \MDP{} and the predicate into an \emph{unrolled} \MDP.
Here, we encode the step bounds in the state space as follows:
we copy the state space $\States$ a number of $k_\mathrm{max}+1$ times to a new state space $\States_\mathrm{unrolled} = \dot{\bigcup}_{i\in \{0, \ldots, k_\mathrm{max}\}} \States_i$.
We call each set of states $\States_i$ a \emph{layer}.
For each state $s\in \States$ and $i\in \{0, \ldots, k_\mathrm{max}\}$ we have $s_i \in \States_i$.
If we have a transition from a state $s$ to a state $s'$, in the unrolled \MDP{} for all $i\in \{0, \ldots, k_\mathrm{max}-1\}$ we have an according transition from $s_i$ to $s_{i+1}$ instead.
We also have a transition from $s_{k_\mathrm{max}}$ to $s'_{k_\mathrm{max}}$.
Formally, for $i < k_\mathrm{max}$ we have $\intTransitionProbability^\mathrm{unrolled}(s_i,a,s_{i+1}') = \intTransitionProbability(s,a,s')$ for some states $s, s'$ and some action $a$ and zero else,
and then $\intTransitionProbability^\mathrm{unrolled}(s_{k_\mathrm{max}},a,s_{k_\mathrm{max}}') = \intTransitionProbability(s,a,s')$.
Thus, there are only transitions from a one layer to the next layer, except for layer $k_\mathrm{max}$ which behaves like the original \MDP.

Reward structures are defined as follows.
We assume that each reward property uses a different reward structure.
For unbounded reward properties using reward structure $\rew$, we just let $\rew^\mathrm{unrolled}(s_i,a) = \rew(s,a)$ for all $i$ and states $s$.
For a step bounded reward property with bound $k$ we define a modified reward structure as follows:
for layers $0$ to $k-1$, the reward is obtained as usual, that is $\rew^\mathrm{unrolled}(s_i,a) = \rew(s,a)$ for $i \in \{0,\ldots, k-1\}$.
However, to simulate the step bound, we let $\rew(s_i,a) = 0$ for $i\geq k$.

By removing the step bound from predicate, we can now analyse the unrolled \MDP{} and obtain the same result as in the original \MDP{} using the original step bounded predicate.
As we are considering only unbounded properties, we obtain a set of memoryless deterministic strategies.
We can than construct a counting scheduler for the original model by mapping the layer number to the step number, that is
$\str(s,i,a) = \str^\mathrm{unrolled}(s_i,a)$.
In this way, we can show the correctness of the above scheduler computation for the step bounded case, because then also the values for the state action frequencies carry over, that is e.g. $y(s,i,a) = y^\mathrm{unrolled}(s_i,a)$.
Note that for $i < k_\mathrm{max}$ in $y^{\mathrm{unrolled},\str}(s_i,a) = \sum_{j=0}^{\infty} x_{j}^{\str}(s_i) \str(s_i,a)$ only the summand for $j=i$ is relevant.
This is the case because by construction of the unrolled \MDP{} for the other $j$ with $j\neq i$ it is $x_{j}^{\str}(s_i) = 0$.
Thus, $y^{\mathrm{unrolled},\str}(s_i,a) = x_{i}^{\str}(s_i) \str(s_i,a)$.
Accordingly, for
$y^{\mathrm{unrolled},\str}(s_{k_\mathrm{max}},a) = \sum_{j=0}^{\infty} x_{j}^{\str}(s_{k_\mathrm{max}}) \str(s_{k_\mathrm{max}},a)$ only $j$ with $j \geq k_\mathrm{max}$ are relevant and thus
\sloppy$y^{\mathrm{unrolled},\str}(s_{k_\mathrm{max}},a) = \sum_{j\geq k_\mathrm{max}}^{\infty} x_{j}^{\str}(s_{k_\mathrm{max}}) \str(s_{k_\mathrm{max}},a)$.

%% file: museum-strategy-fewest.tex
\placeArrowne{0}{0}
\placeArrowne{0}{1}
\placeArrowne{1}{0}
\placeArrowne{0}{2}
\placeArrowne{1}{1}
\placeArrowne{2}{0}
\placeArrowne{0}{3}
\placeArrowne{1}{2}
\placeArrowne{2}{1}
\placeArrowne{3}{0}
\placeArrowne{0}{4}
\placeArrowne{1}{3}
\placeArrowne{2}{2}
\placeArrowne{3}{1}
\placeArrowne{4}{0}
\placeArrowne{0}{5}
\placeArrowne{1}{4}
\placeArrowne{2}{3}
\placeArrowne{3}{2}
\placeArrowne{4}{1}
\placeArrowne{5}{0}
\placeArrowne{0}{6}
\placeArrowne{1}{5}
\placeArrowne{2}{4}
\placeArrowne{3}{3}
\placeArrowne{4}{2}
\placeArrowne{5}{1}
\placeArrowne{6}{0}
\placeArrowne{0}{7}
\placeArrowne{1}{6}
\placeArrowne{2}{5}
\placeArrowne{3}{4}
\placeArrowne{4}{3}
\placeArrowne{5}{2}
\placeArrowne{6}{1}
\placeArrowne{7}{0}
\placeArrowse{0}{8}
\placeArrowne{1}{7}
\placeArrowne{2}{6}
\placeArrowne{3}{5}
\placeArrowne{4}{4}
\placeArrowne{5}{3}
\placeArrowne{6}{2}
\placeArrowne{7}{1}
\placeArrownw{8}{0}
\placeArrowse{0}{9}
\placeArrowse{1}{8}
\placeArrowne{2}{7}
\placeArrowne{3}{6}
\placeArrowne{4}{5}
\placeArrowne{5}{4}
\placeArrowne{6}{3}
\placeArrowne{7}{2}
\placeArrownw{8}{1}
\placeArrownw{9}{0}
\placeArrowse{0}{10}
\placeArrowse{1}{9}
\placeArrowse{2}{8}
\placeArrowne{3}{7}
\placeArrowne{4}{6}
\placeArrowne{5}{5}
\placeArrowne{6}{4}
\placeArrowne{7}{3}
\placeArrownw{8}{2}
\placeArrownw{9}{1}
\placeArrownw{10}{0}
\placeArrowse{0}{11}
\placeArrowse{1}{10}
\placeArrowse{2}{9}
\placeArrowse{3}{8}
\placeArrowne{4}{7}
\placeArrowne{5}{6}
\placeArrowne{6}{5}
\placeArrowne{7}{4}
\placeArrownw{8}{3}
\placeArrownw{9}{2}
\placeArrownw{10}{1}
\placeArrownw{11}{0}
\placeArrowse{0}{12}
\placeArrowse{1}{11}
\placeArrowse{2}{10}
\placeArrowse{3}{9}
\placeArrowne{4}{8}
\placeArrowne{5}{7}
\placeArrowne{6}{6}
\placeArrowne{7}{5}
\placeArrowne{8}{4}
\placeArrownw{9}{3}
\placeArrownw{10}{2}
\placeArrownw{11}{1}
\placeArrownw{12}{0}
\placeArrowse{0}{13}
\placeArrowse{1}{12}
\placeArrowse{2}{11}
\placeArrowse{3}{10}
\placeArrowne{4}{9}
\placeArrowne{5}{8}
\placeArrowne{6}{7}
\placeArrowne{7}{6}
\placeArrowne{8}{5}
\placeArrowne{9}{4}
\placeArrownw{10}{3}
\placeArrownw{11}{2}
\placeArrownw{12}{1}
\placeArrownw{13}{0}
\placeArrowse{1}{13}
\placeArrowse{2}{12}
\placeArrowse{3}{11}
\placeArrowne{4}{10}
\placeArrowne{5}{9}
\placeArrowne{6}{8}
\placeArrowne{7}{7}
\placeArrowne{8}{6}
\placeArrowne{9}{5}
\placeArrowne{10}{4}
\placeArrownw{11}{3}
\placeArrownw{12}{2}
\placeArrownw{13}{1}
\placeArrowse{2}{13}
\placeArrowse{3}{12}
\placeArrowne{4}{11}
\placeArrowne{5}{10}
\placeArrowne{6}{9}
\placeArrowne{7}{8}
\placeArrowne{8}{7}
\placeArrowne{9}{6}
\placeArrowne{10}{5}
\placeArrowne{11}{4}
\placeArrownw{12}{3}
\placeArrownw{13}{2}
\placeArrowse{3}{13}
\placeArrowne{4}{12}
\placeArrowne{5}{11}
\placeArrowne{6}{10}
\placeArrowne{7}{9}
\placeArrowne{8}{8}
\placeArrowne{9}{7}
\placeArrowne{10}{6}
\placeArrowne{11}{5}
\placeArrowne{12}{4}
\placeArrownw{13}{3}
\placeArrowse{4}{13}
\placeArrowne{5}{12}
\placeArrowne{6}{11}
\placeArrowne{7}{10}
\placeArrowne{8}{9}
\placeArrowne{9}{8}
\placeArrowne{10}{7}
\placeArrowne{11}{6}
\placeArrowne{12}{5}
\placeArrownw{13}{4}
\placeArrowse{5}{13}
\placeArrowne{6}{12}
\placeArrowne{7}{11}
\placeArrowne{8}{10}
\placeArrowne{9}{9}
\placeArrowne{10}{8}
\placeArrowne{11}{7}
\placeArrowne{12}{6}
\placeArrownw{13}{5}
\placeArrowse{6}{13}
\placeArrowne{7}{12}
\placeArrowne{8}{11}
\placeArrowne{9}{10}
\placeArrowne{10}{9}
\placeArrowne{11}{8}
\placeArrowne{12}{7}
\placeArrownw{13}{6}
\placeArrowse{7}{13}
\placeArrowne{8}{12}
\placeArrowne{9}{11}
\placeArrowne{10}{10}
\placeArrowne{11}{9}
\placeArrowne{12}{8}
\placeArrownw{13}{7}
\placeArrowse{8}{13}
\placeArrowne{9}{12}
\placeArrowne{10}{11}
\placeArrowne{11}{10}
\placeArrowne{12}{9}
\placeArrownw{13}{8}
\placeArrowse{9}{13}
\placeArrowne{10}{12}
\placeArrowne{11}{11}
\placeArrowne{12}{10}
\placeArrownw{13}{9}
\placeArrowse{10}{13}
\placeArrowne{11}{12}
\placeArrowne{12}{11}
\placeArrownw{13}{10}
\placeArrowse{11}{13}
\placeArrowne{12}{12}
\placeArrownw{13}{11}
\placeArrowse{12}{13}
\placeArrownw{13}{12}

%% file: museum-strategy-31-1.tex
\placeArrowne{0}{0}
\placeArrowne{0}{1}
\placeArrowne{1}{0}
\placeArrowne{0}{2}
\placeArrowne{1}{1}
\placeArrowne{2}{0}
\placeArrowne{0}{3}
\placeArrowne{1}{2}
\placeArrowne{2}{1}
\placeArrowne{3}{0}
\placeArrowne{0}{4}
\placeArrowne{1}{3}
\placeArrowne{2}{2}
\placeArrowne{3}{1}
\placeArrowne{4}{0}
\placeArrowne{0}{5}
\placeArrowne{1}{4}
\placeArrowne{2}{3}
\placeArrowne{3}{2}
\placeArrowne{4}{1}
\placeArrowne{5}{0}
\placeArrowne{0}{6}
\placeArrowne{1}{5}
\placeArrowne{2}{4}
\placeArrowne{3}{3}
\placeArrowne{4}{2}
\placeArrowne{5}{1}
\placeArrowne{6}{0}
\placeArrowne{0}{7}
\placeArrowne{1}{6}
\placeArrowne{2}{5}
\placeArrowne{3}{4}
\placeArrowne{4}{3}
\placeArrowne{5}{2}
\placeArrowne{6}{1}
\placeArrowne{7}{0}
\placeArrowse{0}{8}
\placeArrowne{1}{7}
\placeArrowne{2}{6}
\placeArrowne{3}{5}
\placeArrowne{4}{4}
\placeArrowne{5}{3}
\placeArrowne{6}{2}
\placeArrowne{7}{1}
\placeArrownw{8}{0}
\placeArrowse{0}{9}
\placeArrowse{1}{8}
\placeArrowne{2}{7}
\placeArrowne{3}{6}
\placeArrowne{4}{5}
\placeArrowne{5}{4}
\placeArrowne{6}{3}
\placeArrowne{7}{2}
\placeArrownw{8}{1}
\placeArrownw{9}{0}
\placeArrowse{0}{10}
\placeArrowse{1}{9}
\placeArrowse{2}{8}
\placeArrowne{3}{7}
\placeArrowne{4}{6}
\placeArrowne{5}{5}
\placeArrowne{6}{4}
\placeArrowne{7}{3}
\placeArrownw{8}{2}
\placeArrownw{9}{1}
\placeArrownw{10}{0}
\placeArrowse{0}{11}
\placeArrowse{1}{10}
\placeArrowse{2}{9}
\placeArrowse{3}{8}
\placeArrowne{4}{7}
\placeArrowne{5}{6}
\placeArrowne{6}{5}
\placeArrowne{7}{4}
\placeArrownw{8}{3}
\placeArrownw{9}{2}
\placeArrownw{10}{1}
\placeArrownw{11}{0}
\placeArrowse{0}{12}
\placeArrowse{1}{11}
\placeArrowse{2}{10}
\placeArrowse{3}{9}
\placeArrowne{4}{8}
\placeArrowne{5}{7}
\placeArrowne{6}{6}
\placeArrowne{7}{5}
\placeArrowne{8}{4}
\placeArrownw{9}{3}
\placeArrownw{10}{2}
\placeArrownw{11}{1}
\placeArrownw{12}{0}
\placeArrowse{0}{13}
\placeArrowse{1}{12}
\placeArrowse{2}{11}
\placeArrowse{3}{10}
\placeArrowne{4}{9}
\placeArrowne{5}{8}
\placeArrowne{6}{7}
\placeArrowne{7}{6}
\placeArrowne{8}{5}
\placeArrowne{9}{4}
\placeArrownw{10}{3}
\placeArrownw{11}{2}
\placeArrownw{12}{1}
\placeArrownw{13}{0}
\placeArrowse{1}{13}
\placeArrowse{2}{12}
\placeArrowse{3}{11}
\placeArrowne{4}{10}
\placeArrowne{5}{9}
\placeArrowne{6}{8}
\placeArrowne{7}{7}
\placeArrowne{8}{6}
\placeArrowne{9}{5}
\placeArrowne{10}{4}
\placeArrownw{11}{3}
\placeArrownw{12}{2}
\placeArrownw{13}{1}
\placeArrowse{2}{13}
\placeArrowse{3}{12}
\placeArrowne{4}{11}
\placeArrowne{5}{10}
\placeArrowne{6}{9}
\placeArrowne{7}{8}
\placeArrowne{8}{7}
\placeArrowne{9}{6}
\placeArrowne{10}{5}
\placeArrowne{11}{4}
\placeArrownw{12}{3}
\placeArrownw{13}{2}
\placeArrowse{3}{13}
\placeArrowne{4}{12}
\placeArrowne{5}{11}
\placeArrowne{6}{10}
\placeArrowne{7}{9}
\placeArrowne{8}{8}
\placeArrowne{9}{7}
\placeArrowne{10}{6}
\placeArrowne{11}{5}
\placeArrowne{12}{4}
\placeArrownw{13}{3}
\placeArrowse{4}{13}
\placeArrowne{5}{12}
\placeArrowne{6}{11}
\placeArrowne{7}{10}
\placeArrowne{8}{9}
\placeArrowne{9}{8}
\placeArrowne{10}{7}
\placeArrowne{11}{6}
\placeArrowne{12}{5}
\placeArrownw{13}{4}
\placeArrowse{5}{13}
\placeArrowne{6}{12}
\placeArrowne{7}{11}
\placeArrowne{8}{10}
\placeArrowne{9}{9}
\placeArrowne{10}{8}
\placeArrowne{11}{7}
\placeArrowne{12}{6}
\placeArrownw{13}{5}
\placeArrowse{6}{13}
\placeArrowne{7}{12}
\placeArrowne{8}{11}
\placeArrowne{9}{10}
\placeArrowne{10}{9}
\placeArrowne{11}{8}
\placeArrowne{12}{7}
\placeArrownw{13}{6}
\placeArrowse{7}{13}
\placeArrowne{8}{12}
\placeArrowne{9}{11}
\placeArrowne{10}{10}
\placeArrowne{11}{9}
\placeArrowne{12}{8}
\placeArrownw{13}{7}
\placeArrowse{8}{13}
\placeArrowne{9}{12}
\placeArrowne{10}{11}
\placeArrowne{11}{10}
\placeArrowne{12}{9}
\placeArrownw{13}{8}
\placeArrowse{9}{13}
\placeArrowne{10}{12}
\placeArrowne{11}{11}
\placeArrowne{12}{10}
\placeArrownw{13}{9}
\placeArrowse{10}{13}
\placeArrowne{11}{12}
\placeArrowne{12}{11}
\placeArrownw{13}{10}
\placeArrowse{11}{13}
\placeArrowne{12}{12}
\placeArrownw{13}{11}
\placeArrowse{12}{13}
\placeArrownw{13}{12}

%% file: museum-strategy-31-2.tex
\placeArrowne{0}{0}
\placeArrowne{0}{1}
\placeArrowne{1}{0}
\placeArrowne{0}{2}
\placeArrowne{1}{1}
\placeArrowne{2}{0}
\placeArrowne{0}{3}
\placeArrowne{1}{2}
\placeArrowne{2}{1}
\placeArrowne{3}{0}
\placeArrowne{0}{4}
\placeArrowne{1}{3}
\placeArrowne{2}{2}
\placeArrowne{3}{1}
\placeArrowne{4}{0}
\placeArrowne{0}{5}
\placeArrowne{1}{4}
\placeArrowne{2}{3}
\placeArrowne{3}{2}
\placeArrowne{4}{1}
\placeArrowne{5}{0}
\placeArrowne{0}{6}
\placeArrowne{1}{5}
\placeArrowne{2}{4}
\placeArrowne{3}{3}
\placeArrowne{4}{2}
\placeArrowne{5}{1}
\placeArrowne{6}{0}
\placeArrowne{0}{7}
\placeArrowne{1}{6}
\placeArrowne{2}{5}
\placeArrowne{3}{4}
\placeArrowne{4}{3}
\placeArrowne{5}{2}
\placeArrowne{6}{1}
\placeArrowne{7}{0}
\placeArrowne{0}{8}
\placeArrowne{1}{7}
\placeArrowne{2}{6}
\placeArrowne{3}{5}
\placeArrowne{4}{4}
\placeArrowne{5}{3}
\placeArrowne{6}{2}
\placeArrowne{7}{1}
\placeArrowne{8}{0}
\placeArrowse{0}{9}
\placeArrowne{1}{8}
\placeArrowne{2}{7}
\placeArrowne{3}{6}
\placeArrowne{4}{5}
\placeArrowne{5}{4}
\placeArrowne{6}{3}
\placeArrowne{7}{2}
\placeArrowne{8}{1}
\placeArrowne{9}{0}
\placeArrowse{0}{10}
\placeArrowse{1}{9}
\placeArrowne{2}{8}
\placeArrowne{3}{7}
\placeArrowne{4}{6}
\placeArrowne{5}{5}
\placeArrowne{6}{4}
\placeArrowne{7}{3}
\placeArrowne{8}{2}
\placeArrowne{9}{1}
\placeArrowne{10}{0}
\placeArrowse{0}{11}
\placeArrowse{1}{10}
\placeArrowse{2}{9}
\placeArrowse{3}{8}
\placeArrowne{4}{7}
\placeArrowne{5}{6}
\placeArrowne{6}{5}
\placeArrowse{7}{4}
\placeArrowse{8}{3}
\placeArrowne{9}{2}
\placeArrowne{10}{1}
\placeArrowne{11}{0}
\placeArrowse{0}{12}
\placeArrowse{1}{11}
\placeArrowse{2}{10}
\placeArrowse{3}{9}
\placeArrowne{4}{8}
\placeArrowne{5}{7}
\placeArrowne{6}{6}
\placeArrowne{7}{5}
\placeArrowse{8}{4}
\placeArrowne{9}{3}
\placeArrowne{10}{2}
\placeArrowne{11}{1}
\placeArrowne{12}{0}
\placeArrowse{0}{13}
\placeArrowse{1}{12}
\placeArrowse{2}{11}
\placeArrowse{3}{10}
\placeArrowne{4}{9}
\placeArrowne{5}{8}
\placeArrowne{6}{7}
\placeArrowne{7}{6}
\placeArrowse{8}{5}
\placeArrowne{9}{4}
\placeArrowne{10}{3}
\placeArrowne{11}{2}
\placeArrowne{12}{1}
\placeArrownw{13}{0}
\placeArrowse{1}{13}
\placeArrowse{2}{12}
\placeArrowse{3}{11}
\placeArrowne{4}{10}
\placeArrowne{5}{9}
\placeArrowne{6}{8}
\placeArrowne{7}{7}
\placeArrowne{8}{6}
\placeArrowse{9}{5}
\placeArrowne{10}{4}
\placeArrowne{11}{3}
\placeArrowne{12}{2}
\placeArrownw{13}{1}
\placeArrowse{2}{13}
\placeArrowse{3}{12}
\placeArrowne{4}{11}
\placeArrowne{5}{10}
\placeArrowne{6}{9}
\placeArrowne{7}{8}
\placeArrowne{8}{7}
\placeArrowse{9}{6}
\placeArrowse{10}{5}
\placeArrowne{11}{4}
\placeArrowne{12}{3}
\placeArrownw{13}{2}
\placeArrowse{3}{13}
\placeArrowne{4}{12}
\placeArrowne{5}{11}
\placeArrowne{6}{10}
\placeArrowne{7}{9}
\placeArrowne{8}{8}
\placeArrowne{9}{7}
\placeArrowse{10}{6}
\placeArrowse{11}{5}
\placeArrowne{12}{4}
\placeArrownw{13}{3}
\placeArrowse{4}{13}
\placeArrowne{5}{12}
\placeArrowne{6}{11}
\placeArrowne{7}{10}
\placeArrowne{8}{9}
\placeArrowne{9}{8}
\placeArrowne{10}{7}
\placeArrowse{11}{6}
\placeArrowne{12}{5}
\placeArrownw{13}{4}
\placeArrowse{5}{13}
\placeArrowne{6}{12}
\placeArrowne{7}{11}
\placeArrowne{8}{10}
\placeArrowne{9}{9}
\placeArrowne{10}{8}
\placeArrowne{11}{7}
\placeArrowne{12}{6}
\placeArrownw{13}{5}
\placeArrowse{6}{13}
\placeArrowne{7}{12}
\placeArrowne{8}{11}
\placeArrowne{9}{10}
\placeArrowne{10}{9}
\placeArrowne{11}{8}
\placeArrowne{12}{7}
\placeArrownw{13}{6}
\placeArrowse{7}{13}
\placeArrowne{8}{12}
\placeArrowne{9}{11}
\placeArrowne{10}{10}
\placeArrowne{11}{9}
\placeArrowne{12}{8}
\placeArrownw{13}{7}
\placeArrowse{8}{13}
\placeArrowne{9}{12}
\placeArrowne{10}{11}
\placeArrowne{11}{10}
\placeArrowne{12}{9}
\placeArrownw{13}{8}
\placeArrowse{9}{13}
\placeArrowne{10}{12}
\placeArrowne{11}{11}
\placeArrowne{12}{10}
\placeArrownw{13}{9}
\placeArrowse{10}{13}
\placeArrowne{11}{12}
\placeArrowne{12}{11}
\placeArrownw{13}{10}
\placeArrowse{11}{13}
\placeArrowne{12}{12}
\placeArrownw{13}{11}
\placeArrowse{12}{13}
\placeArrownw{13}{12}

%% file: museum-strategy-40-1.tex
\placeArrowne{0}{0}
\placeArrowne{0}{1}
\placeArrowne{1}{0}
\placeArrowne{0}{2}
\placeArrowne{1}{1}
\placeArrowne{2}{0}
\placeArrowne{0}{3}
\placeArrowne{1}{2}
\placeArrowne{2}{1}
\placeArrowne{3}{0}
\placeArrowne{0}{4}
\placeArrowne{1}{3}
\placeArrowne{2}{2}
\placeArrowne{3}{1}
\placeArrowne{4}{0}
\placeArrowse{0}{5}
\placeArrowne{1}{4}
\placeArrowne{2}{3}
\placeArrowne{3}{2}
\placeArrowne{4}{1}
\placeArrowne{5}{0}
\placeArrowse{0}{6}
\placeArrowse{1}{5}
\placeArrowne{2}{4}
\placeArrowne{3}{3}
\placeArrowne{4}{2}
\placeArrowne{5}{1}
\placeArrowne{6}{0}
\placeArrowse{0}{7}
\placeArrowse{1}{6}
\placeArrowse{2}{5}
\placeArrowne{3}{4}
\placeArrowne{4}{3}
\placeArrowne{5}{2}
\placeArrowne{6}{1}
\placeArrowne{7}{0}
\placeArrowse{0}{8}
\placeArrowse{1}{7}
\placeArrowse{2}{6}
\placeArrowse{3}{5}
\placeArrowse{4}{4}
\placeArrowne{5}{3}
\placeArrowne{6}{2}
\placeArrowne{7}{1}
\placeArrowne{8}{0}
\placeArrowse{0}{9}
\placeArrowse{1}{8}
\placeArrowse{2}{7}
\placeArrowse{3}{6}
\placeArrowse{4}{5}
\placeArrowse{5}{4}
\placeArrowne{6}{3}
\placeArrowne{7}{2}
\placeArrowne{8}{1}
\placeArrowne{9}{0}
\placeArrowse{0}{10}
\placeArrowse{1}{9}
\placeArrowne{2}{8}
\placeArrowse{3}{7}
\placeArrowse{4}{6}
\placeArrowse{5}{5}
\placeArrowse{6}{4}
\placeArrowne{7}{3}
\placeArrowne{8}{2}
\placeArrowne{9}{1}
\placeArrowne{10}{0}
\placeArrowse{0}{11}
\placeArrowse{1}{10}
\placeArrowse{2}{9}
\placeArrowse{3}{8}
\placeArrowse{4}{7}
\placeArrowse{5}{6}
\placeArrowse{6}{5}
\placeArrowse{7}{4}
\placeArrowse{8}{3}
\placeArrowne{9}{2}
\placeArrowne{10}{1}
\placeArrowne{11}{0}
\placeArrowse{0}{12}
\placeArrowse{1}{11}
\placeArrowse{2}{10}
\placeArrowse{3}{9}
\placeArrowne{4}{8}
\placeArrowne{5}{7}
\placeArrowse{6}{6}
\placeArrowse{7}{5}
\placeArrowse{8}{4}
\placeArrowne{9}{3}
\placeArrowne{10}{2}
\placeArrowne{11}{1}
\placeArrowne{12}{0}
\placeArrowse{0}{13}
\placeArrowse{1}{12}
\placeArrowse{2}{11}
\placeArrowse{3}{10}
\placeArrowse{4}{9}
\placeArrowne{5}{8}
\placeArrowne{6}{7}
\placeArrowse{7}{6}
\placeArrowse{8}{5}
\placeArrowne{9}{4}
\placeArrowne{10}{3}
\placeArrowne{11}{2}
\placeArrowne{12}{1}
\placeArrownw{13}{0}
\placeArrowse{1}{13}
\placeArrowse{2}{12}
\placeArrowse{3}{11}
\placeArrowse{4}{10}
\placeArrowse{5}{9}
\placeArrowne{6}{8}
\placeArrowne{7}{7}
\placeArrowse{8}{6}
\placeArrowse{9}{5}
\placeArrowne{10}{4}
\placeArrowne{11}{3}
\placeArrowne{12}{2}
\placeArrownw{13}{1}
\placeArrowse{2}{13}
\placeArrowse{3}{12}
\placeArrowse{4}{11}
\placeArrowse{5}{10}
\placeArrowse{6}{9}
\placeArrowne{7}{8}
\placeArrowne{8}{7}
\placeArrowse{9}{6}
\placeArrowse{10}{5}
\placeArrowne{11}{4}
\placeArrowne{12}{3}
\placeArrownw{13}{2}
\placeArrowse{3}{13}
\placeArrowse{4}{12}
\placeArrowse{5}{11}
\placeArrowse{6}{10}
\placeArrowse{7}{9}
\placeArrowne{8}{8}
\placeArrowse{9}{7}
\placeArrowsw{10}{6}
\placeArrowse{11}{5}
\placeArrowne{12}{4}
\placeArrownw{13}{3}
\placeArrowse{4}{13}
\placeArrowse{5}{12}
\placeArrowse{6}{11}
\placeArrowse{7}{10}
\placeArrowse{8}{9}
\placeArrowse{9}{8}
\placeArrowsw{10}{7}
\placeArrowse{11}{6}
\placeArrowne{12}{5}
\placeArrownw{13}{4}
\placeArrowse{5}{13}
\placeArrowse{6}{12}
\placeArrowse{7}{11}
\placeArrowse{8}{10}
\placeArrowse{9}{9}
\placeArrowsw{10}{8}
\placeArrowse{11}{7}
\placeArrowne{12}{6}
\placeArrownw{13}{5}
\placeArrowse{6}{13}
\placeArrowse{7}{12}
\placeArrowne{8}{11}
\placeArrowse{9}{10}
\placeArrowsw{10}{9}
\placeArrowne{11}{8}
\placeArrowne{12}{7}
\placeArrownw{13}{6}
\placeArrowse{7}{13}
\placeArrowne{8}{12}
\placeArrowne{9}{11}
\placeArrowne{10}{10}
\placeArrowne{11}{9}
\placeArrowne{12}{8}
\placeArrownw{13}{7}
\placeArrowse{8}{13}
\placeArrowne{9}{12}
\placeArrowne{10}{11}
\placeArrowne{11}{10}
\placeArrowne{12}{9}
\placeArrownw{13}{8}
\placeArrowse{9}{13}
\placeArrowne{10}{12}
\placeArrowne{11}{11}
\placeArrowne{12}{10}
\placeArrownw{13}{9}
\placeArrowse{10}{13}
\placeArrowne{11}{12}
\placeArrowne{12}{11}
\placeArrownw{13}{10}
\placeArrowse{11}{13}
\placeArrowne{12}{12}
\placeArrownw{13}{11}
\placeArrowse{12}{13}
\placeArrownw{13}{12}

%% file: museum-strategy-40-2.tex
\placeArrowne{0}{0}
\placeArrowne{0}{1}
\placeArrowne{1}{0}
\placeArrowne{0}{2}
\placeArrowne{1}{1}
\placeArrowne{2}{0}
\placeArrowne{0}{3}
\placeArrowne{1}{2}
\placeArrowne{2}{1}
\placeArrowne{3}{0}
\placeArrowne{0}{4}
\placeArrowne{1}{3}
\placeArrowne{2}{2}
\placeArrowne{3}{1}
\placeArrowne{4}{0}
\placeArrowne{0}{5}
\placeArrowne{1}{4}
\placeArrowne{2}{3}
\placeArrowne{3}{2}
\placeArrowne{4}{1}
\placeArrowne{5}{0}
\placeArrowne{0}{6}
\placeArrowne{1}{5}
\placeArrowne{2}{4}
\placeArrowne{3}{3}
\placeArrowne{4}{2}
\placeArrowne{5}{1}
\placeArrowne{6}{0}
\placeArrowne{0}{7}
\placeArrowne{1}{6}
\placeArrowne{2}{5}
\placeArrowne{3}{4}
\placeArrowne{4}{3}
\placeArrowne{5}{2}
\placeArrowne{6}{1}
\placeArrowne{7}{0}
\placeArrowne{0}{8}
\placeArrowne{1}{7}
\placeArrowne{2}{6}
\placeArrowne{3}{5}
\placeArrowne{4}{4}
\placeArrowne{5}{3}
\placeArrowne{6}{2}
\placeArrowne{7}{1}
\placeArrowne{8}{0}
\placeArrowse{0}{9}
\placeArrowne{1}{8}
\placeArrowne{2}{7}
\placeArrowne{3}{6}
\placeArrowne{4}{5}
\placeArrowne{5}{4}
\placeArrowne{6}{3}
\placeArrowne{7}{2}
\placeArrowne{8}{1}
\placeArrowne{9}{0}
\placeArrowse{0}{10}
\placeArrowse{1}{9}
\placeArrowne{2}{8}
\placeArrowne{3}{7}
\placeArrowne{4}{6}
\placeArrowne{5}{5}
\placeArrowne{6}{4}
\placeArrowne{7}{3}
\placeArrowne{8}{2}
\placeArrowne{9}{1}
\placeArrowne{10}{0}
\placeArrowse{0}{11}
\placeArrowse{1}{10}
\placeArrowse{2}{9}
\placeArrowse{3}{8}
\placeArrowne{4}{7}
\placeArrowne{5}{6}
\placeArrowne{6}{5}
\placeArrowse{7}{4}
\placeArrowse{8}{3}
\placeArrowne{9}{2}
\placeArrowne{10}{1}
\placeArrowne{11}{0}
\placeArrowse{0}{12}
\placeArrowse{1}{11}
\placeArrowse{2}{10}
\placeArrowse{3}{9}
\placeArrowne{4}{8}
\placeArrowne{5}{7}
\placeArrowne{6}{6}
\placeArrowne{7}{5}
\placeArrowse{8}{4}
\placeArrowne{9}{3}
\placeArrowne{10}{2}
\placeArrowne{11}{1}
\placeArrowne{12}{0}
\placeArrowse{0}{13}
\placeArrowse{1}{12}
\placeArrowse{2}{11}
\placeArrowse{3}{10}
\placeArrowne{4}{9}
\placeArrowne{5}{8}
\placeArrowne{6}{7}
\placeArrowne{7}{6}
\placeArrowse{8}{5}
\placeArrowne{9}{4}
\placeArrowne{10}{3}
\placeArrowne{11}{2}
\placeArrowne{12}{1}
\placeArrownw{13}{0}
\placeArrowse{1}{13}
\placeArrowse{2}{12}
\placeArrowse{3}{11}
\placeArrowne{4}{10}
\placeArrowne{5}{9}
\placeArrowne{6}{8}
\placeArrowne{7}{7}
\placeArrowne{8}{6}
\placeArrowse{9}{5}
\placeArrowne{10}{4}
\placeArrowne{11}{3}
\placeArrowne{12}{2}
\placeArrownw{13}{1}
\placeArrowse{2}{13}
\placeArrowse{3}{12}
\placeArrowne{4}{11}
\placeArrowne{5}{10}
\placeArrowne{6}{9}
\placeArrowne{7}{8}
\placeArrowne{8}{7}
\placeArrowse{9}{6}
\placeArrowse{10}{5}
\placeArrowne{11}{4}
\placeArrowne{12}{3}
\placeArrownw{13}{2}
\placeArrowse{3}{13}
\placeArrowne{4}{12}
\placeArrowne{5}{11}
\placeArrowne{6}{10}
\placeArrowne{7}{9}
\placeArrowne{8}{8}
\placeArrowne{9}{7}
\placeArrowse{10}{6}
\placeArrowse{11}{5}
\placeArrowne{12}{4}
\placeArrownw{13}{3}
\placeArrowse{4}{13}
\placeArrowne{5}{12}
\placeArrowne{6}{11}
\placeArrowne{7}{10}
\placeArrowne{8}{9}
\placeArrowne{9}{8}
\placeArrowne{10}{7}
\placeArrowse{11}{6}
\placeArrowne{12}{5}
\placeArrownw{13}{4}
\placeArrowse{5}{13}
\placeArrowne{6}{12}
\placeArrowne{7}{11}
\placeArrowne{8}{10}
\placeArrowne{9}{9}
\placeArrowne{10}{8}
\placeArrowne{11}{7}
\placeArrowne{12}{6}
\placeArrownw{13}{5}
\placeArrowse{6}{13}
\placeArrowne{7}{12}
\placeArrowne{8}{11}
\placeArrowne{9}{10}
\placeArrowne{10}{9}
\placeArrowne{11}{8}
\placeArrowne{12}{7}
\placeArrownw{13}{6}
\placeArrowse{7}{13}
\placeArrowne{8}{12}
\placeArrowne{9}{11}
\placeArrowne{10}{10}
\placeArrowne{11}{9}
\placeArrowne{12}{8}
\placeArrownw{13}{7}
\placeArrowse{8}{13}
\placeArrowne{9}{12}
\placeArrowne{10}{11}
\placeArrowne{11}{10}
\placeArrowne{12}{9}
\placeArrownw{13}{8}
\placeArrowse{9}{13}
\placeArrowne{10}{12}
\placeArrowne{11}{11}
\placeArrowne{12}{10}
\placeArrownw{13}{9}
\placeArrowse{10}{13}
\placeArrowne{11}{12}
\placeArrowne{12}{11}
\placeArrownw{13}{10}
\placeArrowse{11}{13}
\placeArrowne{12}{12}
\placeArrownw{13}{11}
\placeArrowse{12}{13}
\placeArrownw{13}{12}

%% file: museum-strategy-maxsteps.tex
\placeArrowne{0}{0}
\placeArrowne{0}{1}
\placeArrownw{1}{0}
\placeArrowne{0}{2}
\placeArrownw{1}{1}
\placeArrowne{2}{0}
\placeArrowne{0}{3}
\placeArrownw{1}{2}
\placeArrowne{2}{1}
\placeArrowne{3}{0}
\placeArrowne{0}{4}
\placeArrownw{1}{3}
\placeArrowne{2}{2}
\placeArrowne{3}{1}
\placeArrowne{4}{0}
\placeArrowne{0}{5}
\placeArrownw{1}{4}
\placeArrowne{2}{3}
\placeArrowne{3}{2}
\placeArrowne{4}{1}
\placeArrowne{5}{0}
\placeArrowne{0}{6}
\placeArrownw{1}{5}
\placeArrowne{2}{4}
\placeArrowne{3}{3}
\placeArrowne{4}{2}
\placeArrowne{5}{1}
\placeArrowne{6}{0}
\placeArrowne{0}{7}
\placeArrownw{1}{6}
\placeArrowne{2}{5}
\placeArrowne{3}{4}
\placeArrowne{4}{3}
\placeArrowne{5}{2}
\placeArrowne{6}{1}
\placeArrownw{7}{0}
\placeArrowne{0}{8}
\placeArrownw{1}{7}
\placeArrowne{2}{6}
\placeArrowse{3}{5}
\placeArrowne{4}{4}
\placeArrowne{5}{3}
\placeArrowne{6}{2}
\placeArrownw{7}{1}
\placeArrowne{8}{0}
\placeArrowse{0}{9}
\placeArrowne{1}{8}
\placeArrowne{2}{7}
\placeArrowse{3}{6}
\placeArrowse{4}{5}
\placeArrowne{5}{4}
\placeArrowne{6}{3}
\placeArrownw{7}{2}
\placeArrowne{8}{1}
\placeArrowne{9}{0}
\placeArrowse{0}{10}
\placeArrowse{1}{9}
\placeArrowne{2}{8}
\placeArrowse{3}{7}
\placeArrowsw{4}{6}
\placeArrowse{5}{5}
\placeArrowne{6}{4}
\placeArrownw{7}{3}
\placeArrowne{8}{2}
\placeArrowne{9}{1}
\placeArrowne{10}{0}
\placeArrowse{0}{11}
\placeArrowse{1}{10}
\placeArrowse{2}{9}
\placeArrowse{3}{8}
\placeArrowsw{4}{7}
\placeArrowse{5}{6}
\placeArrowne{6}{5}
\placeArrownw{7}{4}
\placeArrowse{8}{3}
\placeArrowne{9}{2}
\placeArrowne{10}{1}
\placeArrowne{11}{0}
\placeArrowse{0}{12}
\placeArrowse{1}{11}
\placeArrowse{2}{10}
\placeArrowse{3}{9}
\placeArrowsw{4}{8}
\placeArrowse{5}{7}
\placeArrowne{6}{6}
\placeArrownw{7}{5}
\placeArrowsw{8}{4}
\placeArrowne{9}{3}
\placeArrowne{10}{2}
\placeArrowne{11}{1}
\placeArrowne{12}{0}
\placeArrowse{0}{13}
\placeArrowse{1}{12}
\placeArrowse{2}{11}
\placeArrowse{3}{10}
\placeArrowsw{4}{9}
\placeArrowne{5}{8}
\placeArrowne{6}{7}
\placeArrownw{7}{6}
\placeArrownw{8}{5}
\placeArrowne{9}{4}
\placeArrowne{10}{3}
\placeArrowne{11}{2}
\placeArrowne{12}{1}
\placeArrownw{13}{0}
\placeArrowse{1}{13}
\placeArrowse{2}{12}
\placeArrowse{3}{11}
\placeArrowsw{4}{10}
\placeArrowne{5}{9}
\placeArrowne{6}{8}
\placeArrownw{7}{7}
\placeArrownw{8}{6}
\placeArrowse{9}{5}
\placeArrowne{10}{4}
\placeArrowne{11}{3}
\placeArrowne{12}{2}
\placeArrownw{13}{1}
\placeArrowse{2}{13}
\placeArrowse{3}{12}
\placeArrowsw{4}{11}
\placeArrowse{5}{10}
\placeArrowse{6}{9}
\placeArrowne{7}{8}
\placeArrowne{8}{7}
\placeArrowse{9}{6}
\placeArrowse{10}{5}
\placeArrowne{11}{4}
\placeArrowne{12}{3}
\placeArrownw{13}{2}
\placeArrowse{3}{13}
\placeArrowsw{4}{12}
\placeArrowse{5}{11}
\placeArrowse{6}{10}
\placeArrowse{7}{9}
\placeArrowne{8}{8}
\placeArrowse{9}{7}
\placeArrowsw{10}{6}
\placeArrowse{11}{5}
\placeArrowne{12}{4}
\placeArrownw{13}{3}
\placeArrowsw{4}{13}
\placeArrowse{5}{12}
\placeArrowse{6}{11}
\placeArrowse{7}{10}
\placeArrowse{8}{9}
\placeArrowse{9}{8}
\placeArrowsw{10}{7}
\placeArrowse{11}{6}
\placeArrowne{12}{5}
\placeArrownw{13}{4}
\placeArrowse{5}{13}
\placeArrowse{6}{12}
\placeArrowse{7}{11}
\placeArrowse{8}{10}
\placeArrowse{9}{9}
\placeArrowsw{10}{8}
\placeArrowse{11}{7}
\placeArrowne{12}{6}
\placeArrownw{13}{5}
\placeArrowse{6}{13}
\placeArrowse{7}{12}
\placeArrowse{8}{11}
\placeArrowse{9}{10}
\placeArrowsw{10}{9}
\placeArrowse{11}{8}
\placeArrowne{12}{7}
\placeArrownw{13}{6}
\placeArrowse{7}{13}
\placeArrowse{8}{12}
\placeArrowse{9}{11}
\placeArrowsw{10}{10}
\placeArrowse{11}{9}
\placeArrowne{12}{8}
\placeArrownw{13}{7}
\placeArrowse{8}{13}
\placeArrowse{9}{12}
\placeArrowsw{10}{11}
\placeArrowse{11}{10}
\placeArrowne{12}{9}
\placeArrownw{13}{8}
\placeArrowse{9}{13}
\placeArrowsw{10}{12}
\placeArrowse{11}{11}
\placeArrowne{12}{10}
\placeArrownw{13}{9}
\placeArrowsw{10}{13}
\placeArrowse{11}{12}
\placeArrowne{12}{11}
\placeArrownw{13}{10}
\placeArrowse{11}{13}
\placeArrowne{12}{12}
\placeArrownw{13}{11}
\placeArrowse{12}{13}
\placeArrownw{13}{12}

%% file: master.bbl
\begin{thebibliography}{10}

\bibitem{basset2014compositional}
N.~Basset, M.~Kwiatkowska, and C.~Wiltsche.
\newblock Compositional controller synthesis for stochastic games.
\newblock In {\em {CONCUR}}, pages 173--187. Springer, 2014.

\bibitem{BenediktLW13}
M.~Benedikt, R.~Lenhardt, and J.~Worrell.
\newblock {LTL} model checking of interval {Markov} chains.
\newblock In {\em TACAS}, pages 32--46, 2013.

\bibitem{BT97}
D.~Bertsimas and J.~N. Tsitsiklis.
\newblock {\em Introduction to Linear Optimization}.
\newblock Athena Scientific, 1997.

\bibitem{Billingsley1979}
P.~Billingsley.
\newblock {\em Probability and Measure}.
\newblock John Wiley and Sons, 1979.

\bibitem{boyd2004}
S.~Boyd and L.~Vandenberghe.
\newblock {\em Convex optimization}.
\newblock Cambridge university press, 2004.

\bibitem{CRI07}
A.~S. Cantino, D.~L. Roberts, and C.~L. Isbell.
\newblock Autonomous nondeterministic tour guides: improving quality of
  experience with {TTD-MDP}s.
\newblock In {\em {AAMAS}}, page~22, 2007.

\bibitem{CMH06}
K.~Chatterjee, R.~Majumdar, and T.~A. Henzinger.
\newblock {Markov} decision processes with multiple objectives.
\newblock In {\em {STACS}}, volume 3884 of {\em LNCS}, pages 325--336, 2006.

\bibitem{ChatterjeeSH08}
K.~Chatterjee, K.~Sen, and T.~A. Henzinger.
\newblock Model-checking $\omega$-regular properties of interval {Markov}
  chains.
\newblock In {\em FoSSaCS}, pages 302--317, 2008.

\bibitem{chen2013stochastic}
T.~Chen, V.~Forejt, M.~Kwiatkowska, A.~Simaitis, and C.~Wiltsche.
\newblock On stochastic games with multiple objectives.
\newblock In {\em MFCS}, pages 266--277. Springer, 2013.

\bibitem{CHK13}
T.~Chen, T.~Han, and M.~Kwiatkowska.
\newblock On the complexity of model checking interval-valued discrete time
  {M}arkov chains.
\newblock {\em Inf. Process. Lett.}, 113(7):210--216, 2013.

\bibitem{ehrgott2006}
M.~Ehrgott.
\newblock {\em Multicriteria optimization}.
\newblock Springer Science \& Business Media, 2006.

\bibitem{EKNPY12}
M.-A. Esteve, J.-P. Katoen, V.~Y. Nguyen, B.~Postma, and Y.~Yushtein.
\newblock Formal correctness, safety, dependability and performance analysis of
  a satellite.
\newblock In {\em {ICSE}}, pages 1022--1031, 2012.

\bibitem{etessami2007}
K.~Etessami, M.~Kwiatkowska, M.~Y. Vardi, and M.~Yannakakis.
\newblock Multi-objective model checking of {Markov} decision processes.
\newblock In {\em {TACAS}}, pages 50--65. Springer, 2007.

\bibitem{DBLP:conf/spin/FecherLW06}
H.~Fecher, M.~Leucker, and V.~Wolf.
\newblock Don't know in probabilistic systems.
\newblock In {\em SPIN}, volume 3925 of {\em LNCS}, pages 71--88. Springer,
  2006.

\bibitem{FKNPQ11}
V.~Forejt, M.~Kwiatkowska, G.~Norman, D.~Parker, and H.~Qu.
\newblock Quantitative multi-objective verification for probabilistic systems.
\newblock In {\em {TACAS}}, pages 112--127. Springer, 2011.

\bibitem{FKP12}
V.~Forejt, M.~Kwiatkowska, and D.~Parker.
\newblock Pareto curves for probabilistic model checking.
\newblock In {\em {ATVA}}, pages 317--332. Springer, 2012.

\bibitem{GivanLD00}
R.~Givan, S.~M. Leach, and T.~L. Dean.
\newblock Bounded-parameter {Markov} decision processes.
\newblock {\em Artif. Intell.}, 122(1-2):71--109, 2000.

\bibitem{HahnHZ11nfm}
E.~M. Hahn, T.~Han, and L.~Zhang.
\newblock Synthesis for {PCTL} in parametric {Markov} decision processes.
\newblock In {\em {NFM}}, volume 6617 of {\em LNCS}, pages 146--161, 2011.

\bibitem{hashemi2016reward}
V.~Hashemi, H.~Hermanns, and L.~Song.
\newblock Reward-bounded reachability probability for uncertain weighted
  {MDP}s.
\newblock In {\em {VMCAI}}, pages 351--371. Springer, 2016.

\bibitem{DBLP:conf/lics/JonssonL91}
B.~Jonsson and K.~G. Larsen.
\newblock Specification and refinement of probabilistic processes.
\newblock In {\em LICS}, pages 266--277. IEEE Computer Society, 1991.

\bibitem{DBLP:journals/rc/KozineU02}
I.~Kozine and L.~V. Utkin.
\newblock Interval-valued finite {Markov} chains.
\newblock {\em Reliable Computing}, 8(2):97--113, 2002.

\bibitem{KNPQ13}
M.~Kwiatkowska, G.~Norman, D.~Parker, and H.~Qu.
\newblock Compositional probabilistic verification through multi-objective
  model checking.
\newblock {\em Information and Computation}, 232:38--65, 2013.

\bibitem{Lahijanian:TAC:2015}
M.~Lahijanian, S.~B. Andersson, and C.~Belta.
\newblock Formal verification and synthesis for discrete-time stochastic
  systems.
\newblock {\em IEEE Transactions on Automatic Control}, 60(8):2031--2045, 2015.

\bibitem{Lahijanian:CDC:2016}
M.~Lahijanian and M.~Kwiatkowska.
\newblock Specification revision for {M}arkov decision processes with optimal
  trade-off.
\newblock In {\em Conf. on Decision and Control}, pages 7411--7418. IEEE, Dec.
  2016.

\bibitem{luna:wafr:2014}
R.~Luna, M.~Lahijanian, M.~Moll, and L.~E. Kavraki.
\newblock Asymptotically optimal stochastic motion planning with temporal
  goals.
\newblock In {\em {WAFR}}, pages 335--352, 2014.

\bibitem{luna:icra:2014}
R.~Luna, M.~Lahijanian, M.~Moll, and L.~E. Kavraki.
\newblock Fast stochastic motion planning with optimality guarantees using
  local policy reconfiguration.
\newblock In {\em ICRA}, pages 3013--3019, 2014.

\bibitem{luna:aaai:2014}
R.~Luna, M.~Lahijanian, M.~Moll, and L.~E. Kavraki.
\newblock Optimal and efficient stochastic motion planning in partially-known
  environments.
\newblock In {\em AAAI}, pages 2549--2555, 2014.

\bibitem{Mouaddib04}
A.~Mouaddib.
\newblock Multi-objective decision-theoretic plan problem.
\newblock In {\em ICRA}, volume~3, pages 2814--2819, 2004.

\bibitem{NilimG05}
A.~Nilim and L.~El~Ghaoui.
\newblock Robust control of {Markov} decision processes with uncertain
  transition matrices.
\newblock {\em Operations Research}, 53(5):780--798, 2005.

\bibitem{OgryczakPW13}
W.~Ogryczak, P.~Perny, and P.~Weng.
\newblock A compromise programming approach to multiobjective {Markov} decision
  processes.
\newblock {\em IJITDM}, 12(5):1021--1054, 2013.

\bibitem{PernyWGH13}
P.~Perny, P.~Weng, J.~Goldsmith, and J.~P. Hanna.
\newblock Approximation of {Lorenz}-optimal solutions in multiobjective
  {Markov} decision processes.
\newblock In {\em AAAI}, pages 92--94, 2013.

\bibitem{PuggelliThesis2014}
A.~Puggelli.
\newblock {\em Formal Techniques for the Verification and Optimal Control of
  Probabilistic Systems in the Presence of Modeling Uncertainties}.
\newblock PhD thesis, EECS Department, University of California, Berkeley, Aug
  2014.

\bibitem{PuggelliLSS13}
A.~Puggelli, W.~Li, A.~L. Sangiovanni-Vincentelli, and S.~A. Seshia.
\newblock Polynomial-time verification of {PCTL} properties of {MDP}s with
  convex uncertainties.
\newblock In {\em CAV}, pages 527--542, 2013.

\bibitem{RRS15}
M.~Randour, J.-F. Raskin, and O.~Sankur.
\newblock Percentile queries in multi-dimensional {Markov} decision processes.
\newblock In {\em {CAV}}, pages 123--139. Springer, 2015.

\bibitem{WolffTM12}
E.~M. Wolff, U.~Topcu, and R.~M. Murray.
\newblock Robust control of uncertain {Markov} decision processes with temporal
  logic specifications.
\newblock In {\em CDC}, pages 3372--3379, 2012.

\bibitem{WK08}
D.~Wu and X.~D. Koutsoukos.
\newblock Reachability analysis of uncertain systems using bounded parameter
  {Markov} decision processes.
\newblock {\em Artificial Intelligence}, 172(8-9):945–954, 2008.

\end{thebibliography}
